\documentclass[twocolumn,twoside]{IEEEtran}                 

\usepackage{subfigure}
\usepackage{bm}
\usepackage{graphicx}
\usepackage{amsmath, amsthm, amsfonts, amssymb, amsbsy}
\usepackage{setspace}
\usepackage{graphicx}
\usepackage{pstricks,arydshln}
\usepackage{multirow}
\usepackage{booktabs}
\usepackage{stfloats}
\usepackage{algorithmic,amscd,textcomp,amssymb,epsfig,graphics,epsf,color,amsmath,balance,algorithm,cite}
\usepackage{caption}
\usepackage{cite}
\usepackage{epstopdf}
\usepackage{dsfont}
\usepackage{color}
\usepackage{algorithm}
\usepackage{algorithmic}
\DeclareMathOperator{\cvec}{cvec}
\DeclareMathOperator{\rvec}{rvec}

\newtheorem{thm}{Theorem}

\begin{document}
\title{Multi-mode OAM Radio Waves: Generation, Angle of Arrival Estimation and Reception With UCAs}

\author{Rui Chen,\,\IEEEmembership{Member,\,IEEE,}\,Wen-Xuan Long,\,Xiaodong Wang,\,\IEEEmembership{Fellow,\,IEEE} and Jiandong Li,\,\IEEEmembership{Senior Member,\,IEEE}
\thanks{This work was supported in part by the Fundamental Research Funds for the Central Universities and the Innovation Fund of Xidian University.}
\thanks{R. Chen is with the State Key Laboratory of ISN, Xidian University, Xi'an 710071,
China, and also with the National Mobile Communications Research Laboratory,
Southeast University, Nanjing 210018, China (e-mail: rchen@xidian.edu.cn).}
\thanks{W.-X. Long and J. Li are with the State Key Laboratory of Integrated Service Networks (ISN), Xidian University, Shaanxi 710071,
China (e-mail: wxlong@stu.xidian.edu.cn, jdli@mail.xidian.edu.cn).}
\thanks{X. Wang is with the Electrical Engineering Department, Columbia University, New York, NY 10027 USA (e-mail: wangx@ee.columbia.edu).
}}

\maketitle

\thispagestyle{empty}
\begin{abstract}
Orbital angular momentum (OAM) at radio frequency (RF) provides a novel approach of multiplexing a set of orthogonal modes on the same frequency channel to achieve high spectrum efficiencies. However, there are still big challenges in the multi-mode OAM generation, OAM antenna alignment and OAM signal reception. To solve these problems, we propose an overall scheme of the line-of-sight multi-carrier and multi-mode OAM (LoS MCMM-OAM) communication based on uniform circular arrays (UCAs). First, we verify that UCA can generate multi-mode OAM radio beam with both the RF analog synthesis method and the baseband digital synthesis method. Then, for the considered UCA-based LoS MCMM-OAM communication system, a distance and AoA estimation method is proposed based on the two-dimensional ESPRIT (2-D ESPRIT) algorithm. A salient feature of the proposed LoS MCMM-OAM and LoS MCMM-OAM-MIMO systems is that the channel matrices are completely characterized by three parameters, namely, the azimuth angle, the elevation angle and the distance, independent of the numbers of subcarriers and antennas, which significantly reduces the burden by avoiding estimating large channel matrices, as traditional MIMO-OFDM systems. After that, we propose an OAM reception scheme including the beam steering with the estimated AoA and the amplitude detection with the estimated distance. At last, the proposed methods are extended to the LoS MCMM-OAM-MIMO system equipped with uniform concentric circular arrays (UCCAs). Both mathematical analysis and simulation results validate that the proposed OAM reception scheme can eliminate the effect of the misalignment error of a practical OAM channel and approaches the performance of an ideally aligned OAM channel.
\end{abstract}


\section{Introduction}
Currently, explosive growth of emerging services, such as high-definition (HD) video, virtual reality (VR) and auto-pilot driving applications requires higher and higher wireless data rate. To meet the requirement, more and more high frequency bands such as millimeter wave and terahertz bands are being licensed \cite{WRC}. Since radio frequency (RF) spectrum resources are scarce, besides exploiting more frequency bandwidth, innovative techniques to enhance spectrum efficiency (SE) have been explored, such as advanced coding, cognitive radio (CR) and massive multiple-input multiple-output (MIMO). In essence, all these techniques are based on planar electromagnetic (EM) waves physically. Since the discovery in 1992 that light beams with helical phase fronts can carry orbital angular momentum (OAM) \cite{Allen1992Orbital}, a significant research effort has been focused on vortex EM waves as a novel approach for multiplexing a set of orthogonal OAM modes on the same frequency channel and achieving high SEs \cite{Tamburini2012Encoding,Yan2014High,Ren2017Line,Zhang2017Mode,Chen2018A, Zhang2019Orbital,Chen2020}. However, there are still some technical challenges for practical application of line-of-sight (LoS) OAM wireless communications.

The precondition of achieving mode multiplexing is to generate multiple OAM modes simultaneously. Typical approaches of using planar phase plate \cite{Bennis2013Flat,Cheng2014Generation}, spiral phase plate (SPP) \cite{Turnbull1996The}, helicoidal parabolic antenna \cite{Mari2015Near} and uniform circular array (UCA) \cite{Mohammadi2010system} are easy to generate single-mode OAM wave due to simple antenna structures. However, radiating multi-mode OAM beams from a single aperture still has big challenges. So far, a few antenna structures have been proposed to generate multi-mode OAM waves by SPPs \cite{Yan2014High}, solid cylinder dielectric resonator antenna \cite{Liang2016Orbital}, a tri-mode concentric circularly polarized patch antenna \cite{Zhang2016Generation}, metallic traveling-wave ring-slot structure \cite{Zhang2017Four}, time-switched array structure \cite{Tennant2012Generation}, the uniform concentric circular array (UCCA) \cite{lee2018experimental} and metasurface \cite{Yu2016Generating,Sanming2018Dual}. Since antenna array is popular in modern wireless communication systems (e.g., 4G, 5G and Beyond 5G) and easy to steer the beam direction of OAM \cite{XieG2016,Chen2018Beam}, we focus on the UCA instead of other complicated antenna structures for OAM generation and reception in this paper.

Another challenge for practical application of LoS OAM wireless communication is that OAM requires perfect alignment between the transmit and receive antennas \cite{Xie2015Performance}. At least, the beam directions of the transmit and receive antennas have to be aligned to avoid large performance degradation \cite{Chen2018Beam}. To steer the beam direction of the receive antenna towards the angle of arrived beam, angle of arrival (AoA) estimation has to be performed first. Although traditional AoA estimation of planar EM waves has been well studied, the AoA estimation of vortex EM waves is rarely investigated in the field of wireless communications. Interestingly, a few papers on OAM-based radar claimed that with vortex EM waves radar could obtain azimuthal super-resolution. In \cite{Liu2015Orbital}, an OAM-based radar target detection model with UCA was established for the first time. To accurately perform azimuth estimation with limited number of subcarriers and OAM modes, a 2-D super-resolution OAM radar target detection method is proposed in \cite{Chen2018OAMradar} based on the estimating signal parameter via rotational invariance (ESPRIT) algorithm. Inspired by the OAM radar target detection, in this paper, we propose an accurate AoA estimation method for aligning the transmit and receive beam directions in LoS OAM wireless communications.

It is now generally accepted that the superiority of UCA-based OAM system to the MIMO system is the low transmitter and receiver complexity due to no need of channel state information (CSI) at either transmitter or receiver, and OAM dose not offer any additional gains in channel capacity \cite{Edfors2012Is,Oldoni2015,Zhang2017Mode}. However, in practical applications when the transmit and receive antennas are misaligned with even a small oblique angle, large performance degradation occurs. To deal with the problem, the transmit and receive beam steering approach is proposed in \cite{Chen2018Beam}, which is shown to be able to circumvent the performance deterioration in a single-mode OAM communication system. Therefore, it is reasonable to speculate that good reception performance could also be obtained in a LoS multi-mode OAM communication system by the beam steering method with highly accurate AoA estimate.

\begin{figure}[t]
\setlength{\abovecaptionskip}{0cm}   
\setlength{\belowcaptionskip}{-0.3cm}   
\centering
\subfigure[]{
\includegraphics[width=4.1cm,height=4.1cm]{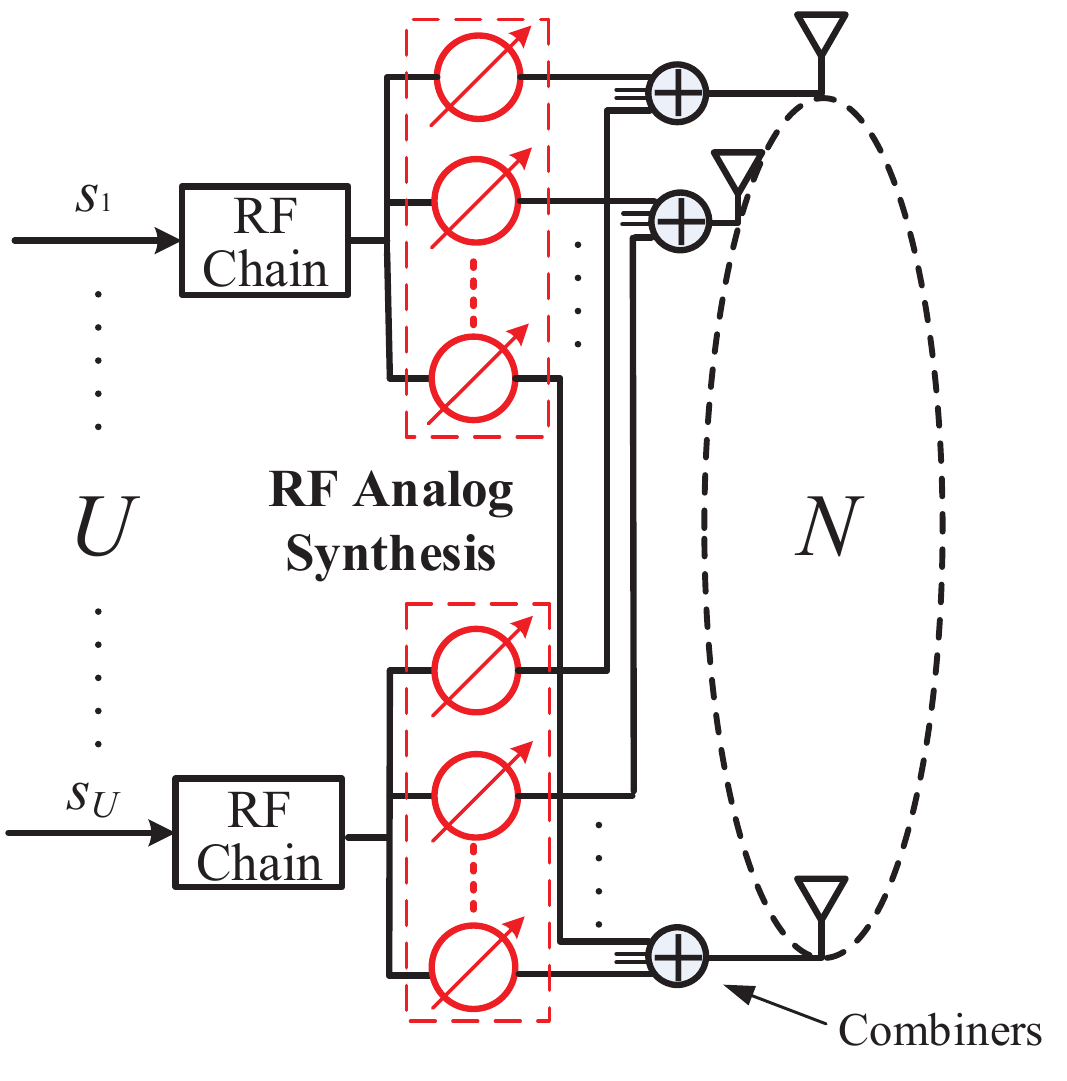}}
\subfigure[]{
\includegraphics[width=3.6cm,height=4.1cm]{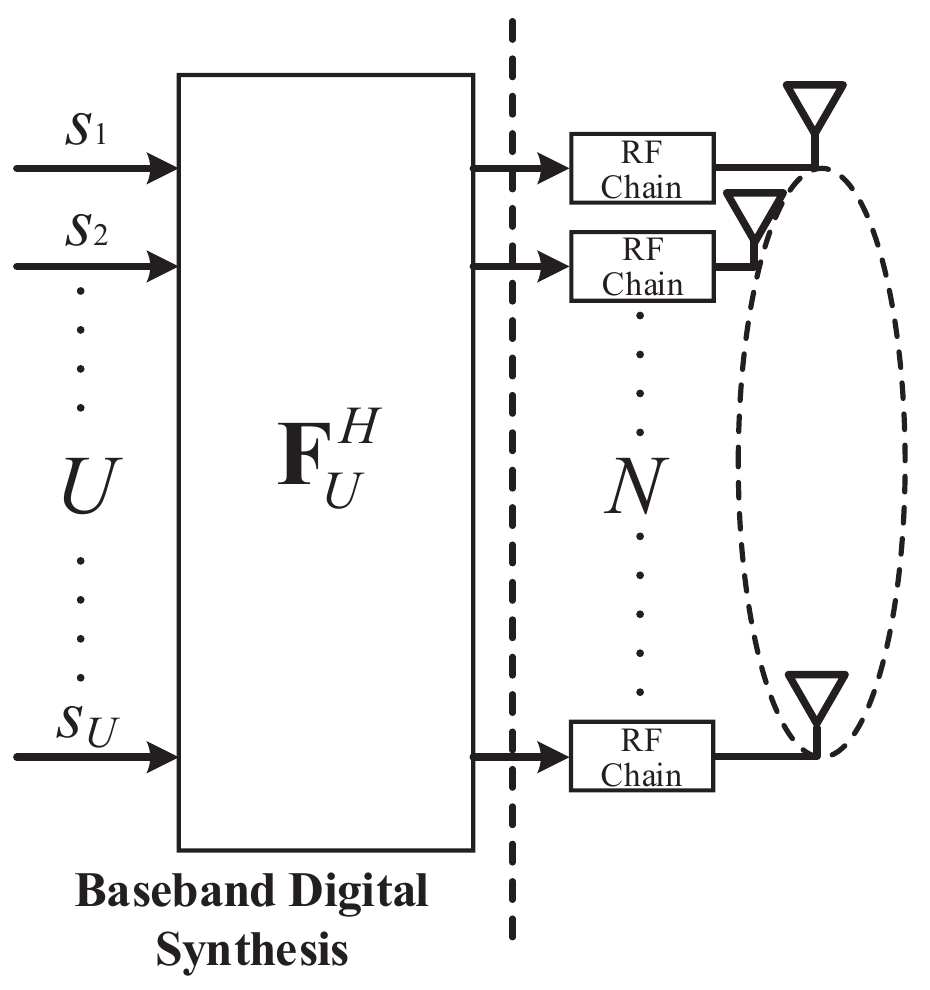}}
\caption{UCA-based multi-mode OAM transmitter implemented by (a) RF analog synthesis with phase shifters and power combiners, and (b) baseband digital synthesis.}
\label{Fig1}
\end{figure}

In this paper, to deal with the aforementioned challenges, we consider the LoS multi-carrier and multi-mode OAM/OAM-MIMO (MCMM-OAM/MCMM-OAM-MIMO) communication schemes including the generation, the distance and AoA estimation and the reception of multi-mode OAM waves.
The novelty and major contributions of this paper are summarized as follows:
\begin{itemize}
\item
First, we verify that UCA can generate multi-mode OAM radio beams with the proposed RF analog synthesis method and baseband digital synthesis method as shown in Fig. \ref{Fig1}. Through EM simulation, we show that phase distributions of UCA-based multi-mode OAM beam coincide with the theoretical superimposed phase distributions of multiple OAM modes.
\item
Second, we propose a UCA-based LoS MCMM-OAM communication scheme, and formulate the signal models of training and data transmission stages. The channel matrix of this  system is parametrized by only three parameters, i.e., azimuth angle, the elevation angle and the distance. Then, we propose an effective method to estimate these three parameters based on the two-dimensional ESPRIT (2-D ESPRIT) algorithm. Hence, compared with the traditional MIMO-OFDM system, there is no need to estimate large channel matrices, greatly reducing the required training overhead especially at sub 6 GHz bands.
\item
Third, we propose an OAM reception scheme including the beam steering with the estimated AoA and the amplitude detection with the estimated distance. We prove that the proposed OAM reception scheme can eliminate the inter-mode interference induced by the misalignment error in a practical OAM channel and approaches the performance of an ideally aligned OAM channel. Furthermore, we extend the proposed methods to the UCCA-based LoS MCMM-OAM-MIMO system, for which again the channel matrices are parameterized by only three parameters, azimuth angle, the elevation angle and the distance.
\end{itemize}

The remainder of this paper is organized as follows. In Section II, we verify the RF analog and baseband digital synthesis methods for multi-mode OAM beam generation. Based on the feasibility of generating multi-mode OAM beam by baseband digital method, we model the UCA-based LoS MCMM-OAM communication system in Section III. In Section IV, the distance and AoA estimation method based on the 2-D ESPRIT algorithm is proposed. With the estimated distance and AoA, an OAM reception scheme is proposed to correctly recover the transmitted data symbols in Section V. In Section VI, the proposed methods are extended to the UCCA-based LoS MCMM-OAM-MIMO system to achieve higher SE. Simulation results are shown in Section VII and conclusions are summarized in Section VIII.


\section{Generation of Muti-mode OAM With UCA}
The feasibility of utilizing UCA to generate OAM waves has been verified in theory \cite{Thid2007Utilization,Mohammadi2010system} and in practice \cite{Gaffoglio2016OAM,Gong2017Generation}. The antenna elements in a UCA are fed with the same input signal, but with a successive phase shift from element to element such that after a full turn the phase has the increment of $2\pi\ell$, where $\ell$ is an unbounded integer termed as topological charge or OAM mode number \cite{Allen1992Orbital}.
In general, the phased UCA can generate only single-mode OAM beam, but with phase shifters network (PSN) and power combiners UCA is expected to generate multi-mode OAM beam \cite{Chen2018A} as shown in Fig.\ref{Fig1}(a).

\subsection{Superimposed Radiation of Multiple OAM Modes}
The transmitter generates OAM beams with the UCA consisting of $N$ equidistant antenna elements on a circle with radius of $R_t$. The phase of the $n$th antenna element $\phi_n=\ell\varphi_{n}$, where $\varphi_{n}=2\pi (n-1)/N$ is the azimuthal angle of the $n$th antenna element. Thus, for the point $T(\bar{r},\bar{\varphi},\bar{\alpha})$ in spherical coordinates in the direction of the OAM beam, the electric field vector can be expressed as \cite{Mohammadi2010system,Yuan2017Beam}
\begin{align}\label{E}
\bm{E}_\ell(\bar{r},\bar{\varphi},\bar{\alpha})&\approx\frac{\beta_t}{4\pi} \sum_{n=1}^{N}e^{i\ell\varphi_n}\frac{e^{ik|\bar{\bm{r}}-\bm{r'}_n|}} {\left|\bar{\bm{r}}-\bm{r'}_n\right|} \nonumber\\
&\approx\frac{\beta_t}{4\pi}\frac{e^{ik\bar{r}}}{\bar{r}}\sum_{n=1}^{N} e^{-i(\bm{k}\cdot\bm{r}_n-\ell\varphi_n)} \nonumber\\
&\approx\frac{\beta_tNe^{ik\bar{r}}e^{i\ell\bar{\varphi}}}{4\pi \bar{r}}i^{-\ell}{J_\ell}(kR_t\sin\bar{\alpha}),
\end{align}
where $\beta_t$ models all the constants relative to each transmit antenna element, $i=\sqrt{-1}$ is the imaginary unit, $\bm{k}$ is the wave vector, $\bar{\bm{r}}$ is the position vector of $T(\bar{r},\bar{\varphi},\bar{\alpha})$, and $J_{\ell}(\cdot)$ is $\ell$th-order Bessel function of the first kind. The far-field approximations are $|\bar{\bm{r}}-\bm{r}_n|\approx \bar{r}$ for amplitudes and $|\bar{\bm{r}}-\bm{r}_n|\approx \bar{r}-\bm{\hat{r}}\cdot\bm{r}_n$ for phases, where $\bm{\hat{r}}$ is the unit vector of $\bar{\bm{r}}$, $\bm{r}_n=R_t(\bm{\hat{x}}\cos\varphi_n+\bm{\hat{y}}\sin\varphi_n)$, $\bm{\hat{x}}$ and $\bm{\hat{y}}$ are the unit vectors of x-axis and y-axis of the coordinate system at the transmitter, respectively.
\begin{figure}[tb]
\setlength{\abovecaptionskip}{0cm}   
\setlength{\belowcaptionskip}{-0.2cm}   
\begin{center}
\includegraphics[width=4.5cm,height=2.0cm]{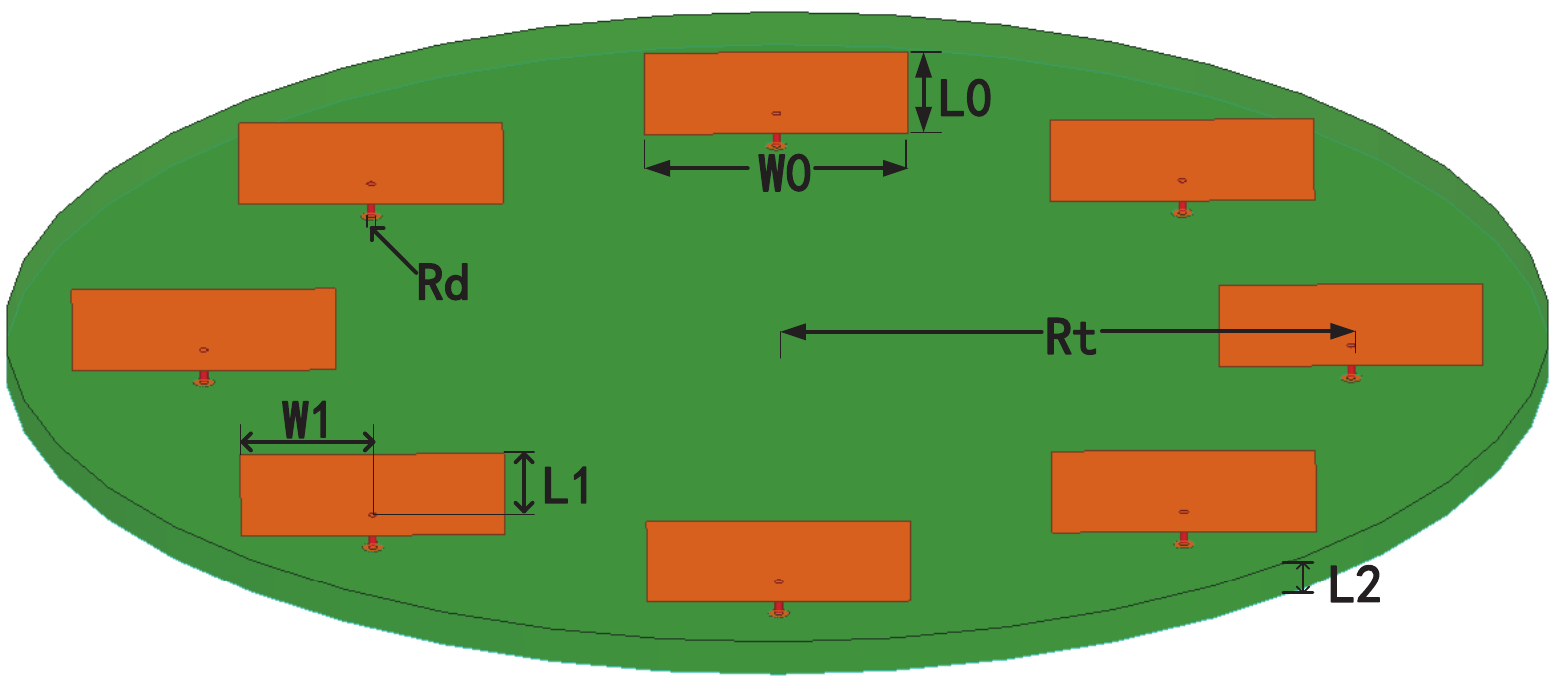}
\end{center}
\caption{The UCA model built in HFSS.}
\label{Fig2}
\end{figure}

According to the pattern multiplication principle, the electric field vector $\bm{E}_\ell(\bar{r},\bar{\varphi},\bar{\alpha})$ can be written as
\begin{align}\label{El}
\bm{E}_\ell(\bar{r},\bar{\varphi},\bar{\alpha})=\bm{E}_{\textrm{dipole}}(\bar{r})\Psi_{\ell}(\bar{\varphi},\bar{\alpha}),
\end{align}
where $\bm{E}_{\textrm{dipole}}(\bar{r})$ corresponds to electric dipole radiation, and the array factor denoted by $\Psi_{\ell}(\bar{\varphi},\bar{\alpha})$ takes the form
\begin{align}\label{Psi}
\Psi_{\ell}(\bar{\varphi},\bar{\alpha})=\sum_{n=1}^{N}e^{-i(\bm{k}\cdot\bm{r}_n-\phi_n)}.
\end{align}
Considering OAM mode multiplexing or multi-mode OAM beam, the electric field at the observation point $T(\bar{r},\bar{\varphi},\bar{\alpha})$ should be the superimposed electric fields of multiple single-mode OAM beams. Thus, for a $U$-mode OAM beam with each mode carrying one information symbol as shown in Fig. 1, the electric field vector $\bm{E}^{(U)}$ can be written as
\begin{align}\label{EL}
\bm{E}^{(U)}(\bar{r},\bar{\varphi},\bar{\alpha})&=\sum_{u=1}^U\bm{E}_{\ell_u}(\bar{r},\bar{\varphi},\bar{\alpha})s_u \nonumber\\
&=\bm{E}_{\textrm{dipole}}(\bar{r})\sum_{u=1}^U\Psi_{\ell_u}(\bar{\varphi},\bar{\alpha})s_u  \nonumber\\
&=\bm{E}_{\textrm{dipole}}(\bar{r})\sum_{n=1}^{N}A_ne^{-i(\bm{k}\cdot\bm{r}_n-\phi'_n)},
\end{align}
where $s_u$ is the data symbol transmitted on the $u$th OAM mode, $A_n$ and $\phi'_n$ satisfies $A_ne^{i\phi'_n}=\sum_{u=1}^Ue^{i\ell_u\varphi_n}s_u$. Comparing \eqref{El} and \eqref{Psi} with \eqref{EL}, we find that radiating $U$-mode OAM beam can be achieved by feeding the $n$th antenna with the amplitude $A_n$ and the phase $\phi'_n$, $n=1,2,\cdots,N$. Thus, the operation of multi-mode synthesis can be fulfilled in RF analog form as shown in Fig.\ref{Fig1}(a) or in baseband digital form as shown in Fig.\ref{Fig1}(b). We note that other RF analog and baseband digital synthesis methods of OAM beams have been applied in \cite{Xu2016Free} and \cite{Zhang2016Orbital}, respectively.

\subsection{Electromagnetic Simulations}
To verify both the synthesis methods of generating multi-mode OAM beam, we first build a UCA model in the high frequency structure simulator (HFSS) from ANSYS as shown in Fig. \ref{Fig2}. In this model, the UCA is formed by $N=8$ linearly polarised patch antennas designed at 2.45 GHz, the radius of the UCA is $0.66\lambda$ and $s_u=1 (u=1,2,\cdots,U)$. To avoid excessive computational burden of finite element calculations of HFSS, the transmission distance from the UCA to the observation plane is set as $2\lambda$. The detailed parameters are listed in Table \ref{table1}.
\par
Then, we compare the phase distributions of the electric fields radiated by the UCA in HFSS with the ideal phase distributions of Laguerre-Gaussian beams in Fig.\ref{Fig3}. It can be seen from the figure that the phase distributions of single-mode OAM beams with mode numbers $\ell=+1,+2,+3$ generated by the designed UCA coincide with those of ideal OAM beams in principle. Thus, the designed UCA in HFSS can be used to verify the superimposed phase distributions of multi-mode OAM beams.

\begin{table}[t]
\setlength{\abovecaptionskip}{0.3cm}   
\setlength{\belowcaptionskip}{-0.2cm}   
  \centering
  \caption{Parameters of the designed uniform circular array}\label{table1}
  \setlength{\tabcolsep}{5mm}{
\begin{tabular}{cc}
  \toprule
  \specialrule{0em}{1pt}{1pt}
  Parameter & Value(mm) \\
  \specialrule{0em}{1pt}{1pt}
  \hline
  \specialrule{0em}{1pt}{1pt}
  $W0$ &  $ 37.26$  \\
  \specialrule{0em}{1pt}{1pt}
  \hline
  \specialrule{0em}{1pt}{1pt}
  $L0$ & $ 28.1 $  \\
  \specialrule{0em}{1pt}{1pt}
  \hline
  \specialrule{0em}{1pt}{1pt}
  $W1$ &$ 18.63 $\\
  \specialrule{0em}{1pt}{1pt}
  \hline
  \specialrule{0em}{1pt}{1pt}
  $L1$ &$ 21.05 $  \\
  \specialrule{0em}{1pt}{1pt}
  \hline
  \specialrule{0em}{1pt}{1pt}
  $L2$ &$ 5 $  \\
  \specialrule{0em}{1pt}{1pt}
  \hline
  \specialrule{0em}{1pt}{1pt}
  $R_t$ &$ 81 $  \\
  \specialrule{0em}{1pt}{1pt}
  \hline
  \specialrule{0em}{1pt}{1pt}
  $R_d$ &$0.6$  \\
  \specialrule{0em}{1pt}{1pt}
  \bottomrule
\end{tabular}}
\end{table}

After that, we feed each antenna port with corresponding amplitude $A_n$ and phase $\phi'_n$ according to \eqref{EL} and compare the phase distributions of the electric fields radiated by the UCA in HFSS with the ideal superimposed phase distributions of two-mode OAM beams in Fig. \ref{Fig4}.
Comparing the subfigures at the left hand side and those at the right hand side, we confirm that UCA with either RF analog synthesis or baseband digital synthesis can generate multi-mode OAM beams. It is known that for the $N$-element UCA, at most $N$ OAM modes can be resolved, i.e., $|\ell| < N/2$ \cite{Mohammadi2010system}. Hence, the multi-mode OAM beam composed of more and higher-oder OAM modes could also be generated by a UCA with enough number of elements theoretically.

\begin{figure}[t]
\setlength{\abovecaptionskip}{0cm}   
\setlength{\belowcaptionskip}{-0.2cm}   
\centering
\subfigure[]{
\includegraphics[width=2.7cm]{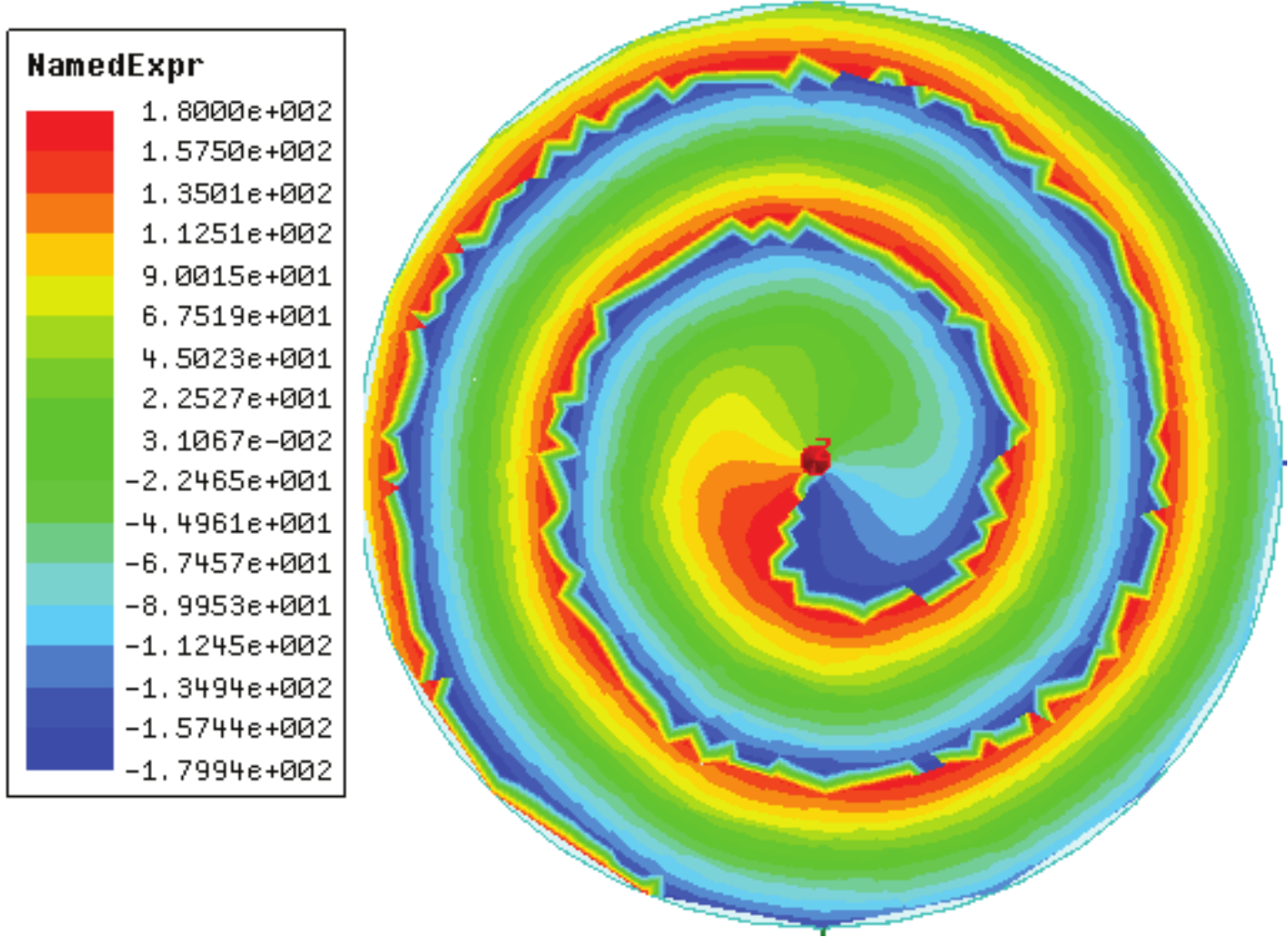}}
\subfigure[]{
\includegraphics[width=2.6cm]{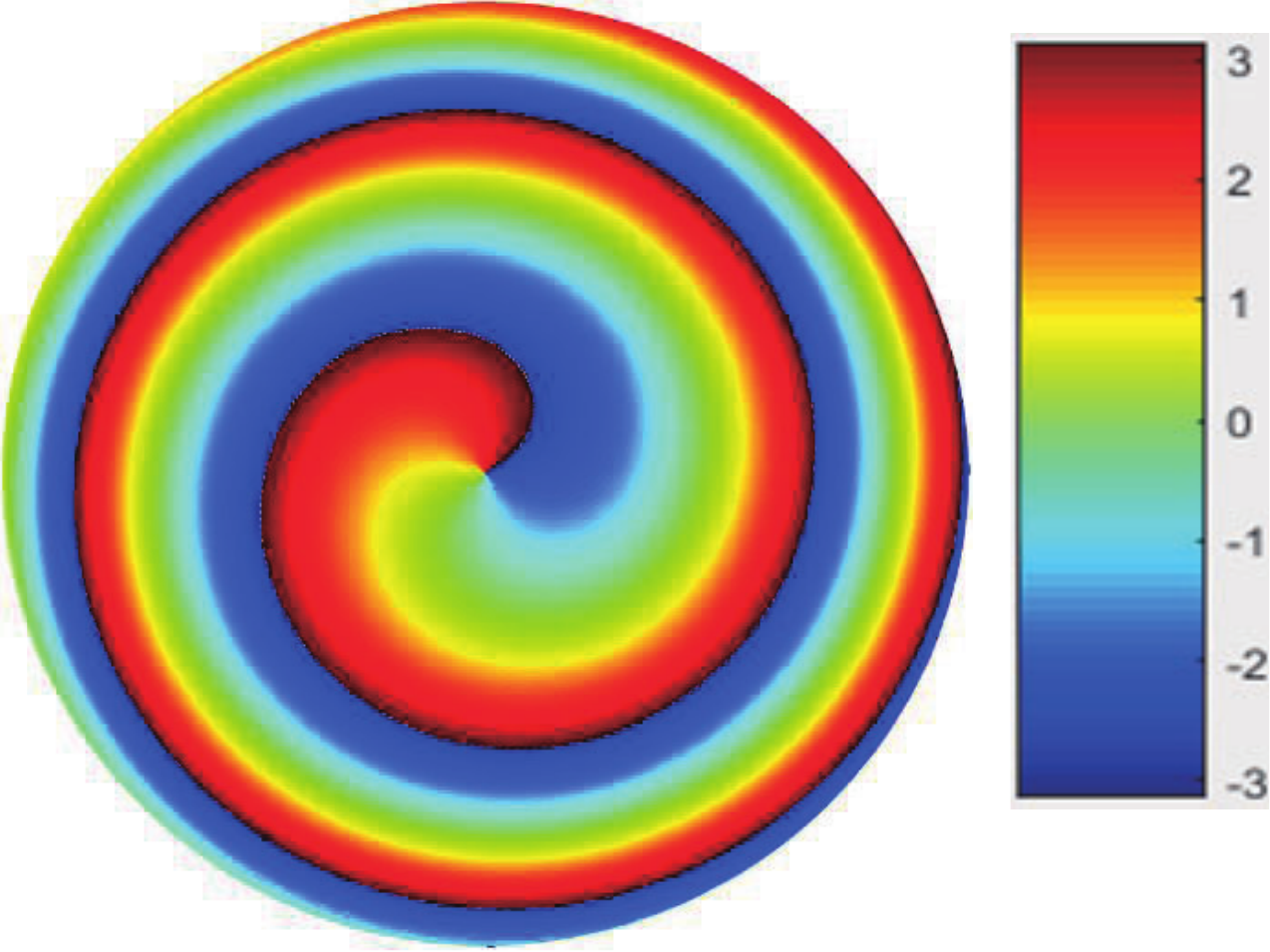}}

\subfigure[]{
\includegraphics[width=2.7cm]{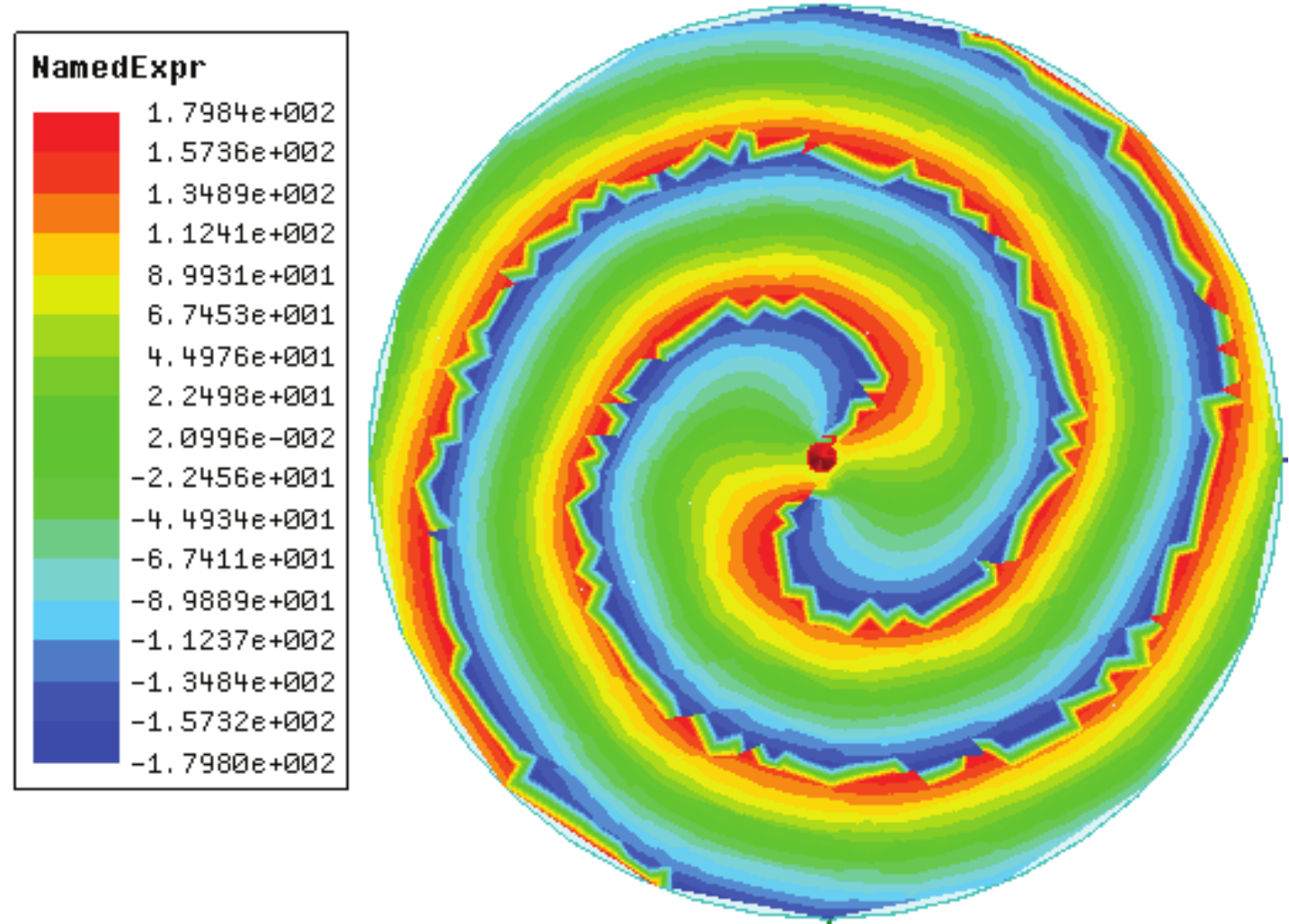}}
\subfigure[]{
\includegraphics[width=2.6cm]{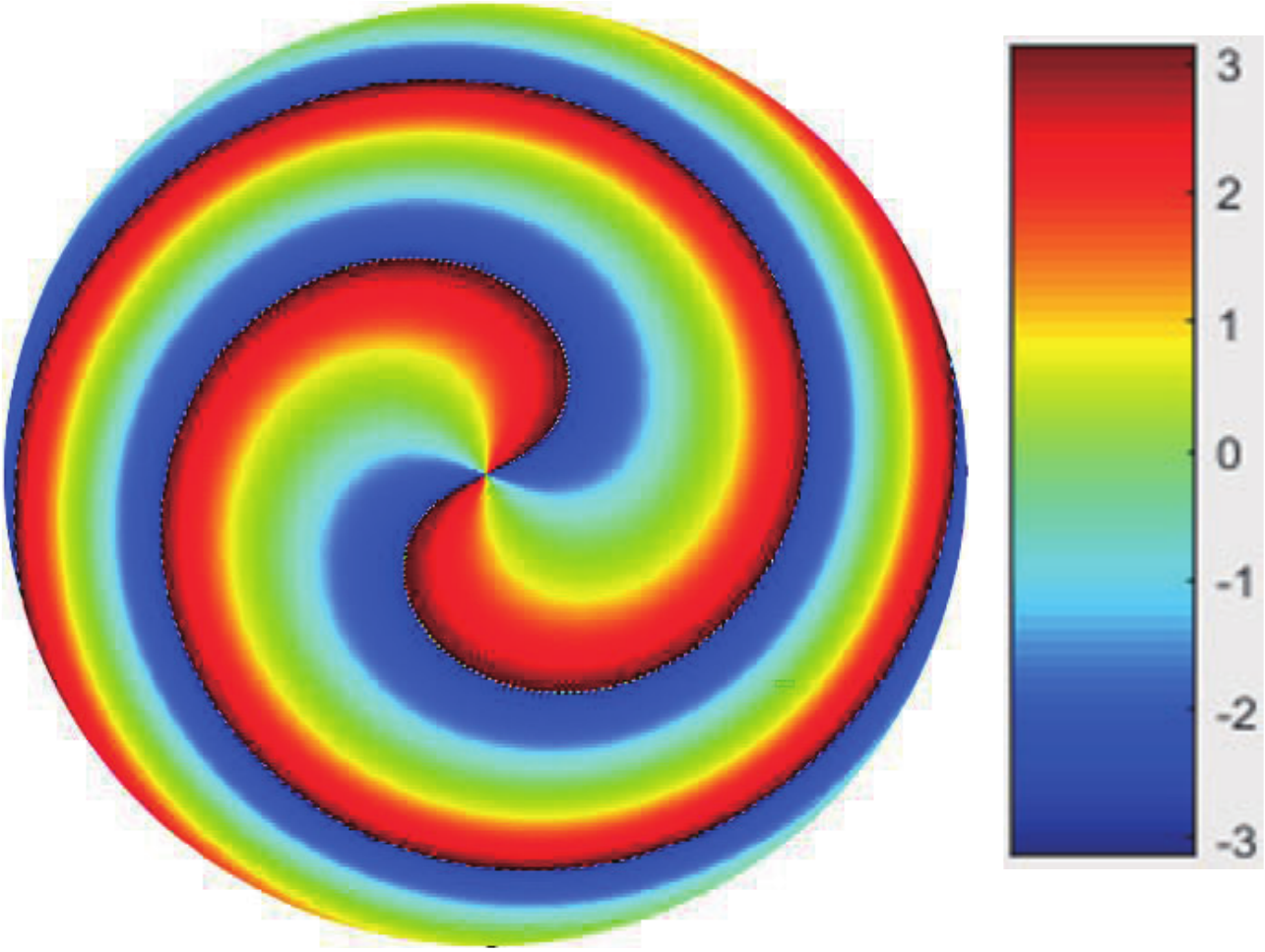}}

\subfigure[]{
\includegraphics[width=2.7cm]{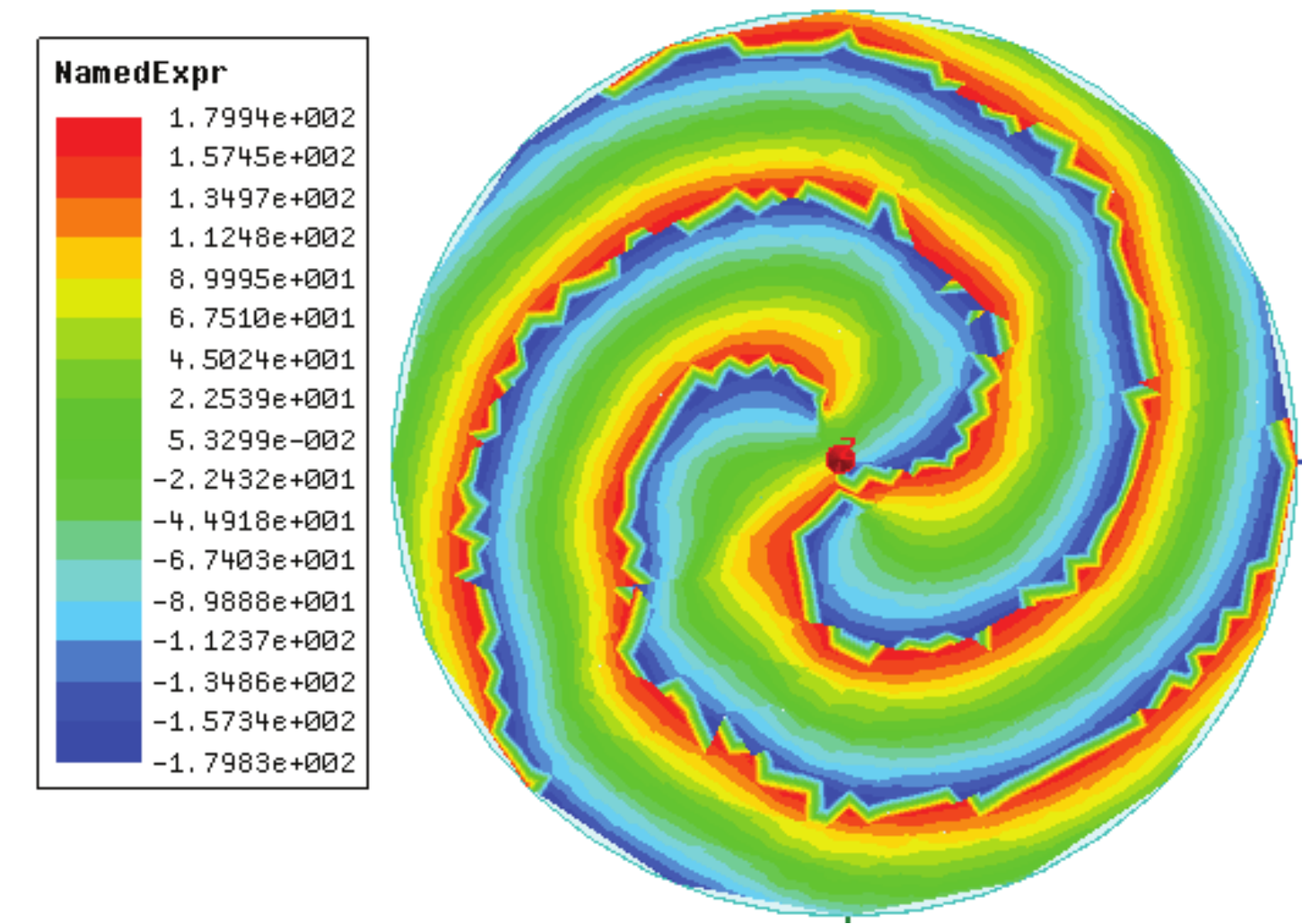}}
\subfigure[]{
\includegraphics[width=2.5cm]{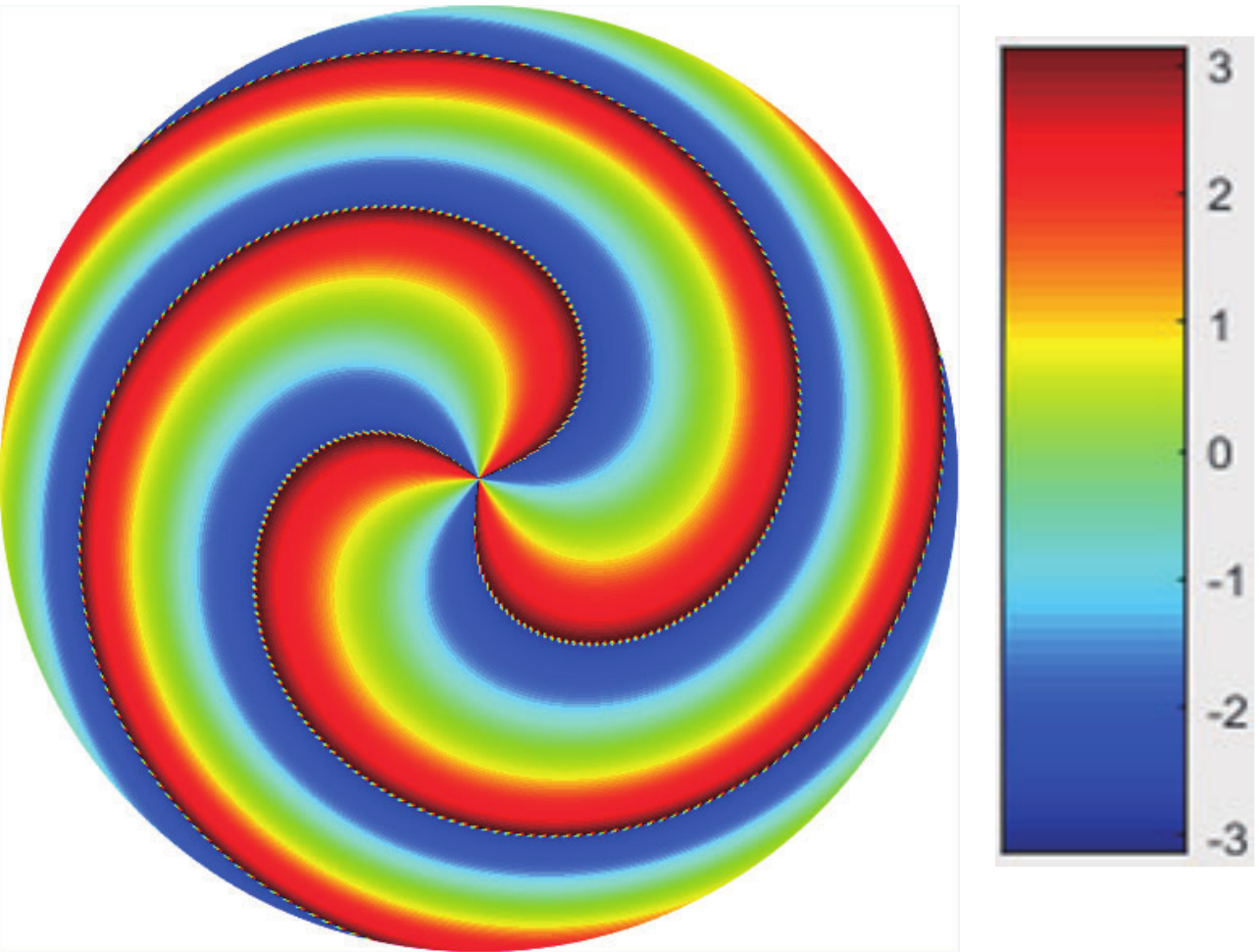}}

\caption{The rotational phase of an OAM beam simulated by HFSS with mode number: (a) $\ell=+1$, (c) $\ell=+2$, (e) $\ell=+3$; and the ideal rotational phase of an OAM beam simulated by MATLAB with mode number: (b) $\ell=+1$, (d) $\ell=+2$, (f) $\ell=+3$. A change in color from blue to green, yellow and red to blue again corresponds to a change in phase of $2\pi$.}
\label{Fig3}
\end{figure}


\section{UCA-Based LoS OAM Communication Systems}
We consider a radio LoS OAM communication system, where a multi-mode OAM beam is generated by an $N$-element UCA at the transmitter and received by another $N$-element UCA at the receiver. In practice, perfect alignment between the transmit and receive UCAs may be difficult to realize. Therefore, we consider more practical misalignment cases. For easier analysis we adopt the non-parallel misalignment case \cite{Chen2018Beam} as shown in Fig. \ref{Fig5} but with a more general model including both elevation angle $\alpha$ and azimuth angle $\varphi$.
\begin{figure}[t]
\setlength{\abovecaptionskip}{0cm}   
\setlength{\belowcaptionskip}{-0.3cm}   
\centering
\subfigure[]{
\includegraphics[width=2.9cm]{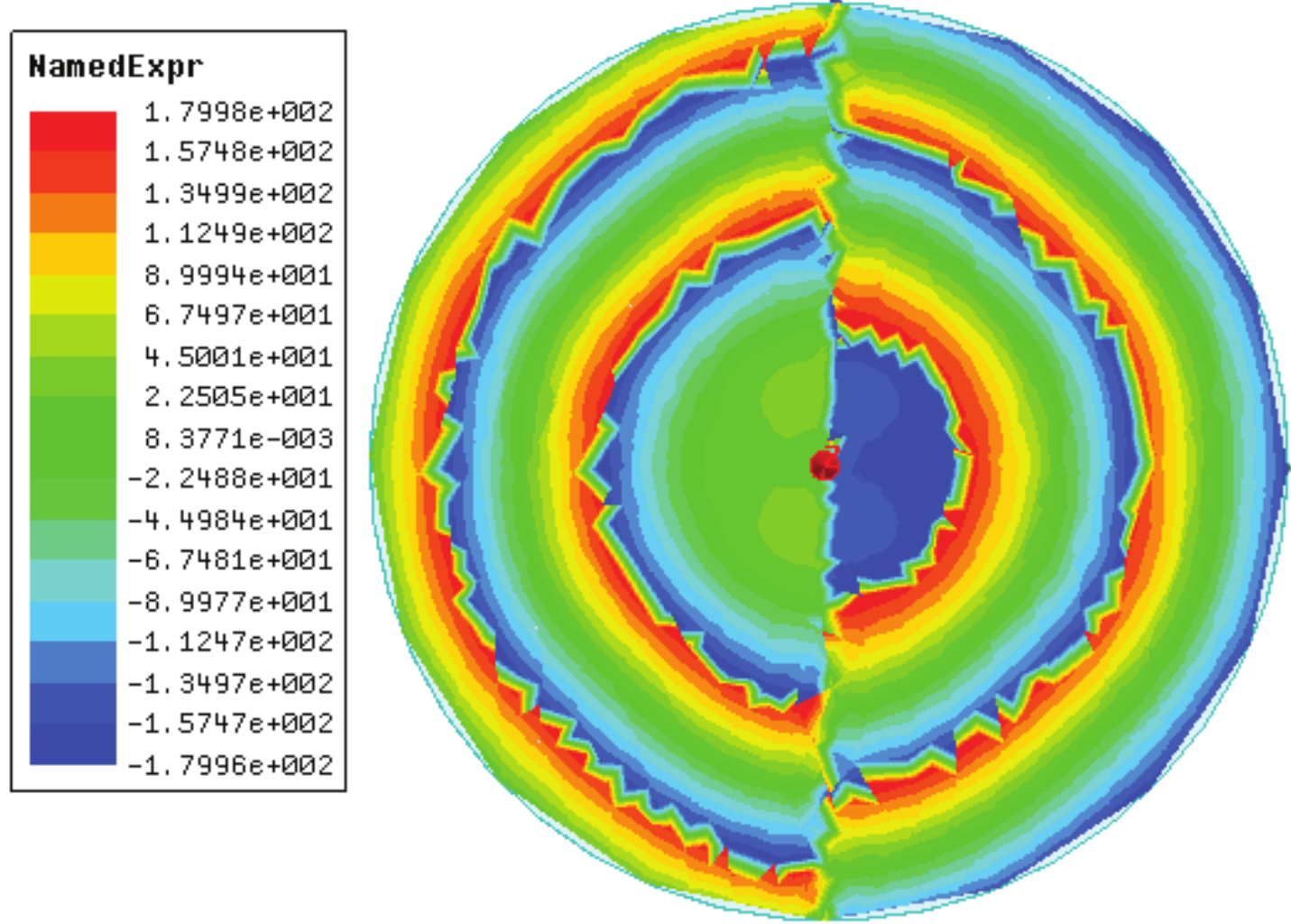}}
\subfigure[]{
\includegraphics[width=2.8cm]{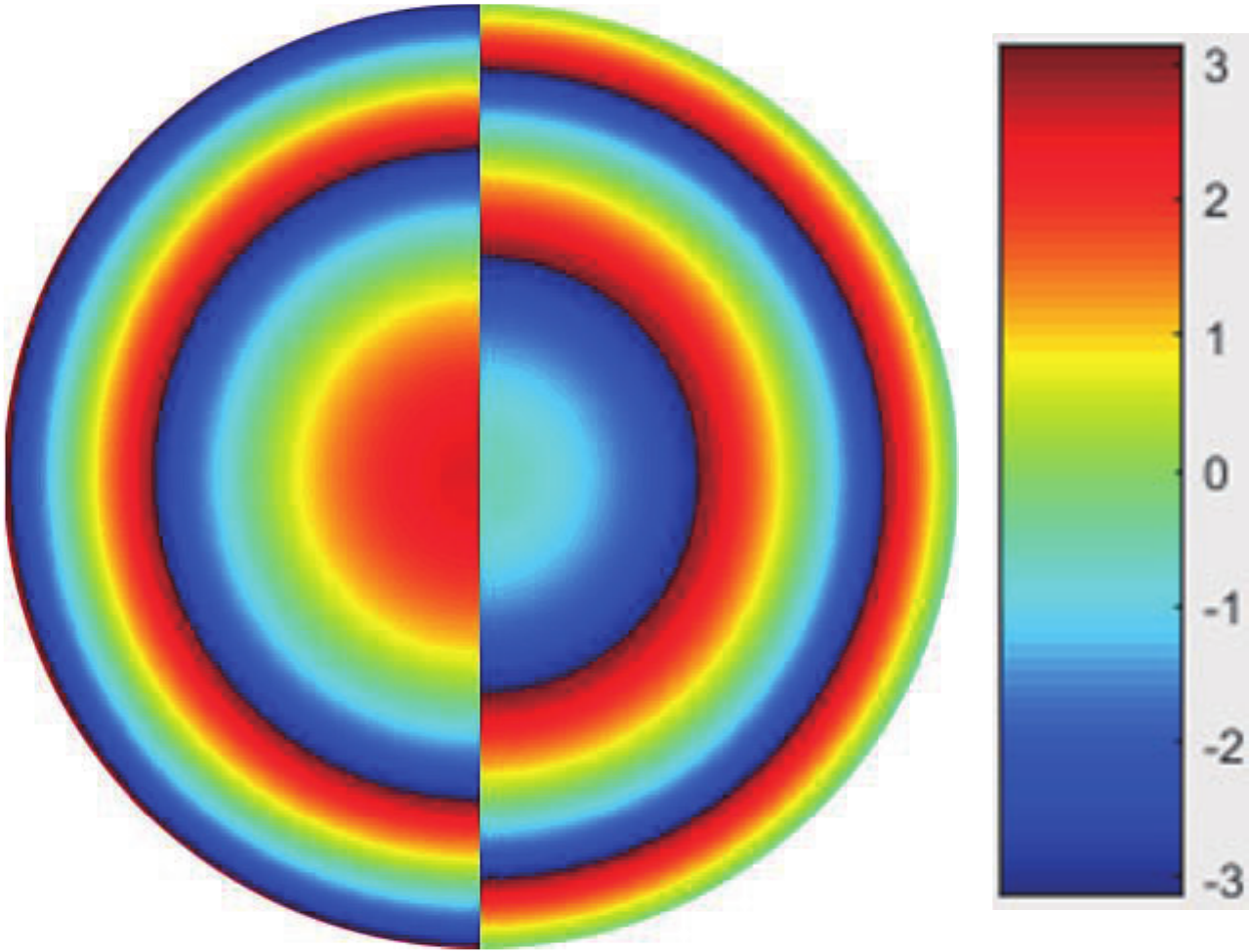}}

\subfigure[]{
\includegraphics[width=2.9cm]{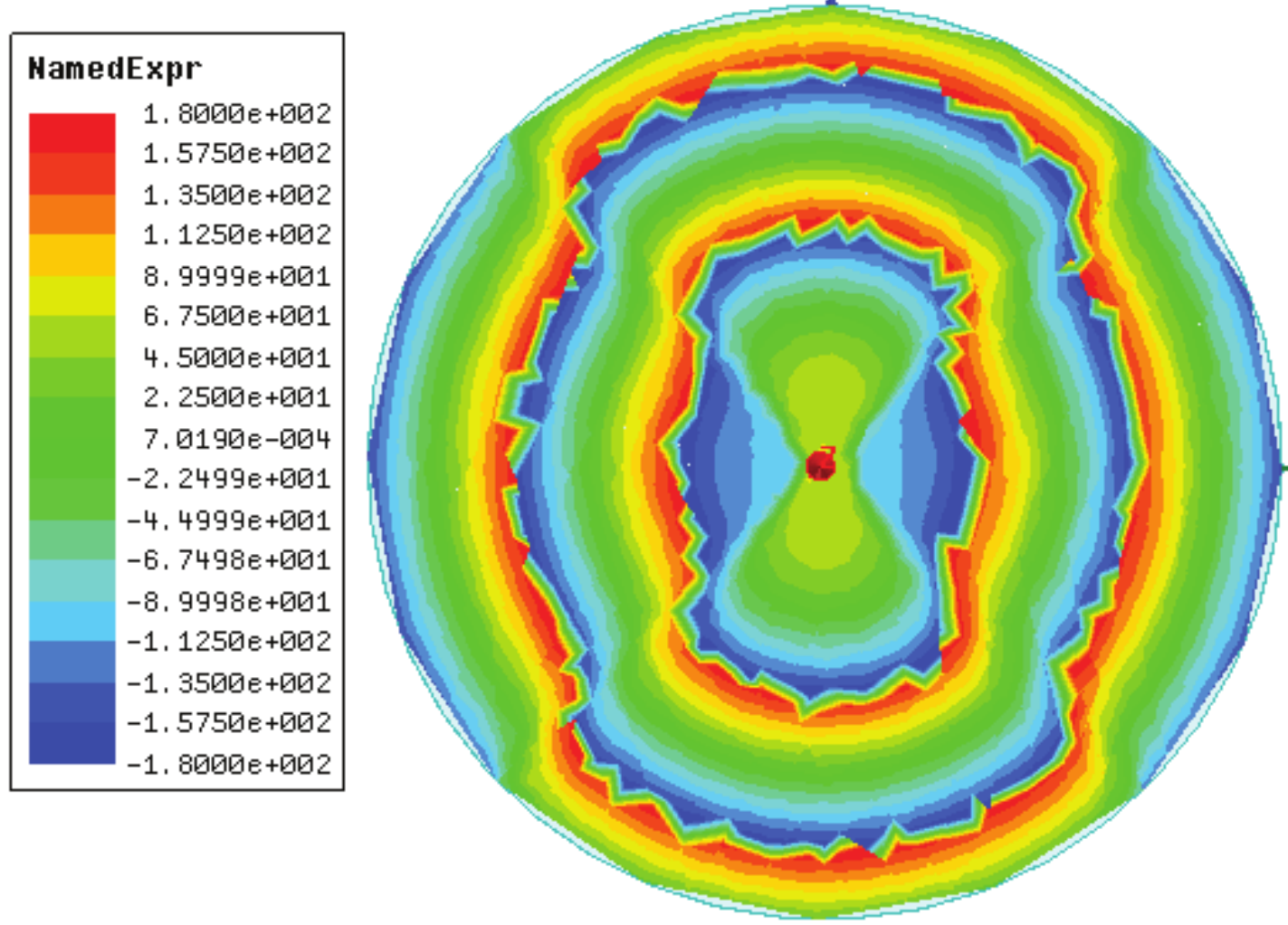}}
\subfigure[]{
\includegraphics[width=2.8cm]{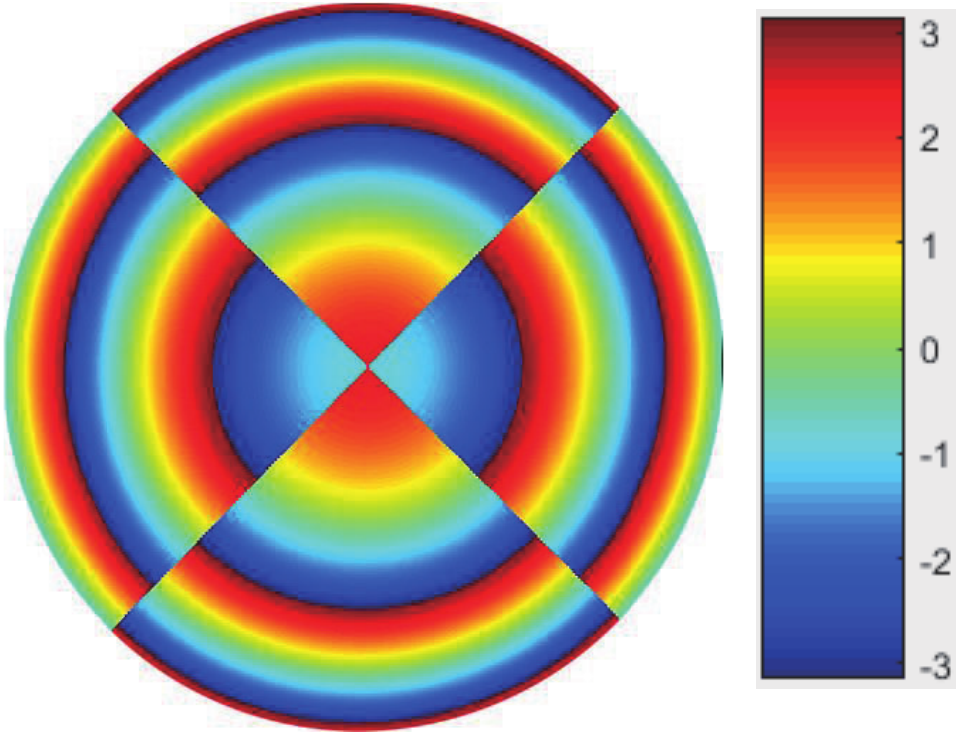}}

\subfigure[]{
\includegraphics[width=2.9cm]{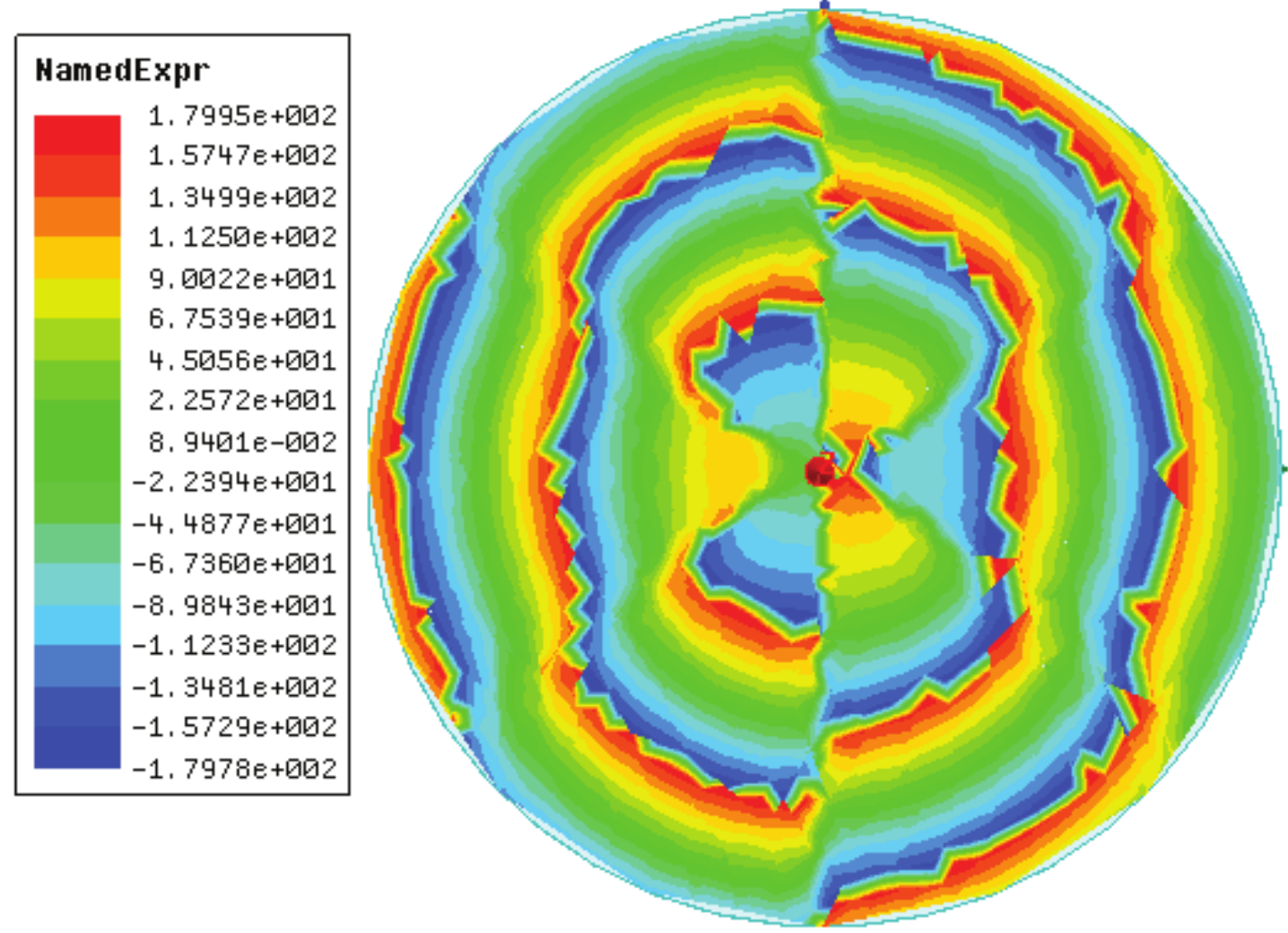}}
\subfigure[]{
\includegraphics[width=2.8cm]{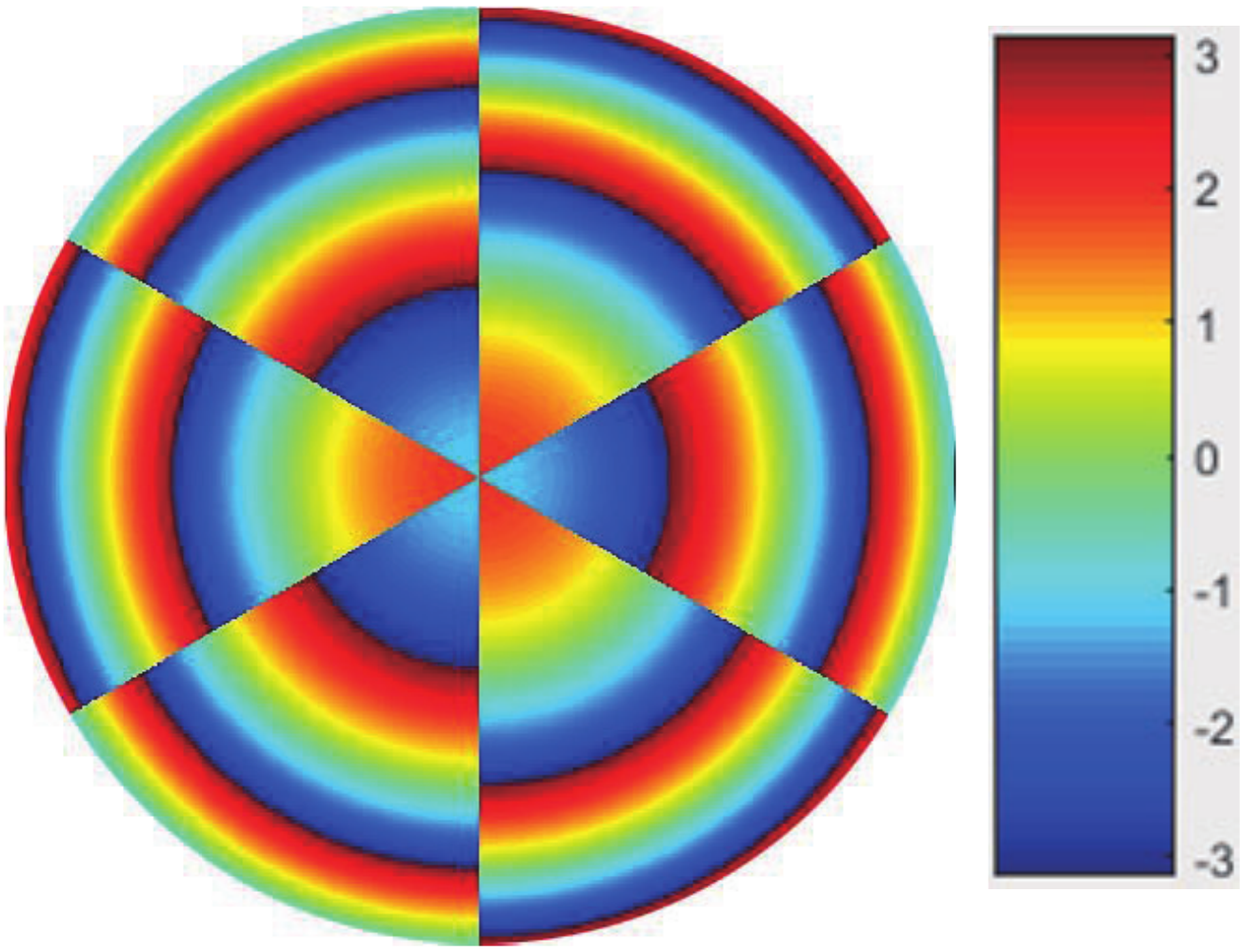}}

\subfigure[]{
\includegraphics[width=2.9cm]{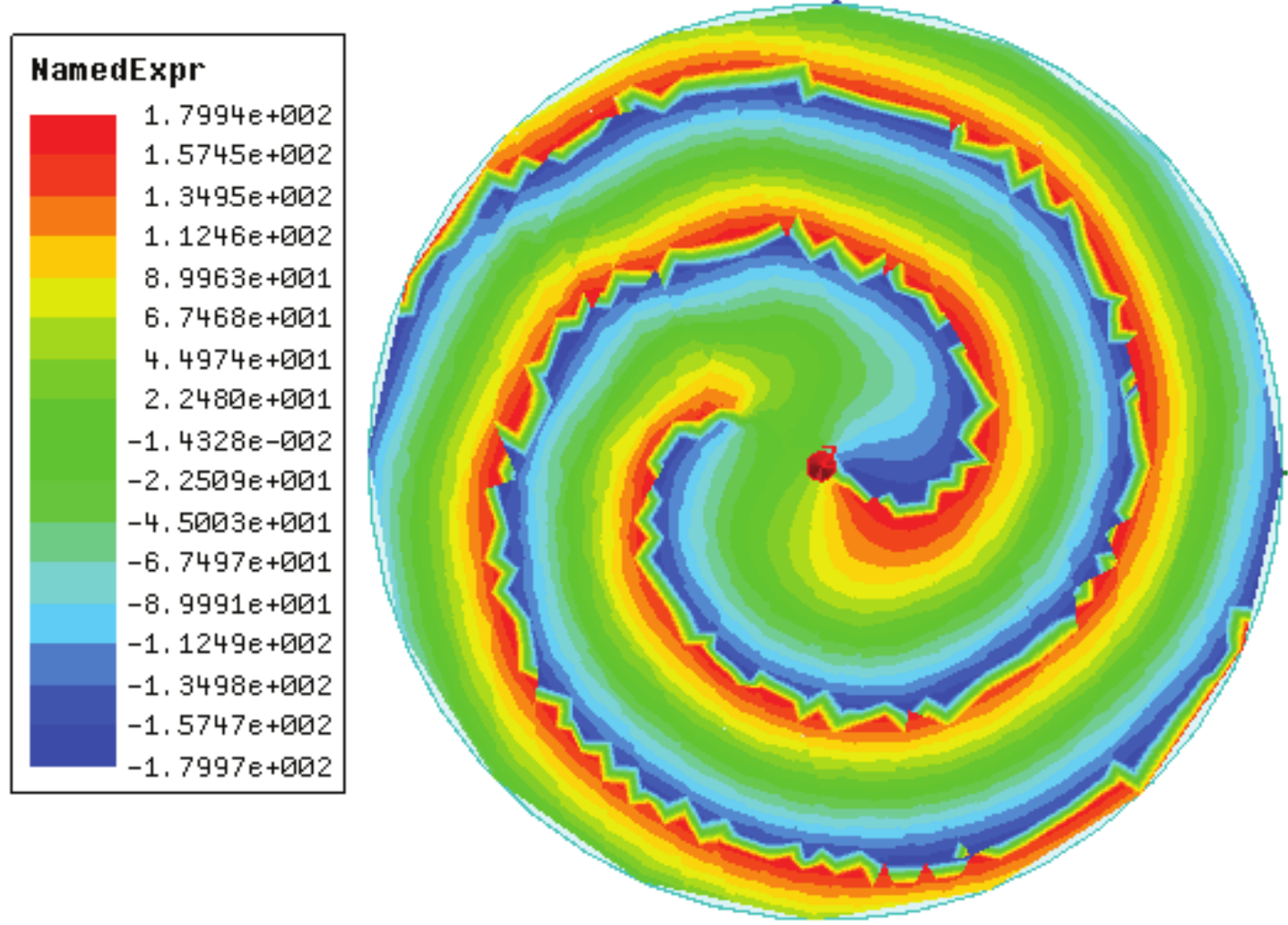}}
\subfigure[]{
\includegraphics[width=2.8cm]{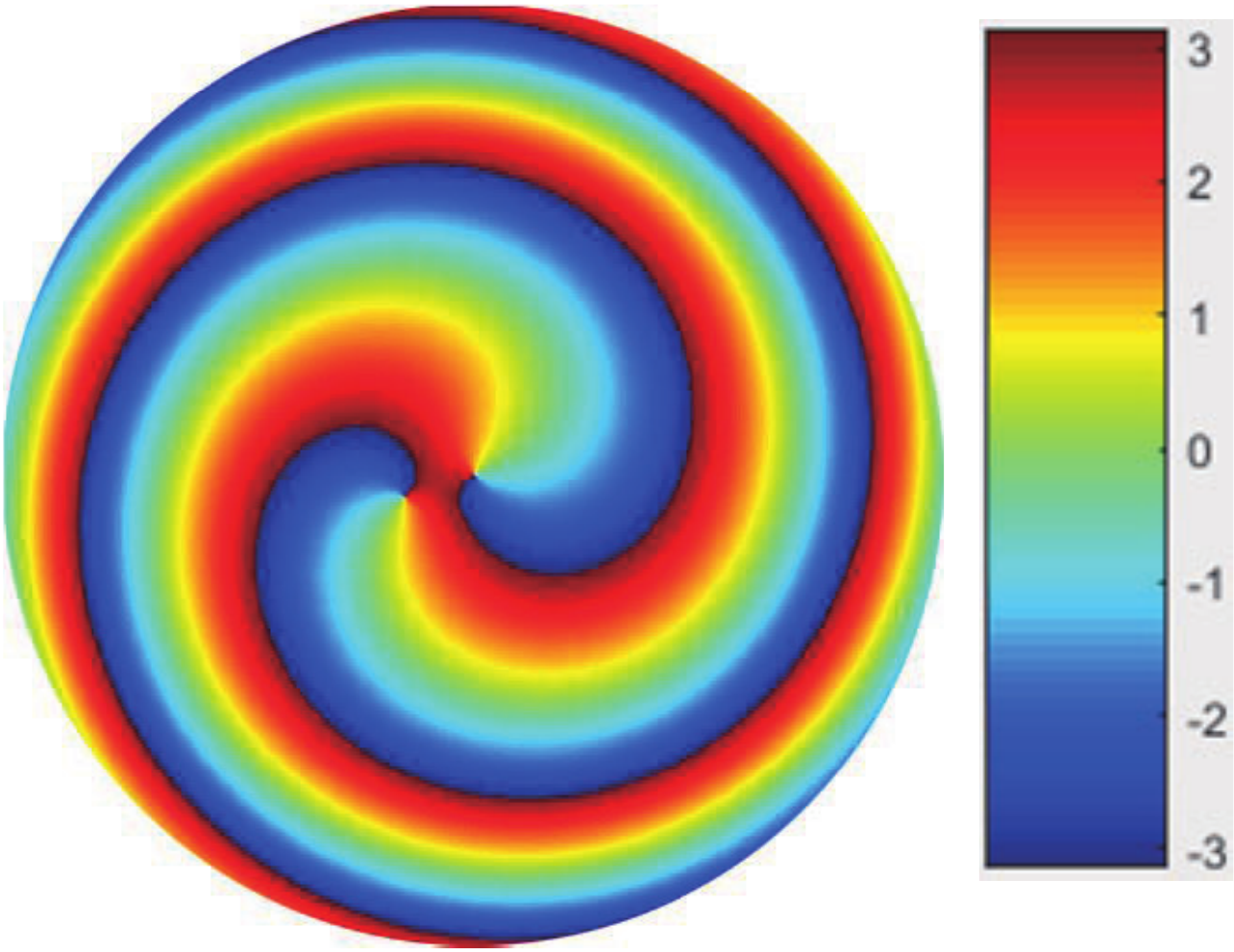}}

\subfigure[]{
\includegraphics[width=2.9cm]{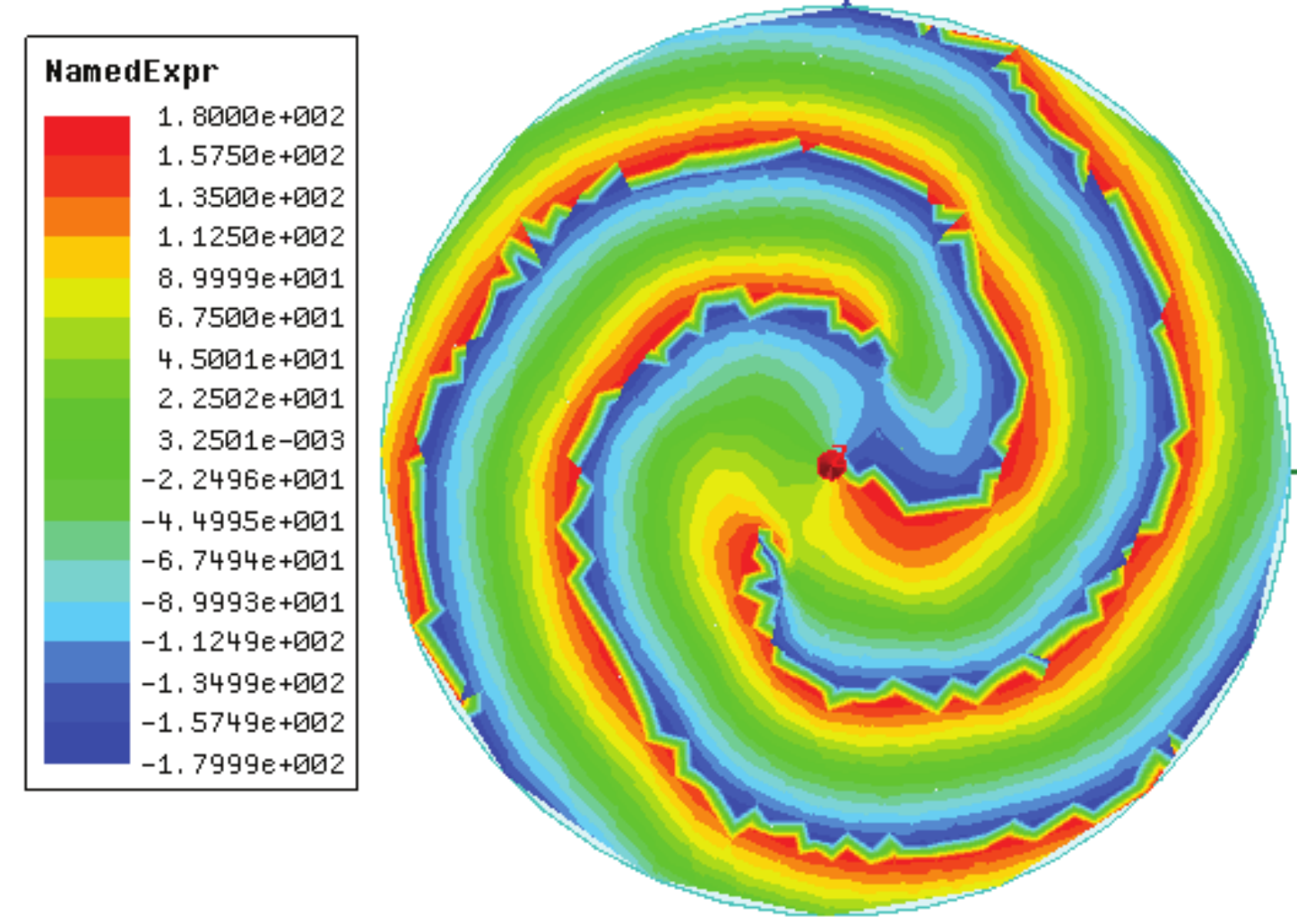}}
\subfigure[]{
\includegraphics[width=2.8cm]{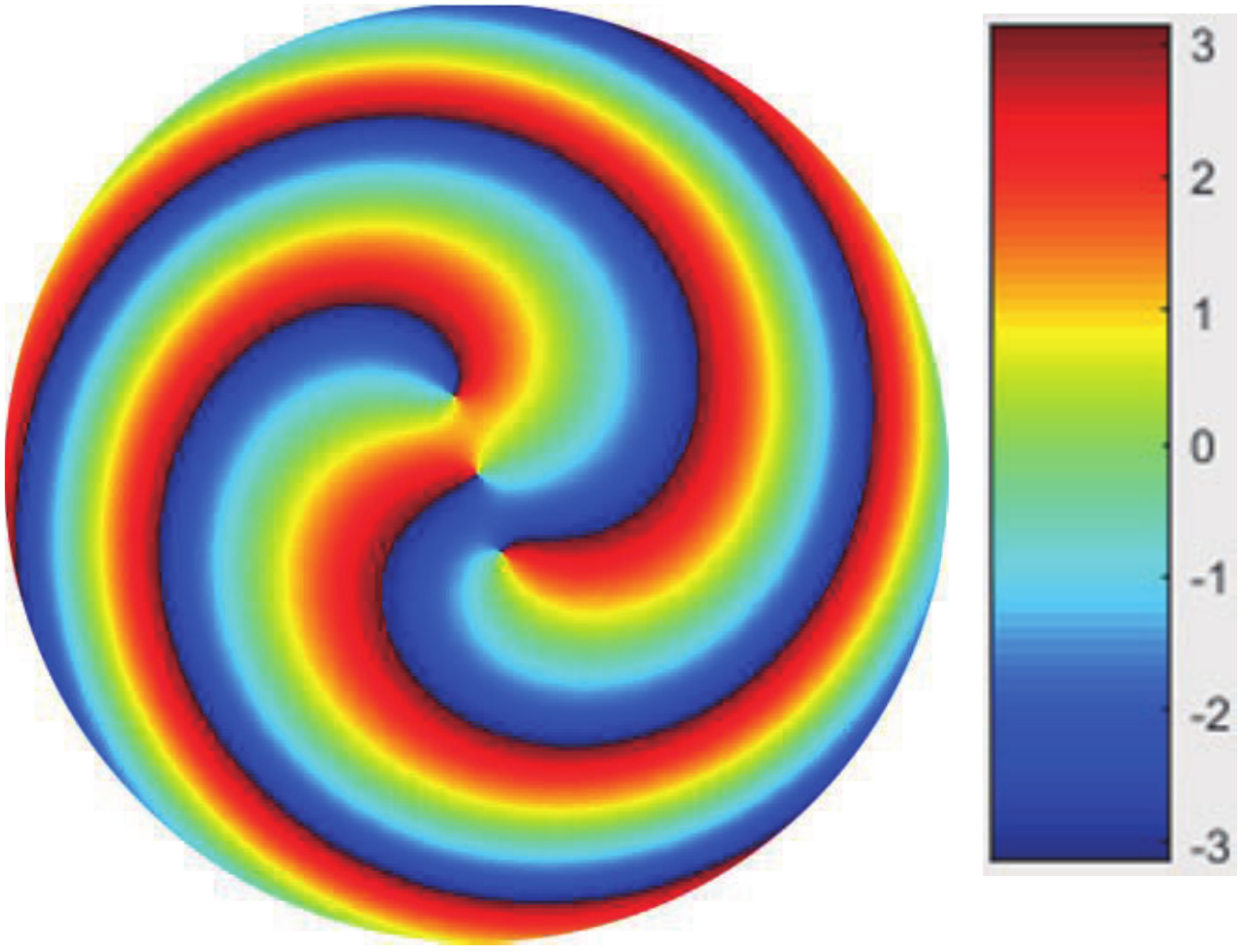}}

\caption{The phase distributions of two-mode OAM beam simulated by HFSS with mode numbers: (a) $\ell_1=-1$, $\ell_2=+1$, (c) $\ell_1=-2$, $\ell_2=+2$, (e) $\ell_1=-3$, $\ell_2=+3$, (g) $\ell_1=+1$, $\ell_2=+2$, (i) $\ell_1=+1$, $\ell_2=+3$; and the phase distributions of two superimposed ideal OAM beams simulated by MATLAB with mode numbers: (b) $\ell_1=-1$, $\ell_2=+1$, (d) $\ell_1=-2$, $\ell_2=+2$, (f) $\ell_1=-3$, $\ell_2=+3$, (h) $\ell_1=+1$, $\ell_2=+2$, (j) $\ell_1=+1$, $\ell_2=+3$.}
\label{Fig4}
\end{figure}

\begin{figure}[t] 
\setlength{\abovecaptionskip}{0cm}   
\setlength{\belowcaptionskip}{-0.1cm}   
\begin{center}
\includegraphics[scale=0.37]{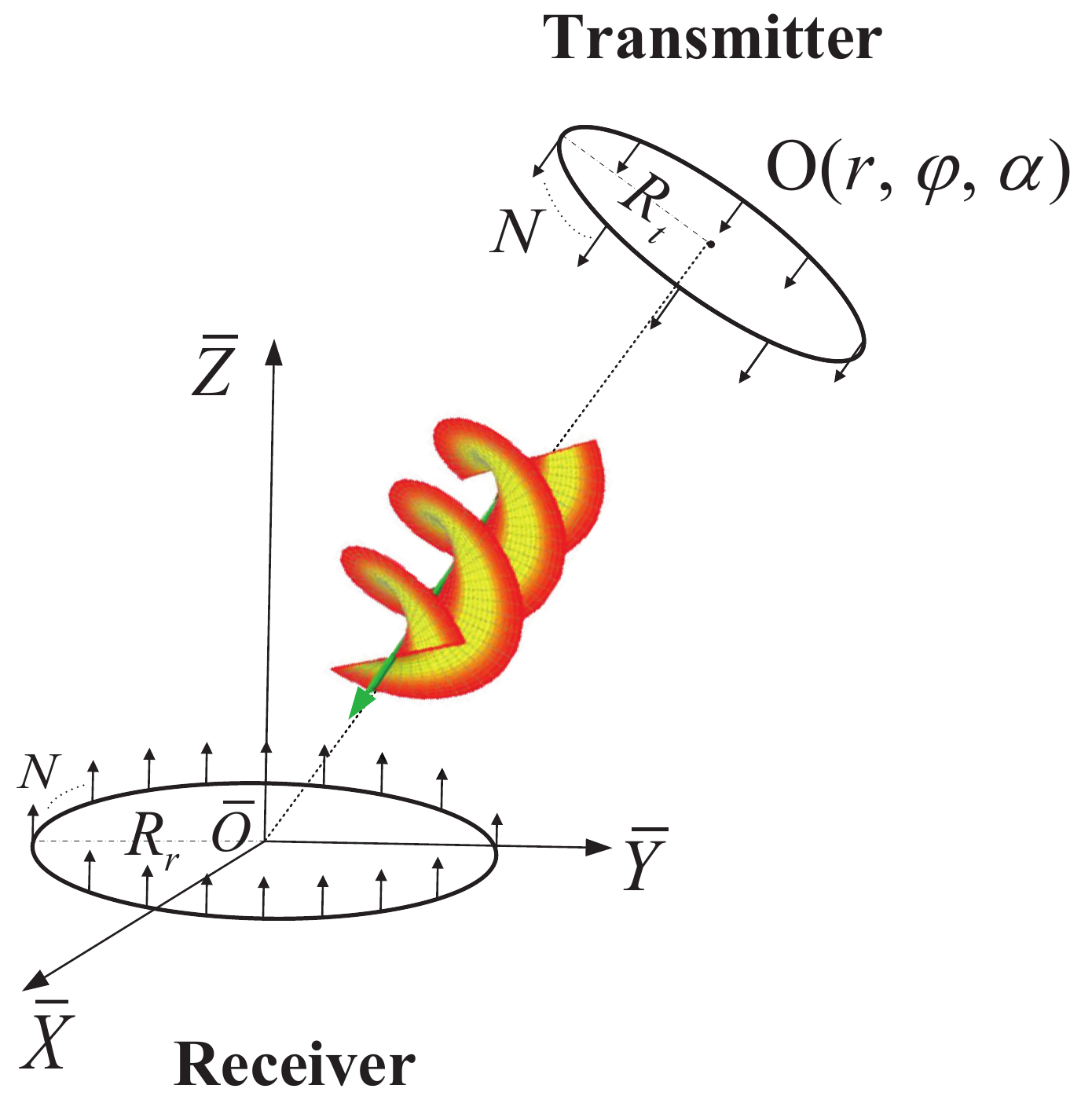}
\end{center}
\caption{UCA-based LoS OAM communication system.}
\label{Fig5}
\end{figure}

\subsection{Channel Model}%
In free space communications, propagation through the RF channel leads to attenuation and phase rotation of the transmitted signal. This effect is modelled through multiplying by a complex constant $h$, whose value depends on the frequency and the distance $d$ between the transmit and receive antenna \cite{Edfors2012Is}:
\begin{equation} \label{FreeSpaceChannel}
h(k,d)=\frac{\beta}{2kd}\textrm{exp}\left(-ikd\right),
\end{equation}
where $\lambda$ is the wavelength, $k=2\pi/\lambda$ is the wave number, and $1/(2kd)$ denotes the degradation of amplitude, $\beta=\beta_r\beta_t$, $\beta_r$ models all the constants relative to each receive antenna element, and the complex exponential term is the phase difference due to the propagation distance. It is worth noting that in order to mainly analyze the factors we concerned, the transmission loss caused by the polarization mismatch between the transmit and receive UCAs is ignored here.

\begin{figure}[!t]
\setlength{\abovecaptionskip}{0cm}   
\setlength{\belowcaptionskip}{-0.1cm}   
\begin{center}
\includegraphics[scale=0.53]{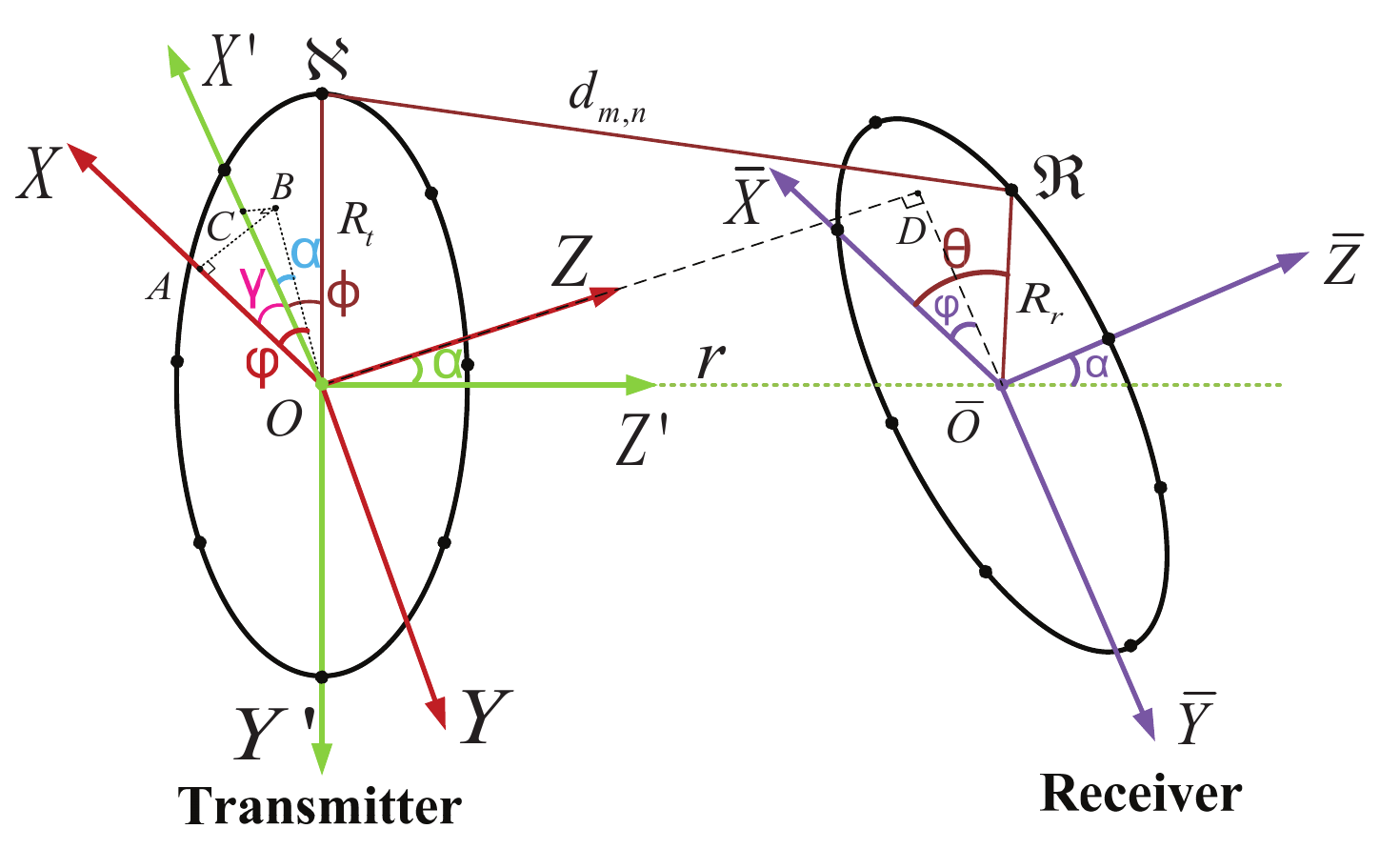}
\end{center}
\caption{The geometrical model of the transmit and receive UCAs in the non-parallel misalignment case.}
\label{Fig6}
\end{figure}

The geometrical model of the UCA-based LoS OAM channel is illustrated in Fig. \ref{Fig6}. In the geometrical model, the transmitter coordinate system $\textrm{Z}'-\textrm{X}'\textrm{O}\textrm{Y}'$ is established using the transmit UCA plane as the $\textrm{X}'\textrm{O}\textrm{Y}'$ plane and the axis through the transmit UCA center $\textrm{O}$ and perpendicular to the UCA plane as the $\textrm{Z}'$-axis, and the receiver coordinate system $\bar{\textrm{Z}}-\bar{\textrm{X}}\bar{\textrm{O}}\bar{\textrm{Y}}$ is established on the plane of the receive UCA by the similar approach. Since the transmit UCA is aligned with the center of the receive UCA in the non-parallel misalignment case, the coordinate of the receive UCA center can be denoted as $\bar{\textrm{O}}(r,0,0)$ in $\textrm{Z}'-\textrm{X}'\textrm{O}\textrm{Y}'$ in the non-parallel misalignment case. $\textrm{D}$ is the projection of $\textrm{O}$ on the $\bar{\textrm{X}}\bar{\textrm{O}}\bar{\textrm{Y}}$ plane, and the coordinate of the transmit UCA center is denoted as $\textrm{O}(r,\varphi,\alpha)$ in $\bar{\textrm{Z}}-\bar{\textrm{X}}\bar{\textrm{O}}\bar{\textrm{Y}}$, where $r$ is the distance between the transmit and the receive UCA centers, $\varphi$ is the azimuth angle, $\alpha$ is the elevation angle, and \emph{$\varphi$ and $\alpha$ are defined as the AoA of the received OAM beams}.

To obtain the estimates of $\varphi$ and $\alpha$, we have to build the coordinate system $\textrm{Z}-\textrm{XOY}$ that is at the transmit UCA and parallel to $\bar{\textrm{Z}}-\bar{\textrm{X}}\bar{\textrm{O}}\bar{\textrm{Y}}$.
According to the geometrical model, the angle between $\textrm{Z}'$-axis and $\bar{\textrm{Z}}$-axis is $\alpha$, so is that between $\textrm{Z}$-axis and $\textrm{Z}'$-axis. Let $\textrm{C}$ be a point on the $\textrm{X}'$-axis, $\textrm{B}$ is the projection of $\textrm{C}$ on the XOY plane, and $\textrm{A}$ is the projection of $\textrm{B}$ on the $\textrm{X}$-axis, then $\angle\textrm{BOC}=\alpha$. As the line $\textrm{OB}$ is parallel to the line $\bar{\textrm{O}}\textrm{D}$ and $\textrm{X}$-axis is parallel to $\bar{\textrm{X}}$-axis, $\angle \textrm{AOB}=\varphi$. Denote the angle between $\textrm{X}'$-axis and $\textrm{X}$-axis as $\gamma$, then $\gamma$ can be obtained as
\begin{align}\label{gamma}
\gamma &= \angle \textrm{AOC} = \arccos\left(\cos(\angle\textrm{AOB})\cos(\angle \textrm{BOC})\right)\nonumber\\
&= \arccos\left(\cos\alpha\cos\varphi\right).
\end{align}
%
Define the angle between the line $\textrm{O}\aleph$ and $\textrm{X}'$-axis as $\phi$ and the angle between the line $\bar{\textrm{O}}\Re$ and $\bar{\textrm{X}}$-axis as $\theta$, where $\aleph$ is the position of the $n$th ($1\le n \le N$) element at the transmitter and $\Re$ is the position of the $m$th ($1\le m \le N$) element at the receiver. According to (\ref{FreeSpaceChannel}) and the geometric relationship in Fig. \ref{Fig6}, the channel coefficients from the $n$th transmit antenna element to the $m$th receive antenna element can be expressed as $h_{m,n}=h(k,d_{m,n})$, where the transmission distance $d_{m,n}$ is calculated as
\begin{align} \label{dmn}
d_{m,n}=&\big[R_t^2+R_r^2+r^2-2rR_r\cos\theta\cos\varphi\sin\alpha \nonumber\\
&-2R_tR_r(\cos\phi\cos\theta\cos\gamma+\sin\phi\sin\theta\cos\varphi) \nonumber\\
&-2R_tR_r(\sin\phi\cos\theta\sin\varphi-\cos\phi\sin\theta\sin\varphi\cos\alpha) \nonumber\\
&+2rR_r\sin\theta\sin\varphi\sin\alpha]^{1/2},
\end{align}
where $R_t$ and $R_r$ are respectively the radii of the transmit and receive UCAs, $\phi=[2\pi(n-1)/N+\phi_0]$, $\theta=[2\pi(m-1)/N+\theta_0]$, $\phi_0$ and $\theta_0$ are respectively the corresponding initial angles of the first reference antenna elements in both UCAs. For easier analysis, we assume $\phi_0=0$ and $\theta_0=0$ here.

In the end, the channel matrix of the UCA-based free space communication system can be expressed as $\mathbf{H}=[h_{m,n}]_{N\times N}$.
Note that when $\alpha=0$ and $\varphi=0$, $\mathbf{H}$ is a circulant matrix that can be decomposed by the $N$-dimentional Fourier matrix $\mathbf{F}_N$ as $\mathbf{H}=\mathbf{F}_N^H\mathbf{\Lambda}\mathbf{F}_N$, where $\mathbf{\Lambda}$ is a diagonal matrix with the eigenvalues of $\mathbf{H}$ on its diagonal.

\subsection{Signal Model}

For higher data rate transmission, OFDM is often applied and naturally the OAM-OFDM communication system has been proposed in \cite{Chen2018A}. In an LoS MCMM-OAM communication system, we assume $P$ subcarriers and $U$ OAM modes for data transmission, and $\widetilde{P}$ subcarriers and $\widetilde{U}$ OAM modes for training, $1\leq\widetilde{P}\leq P, 1\leq\widetilde{U}\leq U$.

\begin{figure*}
\setcounter{equation}{11}
\begin{align} \label{xlk}
x'(\ell_{u},k_p)&=\sum_{m=1}^{N} \mathbf{h}_m(k_p)\mathbf{f}^H(\ell_u)s'(\ell_{u},k_p)+z'(\ell_{u},k_p) \nonumber\\
&=\frac{\beta}{2k_p}\sum_{m=1}^{N}\sum_{n=1}^{N} \frac{e^{ik_p|\bm{d}_{m,n}|}}{|\bm{d}_{m,n}|}e^{i\ell_{u}\varphi_n}
 s'(\ell_{u},k_p)+z'(\ell_{u},k_p)\nonumber\\
&=\frac{\beta}{2k_p}\sum_{m=1}^{N}\sum_{n=1}^{N} \frac{e^{ik_p|\bm{r}-\bm{r}'_n+\bm{r}'_m|}} {|\bm{r}-\bm{r}'_n+\bm{r}'_m|}e^{i\ell_{u}\varphi_n}
 s'(\ell_{u},k_p)+z'(\ell_{u},k_p)\nonumber\\
&\approx\frac{\beta}{2k_p}\sum_{m=1}^{N} e^{i\bm{k}_p\cdot\bm{r}_m}\bigg[\frac{e^{ik_pr}}{r}
\sum_{n=1}^{N} e^{-i(\bm{k}_p\cdot\bm{r}_n-\ell_{u}\varphi_n)}\bigg]s'(\ell_{u},k_p) +z'(\ell_{u},k_p)\nonumber\\
&\approx\sigma_{\ell_u,k_p} e^{ik_pr}e^{i\ell_{u}\gamma}i^{-\ell_u}{J_{\ell_{u}}}(k_pR_t\sin\alpha){J_0}(k_pR_r\sin\alpha)+z'(\ell_{u},k_p),
\end{align}
\setcounter{equation}{7}%
\hrulefill
\end{figure*}

As a UCA can generate both single-mode and multi-mode OAM beams with the baseband (partial) discrete Fourier transform (DFT) matrix $\mathbf{F}_U$, the equivalent baseband signal models of transmitting multi-mode OAM data symbols and transmitting multiple single-mode OAM training symbols can be expressed as $\mathbf{F}_U^H\mathbf{s}(k_p)$ and $\mathbf{F}_{\widetilde{U}}^H \mathbf{S}'(k_p)$, respectively, where $\mathbf{s}(k_p)=[s(\ell_1,k_p),s(\ell_2,k_p),\cdots,s(\ell_U,k_p)]^T$ contains the data symbols transmitted on $U$ OAM modes at the $p$th $(1\leq p\leq P)$ subcarrier simultaneously, and $\mathbf{S}'(k_p)=\textrm{diag}\{s'(\ell_1,k_p),$ $s'(\ell_2,k_p),\cdots,s'(\ell_{\widetilde{U}},k_p)\}$ consists of the training symbols transmitted on $\widetilde{U}$ OAM modes at the $p$th $(1\leq p\leq\widetilde{P})$ subcarrier sequentially.

Thus, the received baseband signal vector $\mathbf{y}(k_p)$ can be written as
\begin{align} \label{y}
\mathbf{y}(k_p)=\mathbf{H}(k_p)\mathbf{F}_U^H\mathbf{s}(k_p)+\mathbf{z}(k_p),
\end{align}
where $\mathbf{y}(k_p)$ is the $N$-dimensional received signal vector, $\mathbf{H}(k_p)$ $=[h_{m,n}(k_p)]_{N\times N}$ is the channel matrix at the $p$th subcarrier, $\mathbf{F}_U=[\mathbf{f}^H(\ell_1),$ $\mathbf{f}^H(\ell_2),\cdots,\mathbf{f}^H(\ell_U)]^H$ is an $U\times N$ (partial) DFT matrix, $\mathbf{f}(\ell_u)=\frac{1}{\sqrt{N}}[1,e^{-i\frac{2\pi \ell_u}{N}},\cdots,$ $e^{-i\frac{2\pi \ell_u(N-1)}{N}}]$, $\mathbf{z}(k_p)=[z(\ell_1,k_p),z(\ell_2,k_p),\cdots,z(\ell_U,k_p)]^T$ is the complex Gaussian noise vector with zero mean and covariance matrix $\sigma^2_z\mathbf{I}_{N}$, $1\leq U\leq N$, $1\leq p\leq P$.

Correspondingly, the received baseband training symbol matrix $\mathbf{Y}'(k_p)$ can be written as
\begin{align} \label{Y'}
\mathbf{Y}'(k_p)=\mathbf{H}(k_p)\mathbf{F}_{\widetilde{U}}^H \mathbf{S}'(k_p)+\mathbf{Z}'(k_p),
\end{align}
where $\mathbf{Y'}(k_p)$ is the $N\times \widetilde{U}$ matrix composed by the sequentially received signals of training symbols on $\widetilde{U}$ OAM modes at the $p$th subcarrier, and $\mathbf{Z'}(k_p)$ is the related additive noise matrix, $1\leq \widetilde{U}\leq N$, $1\leq p\leq \widetilde{P}$.

The UCA-based OAM receiver has the similar RF analog or baseband digital structure to the transmitter (see Fig.1 and Fig.3 in \cite{Chen2018A}) but with the opposite operations, i.e., separating different OAM modes and despiralizing each mode. Thus, the detected OAM data symbol vector $\mathbf{x}(k_p)$ can be expressed as
\begin{align} \label{x}
\mathbf{x}(k_p)&=\mathbf{D}(k_p)\mathbf{y}(k_p)\nonumber\\
&= \mathbf{D}(k_p)\left(\mathbf{H}(k_p)\mathbf{F}_U^H\mathbf{s}(k_p)+\mathbf{z}(k_p)\right),
\end{align}
where $\mathbf{x}(k_p)$$=$$[x(\ell_1,k_p), x(\ell_2,k_p), \cdots, x(\ell_U,k_p)]^T$, $x(\ell_u,k_p)$ ($u=1, 2,\cdots, U$, $p=1, 2, \cdots, P$) is the detected data symbol on the $u$th mode OAM at the $p$th subcarrier, and $\mathbf{D}(k_p)$ is the $U\times N$ data signal detection matrix. In the case that the transmit and receive UCAs are in perfect alignment, the effective multi-mode OAM channel defined as $\mathbf{H}_{\textmd{OAM}}(k_p)=\mathbf{F}_U\mathbf{H}(k_p)\mathbf{F}_U^H$ in \cite{Zhang2017Mode} becomes the diagonal matrix $\mathbf{\Lambda}_U(k_p)$ composed of $U$ eigenvalues of $\mathbf{H}(k_p)$, and $\mathbf{D}(k_p)=\mathbf{\Lambda}_U^{-1}(k_p)\mathbf{F}_U$. Thus, $\mathbf{x}(k_p)=\mathbf{s}(k_p)+\mathbf{\Lambda}_U^{-1}(k_p)\mathbf{z}(k_p)$, which shows the transmitted data symbols being recovered and the mode multiplexing gain of OAM communication systems being achieved. Furthermore, how to design $\mathbf{D}(k_p)$ in a more general misalignment case will be specified in Section V.
\par
Corresponding to the training symbol transmitted on each OAM mode at each time slot, the received signals on all the $N$ elements are combined at each time slot, which can be formulated as
\begin{align} \label{xkp}
\mathbf{x'}(k_p)=\mathbf{1}^T\mathbf{Y}'(k_p) =\mathbf{1}^T\left(\mathbf{H}(k_p)\mathbf{F}_{\widetilde{U}}^H\mathbf{S'}(k_p)+\mathbf{Z'}(k_p)\right),
\end{align}
where $\mathbf{x}'(k_p)$ $=$ $[x'(\ell_1,k_p),$ $x'(\ell_2,k_p),$ $\cdots,$ $x'(\ell_{\widetilde{U}},k_p)]$, $x'(\ell_u,k_p)$ is the combined signal corresponding to $s'(\ell_u,k_p)$, $u=1, $$ 2,\cdots,\widetilde{U}$, $p=1, 2,$ $ \cdots, \widetilde{P}$.

\section{ Distance and AoA Estimation for OAM Beams}

\vspace{1em}
\subsection{Problem Formulation}
As in MIMO communications, in OAM the transmitted frames may carry a combination of training and data symbols. In the case of MIMO, pilot data is used to perform channel estimation, and it is well known that the number of training symbols per subcarrier or subchannel has to be larger than the number of transmit antennas. On the other hand, in the case of OAM, pilot data are exploited for distance and AoA estimation, and $\widetilde{U}~(1\leq \widetilde{U} \leq N)$ training symbols per subcarrier or subchannel are required. Accordingly, we assume the training symbols $\{s'(\ell_u,k_p) | u=1, 2, \cdots, \widetilde{U}; p=1, 2, \cdots, \widetilde{P}\}$ are known to the OAM receiver.

Then, based on \eqref{E}, \eqref{xkp} and the coordinate transformation in Section III. A, the combined signal $x'(\ell_{u},k_p)$ in the $\bar{\textrm{Z}}-\bar{\textrm{X}}\bar{\textrm{O}}\bar{\textrm{Y}}$ coordinate system can be derived in (12),
where $\mathbf{h}_m(k_p)$ $=$ $[h_{m,1}(k_p),$ $h_{m,2}(k_p),$ $\cdots,$ $h_{m,N}(k_p)]$, $\sigma_{\ell_{u},k_p}=\frac{\beta N^2}{2k_pr}s'(\ell_{u},k_p)$ is a function of $r$ and unrelated to $\alpha$ and $\varphi$, and $\bm{d}_{m,n}$ is the position vector from the $n$th transmit antenna element to the $m$th receive antenna element, $\bm{r}_m=R_r(\bm{\hat{x}}'\cos\theta+\bm{\hat{y}}'\sin\theta)$, $\bm{\hat{x}}'$ and $\bm{\hat{y}}'$ are the unit vectors of $\bar{\textrm{X}}$-axis and $\bar{\textrm{Y}}$-axis, respectively.

Overall, the aim of the distance and AoA estimation is to obtain the distance $r$, the elevation angle $\alpha$ and the azimuthal angle $\varphi$ from the signals $\{x'(\ell_{u},k_p)|{u}=1,2,\dots,{\widetilde{U}}; p=1,2,\cdots,\widetilde{P}\}$.

\vspace{-0.8em}
\subsection{Distance and AoA Estimation Based on 2-D ESPRIT}
According to the expression in (12), we propose to first estimate $r$ and $\gamma$ with 2-D ESPRIT algorithm, and then extract $\alpha$ and $\varphi$.

\subsubsection{Estimation of $r$ and $\gamma$}
The  ESPRIT algorithm provides an elegant means for estimating the parameters of complex sinusoidal signals embedded in white Gaussian noise.  Accordingly, at the receiver, we need to extract the exponentials containing $r$ and $\gamma$ from the signal $x'(\ell_{u},k_p)$  received on the $u$th mode OAM at the $p$th subcarrier by performing the following operation
\setcounter{equation}{12}
\begin{align}
\tilde{x}(\ell_{u},k_p)&=\frac{x'(\ell_{u},k_p)}{|x'(\ell_{u},k_p)|} \frac{s'(\ell_{u},k_p)^*}{|s'(\ell_{u},k_p)|}i^{\ell_u}{\rm{sign}}({x'}(\ell_{u},k_p))\nonumber\\
&=e^{ik_pr}e^{i\ell_{u}\gamma}+\tilde{z}(\ell_{u},k_p),
\label{align1}
\end{align}
where $\tilde{z}(\ell_{u},k_p)$ is the noise, $k_p=2\pi f_p/c$,  $k_{p+1}-k_p=1$,  $\ell_{{u}+1}-\ell_{{u}}=1$,
$p=1,2,\cdots,\widetilde{P}$ and ${u}=1,2,\cdots,{\widetilde{U}}$.

All the  signals received on the ${\widetilde{U}}$ OAM modes at the  $\widetilde{P}$ subcarriers can be collected in the matrix $\mathbf{\tilde{X}}_{{\widetilde{U}},\widetilde{P}}=\left[\tilde{x}(\ell_u,k_p)\right]_{\widetilde{U}\times\widetilde{P}}$.
%
In the estimation of $r$, we first build the ${\widetilde{U}}\widetilde{P}$-dimensional vector $\mathbf{x}_{C}$ by stacking the   columns of $\mathbf{\tilde{X}}_{{\widetilde{U}},\widetilde{P}}$ as
\begin{align*}
\mathbf{x}_{C}=\cvec\left(\mathbf{\tilde{X}}_{{\widetilde{U}},\widetilde{P}}\right)=[\tilde{x}(\ell_1,k_1), \tilde{x}(\ell_2,k_1),\cdots, \tilde{x}(\ell_{\widetilde{U}},k_{\widetilde{P}})]^T,
\end{align*}
so that we can express the received signal in compact form as
\begin{equation}
\mathbf{x}_{C}=\mathbf{{a}}+\mathbf{z}_{C}.
\end{equation}
where $\mathbf{a}=[e^{ik_1r}e^{i\ell_1\gamma}, e^{ik_1r}e^{i\ell_2\gamma},\cdots,  e^{ik_{\widetilde{P}}r}e^{i\ell_{\widetilde{U}}\gamma}]^T$ and $\mathbf{z}_{C}$ is the  noise vector. The covariance matrix of $\mathbf{x}_{C}$ can be written as
\begin{equation}\label{cx}
\mathbf{R}_{\mathbf{x}_{C}}=\mathbb{E}\left\{\mathbf{x}_{C}\mathbf{x}_{C}^H\right\}
=\mathbf{a}\mathbf{a}^H+\mathbf{R}_{\mathbf{z}_{C}},
\end{equation}
where $\mathbf{R}_{\mathbf{z}_{C}}=\mathbb{E}\left\{\mathbf{z}_C\mathbf{z}_C^H\right\}$. The eigen value decomposition (EVD) of $\mathbf{R}_{\mathbf{x}_{C}}$ is
\begin{equation}\label{EVD}
\mathbf{R}_{\mathbf{x}_{C}}=\mathbf{Q\Omega Q}^{H},
\end{equation}
where $\mathbf{Q}$ is an ${\widetilde{U}}\widetilde{P}\times {\widetilde{U}}\widetilde{P}$ unitary matrix and $\mathbf{\Omega}$$=$$\textrm{diag}(\chi_1,$ $\chi_2,\cdots,\chi_{\widetilde{U}\widetilde{P}})$. Denote $\chi_{max}$$=$$\max\{\chi_m|m=1,2,\cdots,{\widetilde{U}}\widetilde{P}\}$, and the eigenvector corresponding to $\chi_{max}$ as the signal subspace $\mathbf{q}$, which satisfies
\begin{align}
\mathbf{R}_{\mathbf{x}_{C}}\mathbf{q}=\chi_{max}\mathbf{q}.
\end{align}
Now, the subspace spanned by $\mathbf{q}$ is the signal subspace spanned by  $\mathbf{a}$ so that the following relationship holds true
\begin{equation}
\mathbf{a}=\delta\mathbf{q},
\end{equation}
where $\delta$ is a non-zero scalar. If we consider the two vectors $\mathbf{a}_{1}$ and $\mathbf{a}_{2}$ , obtained by taking the first and the last ${\widetilde{U}}\times (\widetilde{P}-1)$ elements of $\mathbf{a}$, respectively, it is $\mathbf{a}_{2} = \mathbf{a}_{1}{\mathbf{\Phi}}$, where $\mathbf{\Phi}=e^{ir}$. To obtain an estimate of  $\mathbf{\Phi}$ and $r$, we construct the two vectors $\mathbf{q}_{1}$ and $\mathbf{q}_{2}$,  composed by the first  and  by the last ${\widetilde{U}}\times (\widetilde{P}-1)$ elements of $\mathbf{q}$, respectively. Then, exploiting the fact that $\mathbf{a}_{1}=\delta\mathbf{q}_{1}$ and $\mathbf{a}_{2}=\delta\mathbf{q}_{2}$, one obtains
\begin{equation}
\mathbf{q}_{2}= \mathbf{q}_{1}{\mathbf{\Phi}},
\end{equation}
which leads to
\begin{equation}\label{Phi}
e^{ir}=\mathbf{\Phi}=\mathbf{q}_{1}^\dagger\mathbf{q}_{2}.
\end{equation}
\par
In the estimation of $\gamma$, we stack the rows of the matrix $\mathbf{\tilde{X}}_{{\widetilde{U}},\widetilde{P}}$  in the  vector denoted by $\mathbf{x}_{R}$,
\begin{align*}
\mathbf{x}_{R}=\rvec\left(\mathbf{\tilde{X}}_{{\widetilde{U}},\widetilde{P}}\right)
=[\tilde{x}(\ell_1,k_1), \tilde{x}(\ell_1,k_2),\cdots, \tilde{x}(\ell_{\widetilde{U}},k_{\widetilde{P}})]^T,
\end{align*}
so that  $\mathbf{x}_{R}$ can be expressed as
\begin{align}
\mathbf{x}_{R}=\mathbf{b}+\mathbf{z}_{R},
\end{align}
where $\mathbf{b}=[e^{ik_1r}e^{i\ell_1\gamma}, e^{ik_2r}e^{i\ell_1\gamma},\cdots,e^{ik_{\widetilde{P}}r}e^{i\ell_{\widetilde{U}}\gamma}]^T$ and $\mathbf{z}_{R}$ is the noise vector.  By computing the correlation matrix $\mathbf{R}_{\mathbf{x}_{R}}=\mathbb{E}\left\{\mathbf{x}_{R}\mathbf{x}_{R}^H\right\}$ and following the same method described for obtaining an estimate of $r$ we can obtain an estimate of $\gamma$.

\subsubsection{Estimation of $\alpha$ and $\varphi$}
After estimating $r$ and $\gamma$ based on 2-D ESPRIT algorithm, we can obtain the elevation angle $\alpha$ with $r$ and the amplitudes $\{|x'(\ell_{u},k_p)|~|p=1,2,\dots,\widetilde{P}; {u}=1,2,\cdots,{\widetilde{U}}\}$ through the equations
\begin{align}
&{J_{\ell_u}}(k_pR_t\sin\alpha){J_0}(k_pR_r\sin\alpha)\approx \delta(\ell_{u},k_p),\nonumber\\
&p=1,2,\dots,\widetilde{P}; \ u=1,2,\cdots,{\widetilde{U}},
\label{amplitude}
\end{align}
where $\delta(\ell_{u},k_p)= \frac{\left|x'(\ell_{u},k_p)\right|} {\left|\sigma_{\ell_{u},k_p}\right|}{\rm{sign}}({x'}(\ell_{u},k_p))$, the approximation neglects the effect of noise at high receive signal-to-noise ratio (SNR). Thus, all the solutions of $\alpha$ can be obtained from \eqref{amplitude} numerically \cite{Bremer2019}.

The receiver can receive signals from the transmitter only when the transmitter is located within the main lobe of the receive UCA. Thus, we assume that $[\alpha_a,\alpha_b]$ is the angle range of the receive UCA's main lobe, and obtain a number of solutions of \eqref{amplitude} denoted as $\{\hat{\alpha}_w(\ell_{u},k_p)| \hat{\alpha}_w(\ell_{u},k_p)\in [\alpha_a,\alpha_b], w=1,2,\ldots,W_{\ell_{\widetilde{U}},k_{\widetilde{P}}}\}$, where $\hat{\alpha}_w(\ell_{u},k_p)$ is the $w$th possible solution of $\alpha$ corresponding to $x'(\ell_{u},k_p)$. In order to determine the final estimate of $\alpha$, we first look for ${\widetilde{U}}\times \widetilde{P}$ solutions $\{\hat{\alpha}_w(\ell_{u},k_p)\}$ by the following method:
\begin{description}
  \item[(i)] Divide $[\alpha_a,\alpha_b]$ into $\mathcal{D}$ intervals denoted as $[\alpha_{a1},$\\$\alpha_{b1}],[\alpha_{a2},\alpha_{b2}],\cdots, [\alpha_{a\mathcal{D}},\alpha_{b\mathcal{D}}]$, where $\alpha_{a1}=\alpha_a$ and $\alpha_{b\mathcal{D}}=\alpha_b$;
  \item[(ii)] Categorize all the solutions $\{\hat{\alpha}_w(\ell_{u},k_p)\}$ into corresponding intervals;
  \item[(iii)] Count the number of solutions in each interval and denote it as $M_{\epsilon}, \epsilon=1,2,\cdots,\mathcal{D}$;
  \item[(iv)] Let $\mathcal{M}=\max\{M_{\epsilon}|\epsilon=1,2,\cdots,\mathcal{D}\}$, if $\mathcal{M}>{\widetilde{U}}\times \widetilde{P}$, $\mathcal{D}=\mathcal{D}+1$, and return to (i);
      else if $\mathcal{M}={\widetilde{U}}\times \widetilde{P}$, $\hat{\alpha}_w(\ell_{u},k_p)$ within $[\alpha_{a\epsilon},\alpha_{b\epsilon}]$ is used as the estimates of $\alpha$, denoted as $\hat{\alpha}(\ell_{u},k_p)$.
\end{description}

In this case, the ${\widetilde{U}}\times \widetilde{P}$ estimates of $\alpha$ can be expressed as
\begin{align}
\hat{\alpha}(\ell_{u},k_p)=&\alpha+\varepsilon(\ell_{u},k_p), p=1,2,\dots,\widetilde{P}; \  {u}=1,2,\cdots,{\widetilde{U}},\nonumber
\end{align}
where $\varepsilon(\ell_1,k_1), \varepsilon(\ell_2,k_1), \cdots, \varepsilon(\ell_{\widetilde{U}},k_{\widetilde{P}})$ represent the estimation errors. Assume $\varepsilon(\ell_1,k_1), \varepsilon(\ell_2,k_1), $ $ \cdots, \varepsilon(\ell_{\widetilde{U}},k_{\widetilde{P}})$ have the average variance $\textrm{Var}(\varepsilon)$. Thus,
\begin{align}
\textrm{Var}\left(\frac{1}{{\widetilde{U}}{\widetilde{P}}}\sum_{p=1}^{\widetilde{P}}\sum_{{u}=1}^{\widetilde{U}}\hat{\alpha}(\ell_{u},k_p)\right) = \frac{\textrm{Var}\left(\varepsilon\right)}{{\widetilde{U}}{\widetilde{P}}}.
\end{align}
Therefore, $\hat{\alpha}=\frac{1}{{\widetilde{U}}{\widetilde{P}}}\sum_{p=1}^{\widetilde{P}}\sum_{{u}=1}^{\widetilde{U}}\hat{\alpha}(\ell_{u},k_p)$ is adopted as the estimate of $\alpha$.
After obtaining $\hat{\alpha}$, the estimated value of $\varphi$ can be calculated from \eqref{gamma} as $\hat{\varphi}=\arccos\left(\cos\hat{\gamma}/\cos\hat{\alpha}\right)$, where $\hat{\gamma}$ is the estimated value of $\gamma$. Thus, the AoA estimation of MCMM-OAM beam is completed.


\section{Reception of Muti-mode OAM Signals}
In this section, we first investigate the effect of $\varphi$ and $\alpha$ on the LoS multi-mode OAM channel, and then find the effect of the proposed beam steering on inter-mode interferences. At last, we specify how to design the signal detection matrix $\mathbf{D}(k_p)$.

\vspace{-0.8em}
\subsection{Effects of $\varphi$ and $\alpha$ on OAM Inter-mode Interferences}
By assuming that the transmit and receive UCAs are in the far-field distance region of each other, i.e. $r\gg R_t$ and $r\gg R_r$, we can approximate $d_{m,n}$ in \eqref{dmn} as
\begin{align} \label{dmn appx}
d_{m,n}&\overset{(a)}{\approx} \sqrt{R_t^2+R_r^2+r^2}-\frac{rR_r\cos\theta\cos\varphi\sin\alpha}{\sqrt{R_t^2+R_r^2+r^2}}\nonumber\\
&\quad -\frac{R_tR_r(\cos\phi\cos\theta\cos\gamma+\sin\phi\sin\theta\cos\varphi)}{\sqrt{R_t^2+R_r^2+r^2}}\nonumber\\
&\quad -\frac{R_tR_r(\sin\phi\cos\theta\sin\varphi-\cos\phi\sin\theta\sin\varphi\cos\alpha)}{\sqrt{R_t^2+R_r^2+r^2}}\nonumber\\
&\quad +\frac{rR_r\sin\theta\sin\varphi\sin\alpha}{\sqrt{R_t^2+R_r^2+r^2}}\nonumber\\
&\overset{(b)}{\approx} r\!-\!\frac{R_tR_r}{r}(\cos\phi\cos\theta\cos\gamma+\sin\phi\sin\theta\cos\varphi)\nonumber\\
&\quad -\frac{R_tR_r}{r}(\sin\phi\cos\theta\sin\varphi-\cos\phi\sin\theta\sin\varphi\cos\alpha)\nonumber\\
&\quad -R_r(\cos\theta\cos\varphi\sin\alpha-\sin\theta\sin\varphi\sin\alpha),
\end{align}
where (a) uses the method of completing a square and the condition $r\gg R_t, R_r$ as same as the simple case $\sqrt{a^2-2b}\approx a-\frac{b}{a}, a\gg b$; (b) is directly obtained from the condition $r\gg R_t, R_r$. Then, substituting \eqref{dmn appx} into \eqref{FreeSpaceChannel} and abbreviating $h(k_p,d_{m,n})$ to $h_{m,n}(k_p)$, we thus have
\begin{align} \label{hmn}
&h_{m,n}(k_p)\!\overset{(a)}\approx\!\frac{\beta}{2k_pr}\!\exp\bigg(-ik_pr+ik_pR_r\cos\theta\cos\varphi\sin\alpha \nonumber\\
&\!+\!i\frac{k_pR_tR_r}{r}(\cos\phi\cos\theta\cos\gamma+\sin\phi\sin\theta\cos\varphi)\nonumber\\
&\!+\!i\frac{k_pR_tR_r}{r}(\sin\phi\cos\theta\sin\varphi\!-\!\cos\phi\sin\theta\sin\varphi\cos\alpha)\nonumber\\
&\!-\!ik_pR_r\sin\theta\sin\varphi\sin\alpha \bigg),
\end{align}
where (a) neglects a few small terms in the denominator and thus only $4\pi r$ is left. Having \eqref{hmn}, the $u$th-row and $v$th-column element $h_{\textrm{OAM},k_p}(u,v)$ in ${\bf{H}}_{\textrm{OAM}}(k_p)$ is calculated as
\begin{align}\label{heffDev}
&h_{\textrm{OAM},k_p}(u,v) =\frac{1}{N}\sum\limits_{m = 1}^N \sum\limits_{n = 1}^N {h_{m,n}}(k_p)\exp \left(-i{\ell_u}\theta+i{\ell_v}\phi \right) \nonumber\\
&=\eta(k_p)\!\!\sum\limits_{m = 1}^N \!\sum\limits_{n = 1}^N\exp\!\bigg(\!-\!i{\ell_u}\theta\!+\!i{\ell_v}\phi\!+\!ik_pR_r\cos\theta\cos\varphi\sin\alpha \nonumber\\
&+i\frac{k_p R_tR_r}{r}\left(\cos\phi\cos\theta\cos\gamma+\sin\phi\sin\theta\cos\varphi\right) \nonumber\\
&+i\frac{k_p R_tR_r}{r}\left(\sin\phi\cos\theta\sin\varphi-\cos\phi\sin\theta\sin\varphi\cos\alpha\right) \nonumber \\
&-ik_pR_r\sin\theta\sin\varphi\sin\alpha \bigg)\nonumber \\
&\overset{(a)}\approx\eta(k_p)\sum\limits_{q = 1}^N\exp\left(i\frac{2\pi q}{N}\ell_v+i\frac{k_p R_tR_r}{r}\cos\frac{2\pi q}{N}\right)\times\nonumber\\
&\sum\limits_{m = 1}^N\sum\limits_{n = 1}^N\exp \bigg(ik_p R_r\cos\varphi\sin\alpha\cos\frac{2\pi (m-1)}{N} \nonumber \\
&+i\frac{k_p R_tR_r}{r}\sin\varphi\sin\frac{2\pi (n-1)}{N}\cos\frac{2\pi (m-1)}{N}\nonumber \\
&-i\frac{k_p R_tR_r}{r}\sin\varphi\cos\frac{2\pi (n-1)}{N}\sin\frac{2\pi (m-1)}{N}\nonumber \\
&-ik_p R_r\sin\varphi\sin\alpha\sin\frac{2\pi (m-1)}{N}-i\frac{2\pi (m-1)}{N}t \bigg),
\end{align}
where $\eta(k_p)=\frac{\beta}{2k_p rN}\exp(-ik_pr)$, $q=n-m\in \mathds{Z}$, $t=\ell_u-\ell_v\in \mathds{Z}$, $u,v=1,2,\cdots,U$, $p=1,2,\cdots,P$, (a) applies the approximation $\cos a\approx1-\frac{a^2}{2}$ for $\cos\alpha$, $\cos\varphi$ and $\cos\gamma$ in the case that $\alpha$, $\varphi$ and $\gamma$ are relatively small, and neglects a few small terms under the condition $r\gg R_r$. Denote the second summation in \eqref{heffDev} as
\begin{small}
\begin{align}\label{H1}
&\xi(\alpha,\varphi,t)=\sum\limits_{m = 1}^N\sum\limits_{n = 1}^N\exp \bigg(ik_p R_r\cos\varphi\sin\alpha\cos\frac{2\pi (m-1)}{N} \nonumber \\
&\quad +i\frac{k_p R_tR_r}{r}\sin\varphi\sin\frac{2\pi (n-1)}{N}\cos\frac{2\pi (m-1)}{N}\nonumber \\
&\quad -i\frac{k_p R_tR_r}{r}\sin\varphi\cos\frac{2\pi (n-1)}{N}\sin\frac{2\pi (m-1)}{N}\nonumber \\
&\quad -ik_p R_r\sin\varphi\sin\alpha\sin\frac{2\pi (m-1)}{N}-i\frac{2\pi (m-1)}{N}t \bigg),
\end{align}
\end{small}
Then, we can observe from \eqref{H1} that if $\varphi=0$, $\alpha=0$ and $t=0$, $\xi=N$; if $\varphi=0$, $\alpha=0$ and $t\neq0$, then $\xi=0$, that is to say, $\mathbf{H}_{\textrm{OAM}}$ is diagonal in perfect alignment case. However, when $\varphi\neq0$ or $\alpha\neq0$ and $t\neq0$, then $\xi\neq0$, which indicates that $\varphi$ and $\alpha$ result in inter-mode interferences. Moreover, even if $\alpha$ and/or $\varphi$ have small values, large interferences occur.

\vspace{-0.2em}
\subsection{Beam Steering for Inter-mode Interference Cancellation}

To alleviate the inter-mode interferences induced by the misalignment, we propose applying the beam steering to the UCA-based LoS multi-mode OAM communication systems. The beam steering approach steers the beam pattern towards the direction of the incident OAM beam and thus compensates the changed phases caused by $\alpha$ and $\varphi$ at the receive UCA.
Through calculating the phase difference between the reference element and the $m$th element of the receive UCA, the receive beam steering matrix $\mathbf{B}(k_p)$ can be designed as $\mathbf{B}(k_p)=\mathbf{1}\otimes\mathbf{\mathfrak{b}}(k_p)$, where $\mathbf{\mathfrak{b}}(k_p)=[e^{iW_1(k_p)}, e^{iW_2(k_p)}, \cdots, e^{iW_N(k_p)}]$,
\begin{equation} \label{W_m}
W_m(k_p)=k_p R_r\left(\sin\theta\sin\varphi\sin\alpha-\cos\theta\cos\varphi\sin\alpha\right),
\end{equation}
$m=1,\cdots,N$, $p=1,\cdots,P$. After involving these phases into the original phases in $\mathbf{F}_U$ at the receive UCA, the effective multi-mode OAM channel matrix at the $p$th subcarrier becomes
\begin{equation} \label{HOAMnew}
\mathbf{H'}_{\textmd{OAM}}(k_p)=\left(\mathbf{F}_U\odot\mathbf{B}(k_p)\right) \mathbf{H}(k_p)\mathbf{F}_U^H.
\end{equation}
\begin{thm}
For a LoS MCMM-OAM communication system composed of an $N$-elements transmit UCA and an $N$-elements receive UCA, beam steering matrix $\mathbf{B}(k_p)$ can eliminate inter-mode interferences induced by $\alpha$ and $\varphi$ in non-parallel misalignment case.
\label{thm:BeamSteering}
\end{thm}

\begin{IEEEproof}
The $u$th-row and $v$th-column element $h'_{\textrm{OAM},k_p}(u,v)$ in $\mathbf{H'}_{\textrm{OAM}}(k_p)$ can be obtained as
\small
\begin{align}\label{heffDev1}
&h'_{\textrm{OAM},k_p}(u,v)\!=
\!\frac{1}{N}\sum\limits_{m = 1}^N \sum\limits_{n = 1}^N {h_{m,n}}(k_p)\exp (-i{\ell_u}\theta\!+\!i{\ell_v}\phi\!+\!iW_m(k_p) ) \nonumber\\
&=\eta(k_p)\sum_{m = 1}^N\sum_{n = 1}^N \exp\bigg(-i{\ell_u}\theta\!+\!i{\ell_v}\phi\!+\!i\frac{k_p R_tR_r}{r}\sin\phi\cos\theta\sin\varphi \nonumber\\
&\quad +i\frac{k_p R_tR_r}{r}(\cos\phi\cos\theta\cos\gamma+\sin\phi\sin\theta\cos\varphi)\nonumber\\
&\quad-i\frac{k_p R_tR_r}{r}\cos\phi\sin\theta\sin\varphi\cos\alpha\bigg)\nonumber\\
&\overset{(a)}\approx \eta(k_p)\sum\limits_{q = 1}^N\exp\bigg(i\frac{2\pi q}{N}\ell_v+i\frac{k_p R_tR_r}{r}\cos\frac{2\pi q}{N}\bigg)\nonumber\\
&\quad \times\sum\limits_{m = 1}^N \exp \left(-i\frac{2\pi (m-1)}{N}t\right),
\end{align}
\normalsize
where (a) applies the approximation $\cos a\approx1-\frac{a^2}{2}$ for $\cos\alpha$, $\cos\varphi$ and $\cos\gamma$, and neglects a few small terms in the condition $r\gg R_r$. From \eqref{heffDev1}, we find that when $t\neq0$, even if $\varphi\neq0$ and $\alpha\neq0$,
\begin{equation*}
\sum_{m=1}^N \exp\left(-i\frac{{2\pi(m-1)t}}{N}\right)=\frac{1-\exp(-i2\pi t)}{1-\exp(-i\frac{2\pi}{N}t)}=0,
\end{equation*}
That is to say, $\mathbf{H'}_{\textmd{OAM}}(k_p)$ $=$ $\textrm{diag}\{h'_{\textrm{OAM},k_p}(1,1),$ $h'_{\textrm{OAM},k_p}(2,2),$ $\cdots,h'_{\textrm{OAM},k_p}(U,U)\}$, which proves that beam steering almost completely eliminates inter-mode interferences.
\end{IEEEproof}

\subsection{Multi-mode OAM Signal Reception}
Since after proper beam steering $\mathbf{H'}_{\textmd{OAM}}(k_p)$ becomes diagonal,
we first consider the behavior of the diagonal elements $h'_{\textrm{OAM},k_p}(u,u), u=1,2,\cdots,U$, $p=1,2,\cdots,P$.

Define $S_{k_p}=k_pR_tR_r/r$, according to \eqref{heffDev1}, the $u$th diagonal element of $\mathbf{H'}_{\textmd{OAM}}(k_p)$ takes the form
\begin{align} \label{diagonal}
h'_{\textrm{OAM},k_p}(u,u)= N\eta(k_p)\sum\limits_{q = 1}^N\exp\bigg(i\frac{2\pi q}{N}\ell_u+i S_{k_p}\cos\frac{2\pi q}{N}\bigg).
\end{align}
\begin{thm}
For a LoS MCMM-OAM communication system composed of an $N$-element transmit UCA with radius $R_t$ and an $N$-element receive UCA with radius $R_r$ at a distance $r~(r\gg R_t,R_r)$ from each other in the non-parallel misalignment case, after beam steering the effective channel gain of each OAM mode $h'_{\textrm{OAM},k_p}(u,u)$ approaches a function of OAM mode number $\ell$ and wave number $k_p$.
\label{thm:CharacDiag}
\end{thm}
\begin{IEEEproof}
Recall that the Taylor series expansion of the exponential function is $e^x=\sum_{g = 0}^{+\infty}(x^g)/g!$. Then, we substitute it into \eqref{diagonal} and have
\begin{align}\label{Taylor}
&h'_{\textrm{OAM},k_p}(u,u) \!= \!N\eta(k_p)\!\sum\limits_{q = 1}^N\!\exp\!\left(\!i\frac{2\pi q}{N}\ell_u\!\right)\!\exp\!\left(\!i S_{k_p}\cos\frac{2\pi q}{N}\!\right)\nonumber\\
&=N\eta(k_p)\sum\limits_{q = 1}^N\exp\left(i\frac{2\pi q}{N}\ell_u\right)\times\nonumber\\
&\quad \sum\limits_{g = 0}^{+\infty}\frac{i^g}{g!}\bigg(\frac{\exp(i\frac{2\pi q}{N})+\exp\left(-i\frac{2\pi q}{N}\right)}{2}\bigg)^gS^g_{k_p} \nonumber\\
&=N\eta(k_p)\sum\limits_{g = 0}^{+\infty}\frac{i^g}{2^g g!}\sum\limits_{q = 1}^N\exp\left(i\frac{2\pi q}{N}\ell_u\right)\times \nonumber\\
&\quad \sum\limits_{l = 0}^{g}\bigg(
\begin{matrix}
g\\
l
\end{matrix}
\bigg)\exp\left(il\frac{2\pi q}{N}\right)\exp\bigg(-i(g-l)\frac{2\pi q}{N}\bigg)S^g_{k_p}\nonumber\\
&=N\eta(k_p)\!\sum\limits_{g = 0}^{+\infty}\frac{i^g}{2^gg!}\sum\limits_{l = 0}^g\!\bigg(
\begin{matrix}
g\\
l
\end{matrix}
\bigg)
\!\sum\limits_{q = 1}^N\exp\!\bigg(\!i(\ell_u+2l\!-\!g)\frac{2\pi q}{N}\!\bigg)S^g_{k_p}.
\end{align}
For the last summation in \eqref{Taylor}, it is easy to verify that
\begin{align} \label{summation}
&\sum\limits_{q = 1}^N\exp\bigg(i(\ell_u+2l-g)\frac{2\pi q}{N}\bigg)=\left\{
\begin{smallmatrix}
0, &\textrm{Mod}_N(\ell_u+2l-g)\neq0;\\
N, &\textrm{Mod}_N(\ell_u+2l-g)=0.
\end{smallmatrix}\right.
\end{align}
When $0\leq g\leq\min\left\{|\ell_u|, N-|\ell_u|\right\}$, we always have (recall that $0 \leq l \leq g$)
\begin{align} \label{ellv}
-N<\ell_u+2l-g< N.
\end{align}
The detailed derivation of \eqref{ellv} is omitted due to limited space. When $0\leq g< \min\left\{|\ell_u|, N-|\ell_u|\right\}$, \eqref{summation} is always zero, i.e., the Taylor series of $h'_{\textrm{OAM},k_p}(u,u)$ does not contain the terms $S^g_{k_p}$, and when $g=\min\left\{|\ell_u|, N-|\ell_u|\right\}$, \eqref{summation} equals to $N$ if and only if
\begin{align}
&l=\left\{
\begin{matrix}
0, &\ell_u\geq0;\\
g, &\ell_u<0.
\end{matrix}\right.
\end{align}
Thus, the coefficient of $S^{\min\left\{|\ell_u|, N-|\ell_u|\right\}}_{k_p}$ in \eqref{Taylor} is written as
\begin{align}
\left\{
\begin{matrix}
\frac{N}{2^{|\ell_u|}}\cdot\frac{i^{|\ell_u|}}{|\ell_u|!}, &g=|\ell_u|; \\
\frac{N}{2^{(N-|\ell_u|)}}\cdot\frac{i^{(N-|\ell_u|)}}{(N-|\ell_u|)!}, &\qquad g=N-|\ell_u|.
\end{matrix}\right.
\end{align}
Furthermore, when $g> \min\left\{|\ell_u|, N-|\ell_u|\right\}$, the coefficient of $S^g_{k_p}$ is much smaller due to the term $\frac{1}{2^gg!}$, and hence the term $\{S^g_{k_p}|g>\min\left\{|\ell_u|, N-|\ell_u|\right\}$ is counted into the higher order infinitesimal of $S^g_{k_p}$ denoted as $o\{S^g_{k_p}\}$ as $S_{k_p}$ goes to zero.

Therefore, for $r\gg R_t, R_r$ the $u$-th diagonal elements of ${\bf{H}}^{'}_{\textrm{OAM}}(k_p)$ can be approximately obtained as
\begin{align} \label{hiord}
h'_{\textrm{OAM},k_p}(u,u)\approx\eta(k_p)\frac{N^2}{2^\tau}\frac{i^\tau}{\tau!}\cdot S^\tau_{k_p},
\end{align}
where $\tau={\min\left\{|\ell_u|, N-|\ell_u|\right\}}$. It follows that in the long-distance communication scenario all diagonal elements $\{h'_{\textrm{OAM},k_p}(u,u), u=1,2,\cdots,U$, $p=1,2,\cdots,P\}$ can be analytically characterized.
\end{IEEEproof}

\begin{figure}[t]
\setlength{\abovecaptionskip}{-0cm}   
\setlength{\belowcaptionskip}{-0.2cm}   
\begin{center}
\includegraphics[width=9.0cm,height=3.2cm]{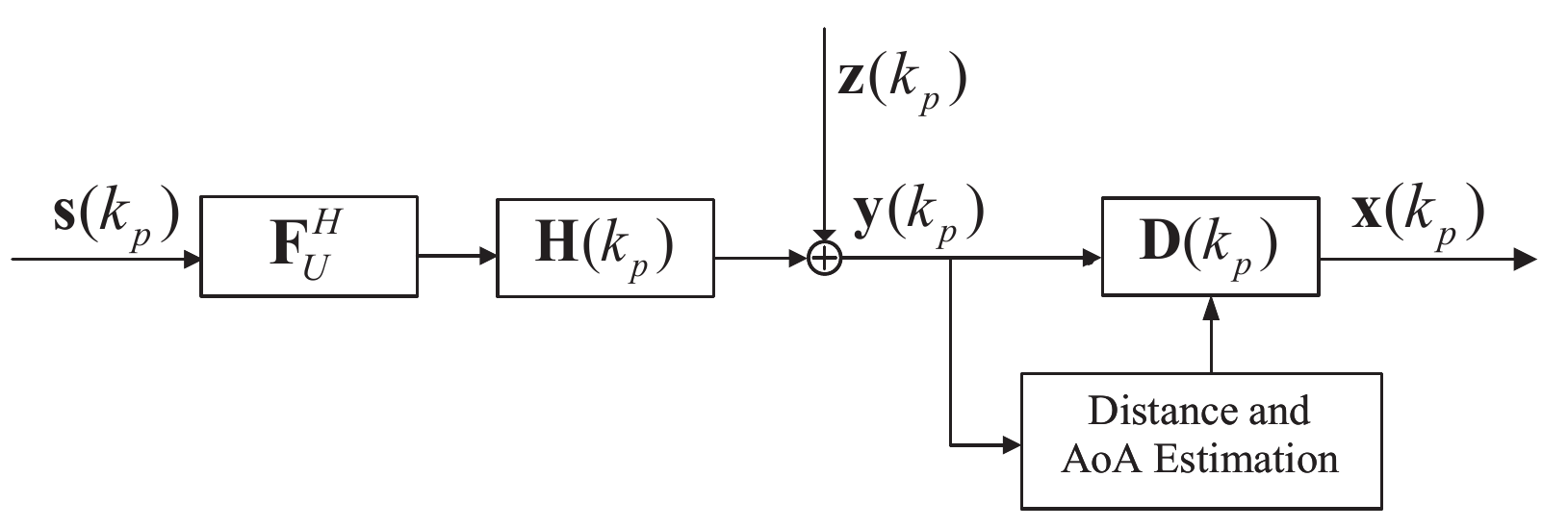}
\end{center}
\caption{The block diagram of a LoS MCMM-OAM communication system with the signal detection assisted by the distance and AoA estimation at the receiver.}
\label{Fig106}
\end{figure}

Given that $k_p$, $R_t$, $R_r$, $N$ and $\ell_u$ are known to the receiver, the effective channel coefficient $h'_{\textrm{OAM}, k_p}(u,u)$ of each OAM mode is only the function of $r$, which leads to very simple signal detection, i.e.,
\begin{align} \label{Ym}
\mathbf{x}(k_p)&=\mathbf{\Gamma}^{-1}(k_p)\mathbf{F}_U\odot\mathbf{B}(k_p)\left(\mathbf{H}(k_p)\mathbf{F}_U^H\mathbf{s}(k_p) +\mathbf{z}(k_p)\right)\nonumber \\
&\approx\mathbf{s}(k_p)+\mathbf{\hat{z}}(k_p), p=1,2,\cdots,P,
\end{align}
where $\mathbf{\hat{z}}(k_p)=\mathbf{\Gamma}^{-1}(k_p)\mathbf{F}_U\odot\mathbf{B}(k_p) \mathbf{z}(k_p)=[\hat{z}(\ell_1,k_p),$ $\hat{z}(\ell_2,k_p),\cdots,\hat{z}(\ell_U,k_p)]^T$, $\mathbf{\Gamma}(k_p)=\textrm{diag}\{\zeta(\ell_1,k_p),\cdots,$ $  \zeta(\ell_U,k_p)\}$ is a diagonal matrix, and
\begin{align} \label{zeta}
\zeta(\ell_u,k_p)&=\eta(k_p)\frac{N^2}{2^\tau}\frac{i^\tau}{\tau!}\cdot \hat{S}^{\tau}_{k_p}, \quad u=1,2,\cdots,U,
\end{align}
$\hat{S}_{k_p}=k_pR_tR_r/\hat{r}$. Following the signal model in \eqref{x}, the detection matrix $\mathbf{D}(k_p)$ can be designed as the cascade of the beam steering matrix $\mathbf{B}(k_p)$, the despiralization matrix $\mathbf{F}_U$ and the amplitude detection matrix $\mathbf{\Gamma}^{-1}(k_p)$, i.e.,
\begin{equation}
\mathbf{D}(k_p)=\mathbf{\Gamma}^{-1}(k_p)\mathbf{F}_U\odot\mathbf{B}(k_p).
\end{equation}
Fig. \ref{Fig106} shows the block diagram of a LoS MCMM-OAM communication system with the signal detection assisted by the distance and AoA estimation at the receiver.

Denote the estimate of $r$ as $\hat{r}$. Then, if $\hat{r}$, $\hat{\alpha}$ and $\hat{\varphi}$ are accurate enough, the effective multi-mode OAM channel matrix $\mathbf{H'}_{\textmd{OAM}}(k_p)$ approaches diagonal according to Theorem 1. Since $\mathbf{F}_U\odot\mathbf{B}(k_p)$ is semi-unitary ($\mathbf{F}_U\odot\mathbf{B}(k_p)(\mathbf{F}_U^H\odot\mathbf{B}^H(k_p))=\mathbf{I}$) and does not change the noise power, beam steering and amplitude detection convert the vector signal detection into multiple independent scalar signal detection greatly reducing the computational complexity. However, considering the inevitable estimation errors of $r$, $\alpha$ and $\varphi$ in practice and the higher order infinitesimal term in \eqref{hiord}, we have to evaluate the effect of these errors on the system SE and detection performance. Following \eqref{Ym}, we have
\begin{align} \label{y2}
x(\ell_u,k_p) & = \frac{1}{\zeta(\ell_u,k_p)}\sum\limits_{v = 1}^U h'_{\textrm{OAM},k_p}(u,v)s(\ell_v,k_p) + \hat{z}(\ell_u,k_p) \nonumber\\
& = s(\ell_u,k_p)+\frac{o\{S^\tau_{k_p}\}s(\ell_u,k_p)}{\zeta(\ell_u,k_p)} + \nonumber\\
&\quad \sum\limits_{v \neq u} h'_{\textrm{OAM},k_p}(u,v)\frac{s(\ell_v,k_p)}{\zeta(\ell_u,k_p)} + \frac{\hat{z}(\ell_u,k_p)}{\zeta(\ell_u,k_p)}.
\end{align}
Therefore, the signal-to-interference-plus-noise ratio (SINR) of the $u$th mode OAM at the $p$th subcarrier can be formulated as
\begin{align} \label{SINR}
&\textrm{SINR}(\ell_u,k_p)= \nonumber\\ &\frac{\zeta(\ell_u,k_p)^2\mathbb{E}\left(\left|s(\ell_u,k_p)\right|^2\right)} {\sum\limits_{v=1}^U|L(u,v,k_p)|^2 \mathbb{E}\left(\left|s(\ell_v,k_p)\right|^2\right) +\mathbb{E}\left(\left|z(\ell_u,k_p)\right|^2\right)},
\end{align}
where
\begin{align}
L(u,v,k_p)=\left\{
\begin{matrix}
 h'_{\textrm{OAM},k_p}(u,v), &v\neq u;\\
o\{S^\tau_{k_p}\}, &v=u.
\end{matrix}\right.
\end{align}
Thus, the SE of the LoS MCMM-OAM communication system described in \eqref{Ym} can be written as
\begin{equation} \label{eq9}
C = \left(1-\frac{T_p\widetilde{P}}{T_cP}\right)\frac{1}{P}\sum_{p=1}^P\sum_{u=1}^U\log_2\left(1 + {\textrm{SINR}}(\ell_u,k_p)\right),
\end{equation}
where $T_c$ is the length of the coherence time over which the channel can be assumed constant, and $T_p$ is the time spent in transmitting training symbols.

\section{UCCA-Based LoS OAM-MIMO Communications}
\vspace{1em}
\subsection{LoS MCMM-OAM-MIMO Systems With UCCAs}
\begin{figure}[t]
\setlength{\abovecaptionskip}{-0.1cm}   
\setlength{\belowcaptionskip}{-0.1cm}   
\begin{center}
\qquad \quad \includegraphics[width=5.7cm,height=3.5cm]{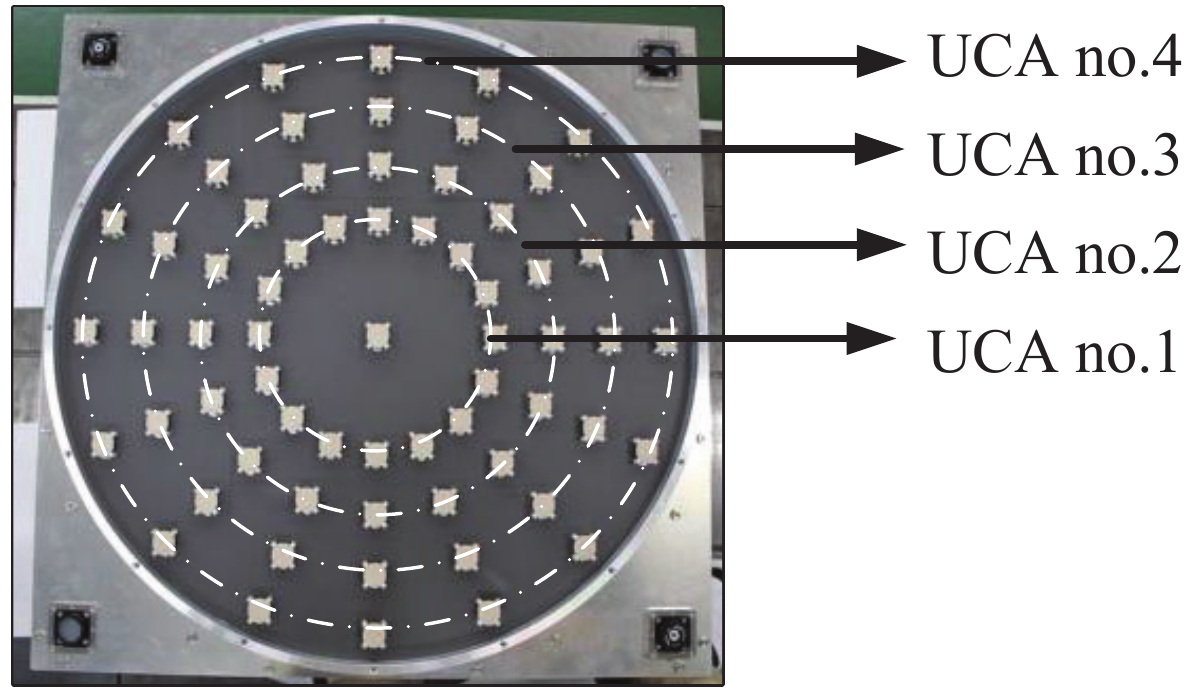}
\end{center}
\caption{The prototype of an uniform concentric circular array with 64 elements for OAM-MIMO multiplexing in \cite{lee2018experimental}.}
\label{figure5}
\end{figure}
In order to further enhance the degree of freedom and increase the multiplexing gain of the UCA-based LoS OAM communication systems, multiple concentric UCAs are exploited in the transmitter and the receiver to achieve 100 Gbps data rate \cite{lee2018experimental}. The prototype of the UCCA for LoS OAM-MIMO communications is shown in Fig. \ref{figure5}.

For the aforementioned LoS MCMM-OAM system combined with MIMO (i.e, replace UCAs with UCCAs), we assume that the UCCAs at the transmitter and the receiver both consist of $\mathfrak{N}$ concentric UCAs with radii $R_{t_\mathfrak{n}}$ and $R_{r_\mathfrak{m}}$, $\mathfrak{m},\mathfrak{n}=1,2,\cdots,\mathfrak{N}$. Then, the channel matrix of the UCCA-based communication system can be written as $\bar{\mathbf{H}}(k_p)=\left[\mathbf{H}_{\mathfrak{m},\mathfrak{n}}(k_p)\right]_{\mathfrak{N}\times\mathfrak{N}}$,
%
%
where $\mathbf{H}_{\mathfrak{m},\mathfrak{n}}(k_p) (\mathfrak{m},\mathfrak{n}=1,2,\cdots,\mathfrak{N})$ denotes the $N\times N$ channel matrix from the $\mathfrak{n}$th UCA in the transmit UCCA to the $\mathfrak{m}$th UCA in the receive UCCA. When $\alpha=0$ and $\varphi=0$, $\{\mathbf{H}_{\mathfrak{m},\mathfrak{n}}|\mathfrak{m},\mathfrak{n}=1,2,\cdots,\mathfrak{N}\}$ are all circulant matrices.

\subsection{Reception of LoS MCMM-OAM-MIMO Signals}
As the distances and AoAs for every pair of the transmit and receive UCAs are exactly the same, the proposed AoA estimation and beam steering methods based on a single UCA in Section IV and V can be directly utilized in the UCCA-based system. Hence, the required number of training symbols of the LoS MCMM-OAM-MIMO system does not increase with the number of subcarriers, transmit and receive antenna elements, which is opposite in traditional MIMO-OFDM channel estimation. Thus,
the effective channel of the LoS MCMM-OAM-MIMO system can be written as
\begin{align}\label{HOAMMIMO1}
\mathbf{H'}_\textmd{OAM-MIMO}(k_p)&=\mathbf{I_{\mathfrak{N}}}\otimes\mathbf{F}_{U}\odot\bar{\mathbf{B}}(k_p) \bar{\mathbf{H}}(k_p)\mathbf{I_{\mathfrak{N}}}\otimes\mathbf{F}_{U}^H \nonumber \\
&=\left[\mathbf{H'}_{\textrm{OAM}_{\mathfrak{m},\mathfrak{n}}}(k_p)\right]_{\mathfrak{N}\times\mathfrak{N}},
\end{align}
where $\bar{\mathbf{B}}(k_p)$$=$$\textrm{diag} \{ \mathbf{B}_{1}(k_p), \mathbf{B}_{2}(k_p), \cdots, \mathbf{B}_{\mathfrak{N}}(k_p) \}$, $\mathbf{B}_{\mathfrak{m}}(k_p)$ is the receive beam steering matrix corresponding the ${\mathfrak{m}}$-th UCA of the receive UCCA, $\mathbf{H'}_{\textrm{OAM}_{\mathfrak{m},\mathfrak{n}}}(k_p) (\mathfrak{m},\mathfrak{n}=1,2,\cdots,\mathfrak{N})$ denotes the $U\times U$ effective MCMM-OAM channel matrix from the $\mathfrak{n}$th UCA in the transmit UCCA to the $\mathfrak{m}$th UCA in the receive UCCA, and $\mathbf{H'}_{{\textmd{OAM}}_{\mathfrak{m},\mathfrak{n}}}(k_p) =\left(\mathbf{F}_{U}\odot\mathbf{B}_{\mathfrak{m}}(k_p)\right) \mathbf{H}_{\mathfrak{m},\mathfrak{n}}(k_p)\mathbf{F}_{U}^H$ is diagonal. It follows that after beam steering only inter-mode interferences are eliminated, while co-mode interferences between different UCAs are remaining. Similar to \eqref{hiord}, the elements of $\mathbf{H'}_{{\textmd{OAM}}_{\mathfrak{m},\mathfrak{n}}}(k_p)$ are functions of $r$ given that $k_p$, $R_{t_\mathfrak{n}}$, $R_{r_\mathfrak{m}}$, $N$ and $\ell_u$ are known.

Then, the signal detection for the LoS MCMM-OAM-MIMO system could be expressed as
\begin{align} \label{Ym2}
&\mathbf{\bar{x}}(k_p)=\mathbf{\bar{D}}(k_p)\mathbf{\bar{y}}(k_p)\nonumber \\
&=\mathbf{\bar{\Gamma}}^{-1}(k_p)\mathbf{I_{\mathfrak{N}}}\otimes\mathbf{F}_{U}\odot\bar{\mathbf{B}}(k_p)
\left(\bar{\mathbf{H}}(k_p)\mathbf{I_{\mathfrak{N}}}\otimes\mathbf{F}_{U}^H \mathbf{\bar{s}}(k_p)+\mathbf{\bar{z}}(k_p)\right)\nonumber \\
&=\mathbf{\bar{s}}(k_p)+\bm{\mathcal{I}}(k_p)+\mathbf{\hat{\bar{z}}}(k_p),
\end{align}
where $\mathbf{\bar{s}}(k_p)=[\mathbf{s}^T_1(k_p)$,$\mathbf{s}^T_2(k_p)$,$\cdots$,$ \mathbf{s}^T_{\mathfrak{N}}(k_p)]^T$, $\mathbf{\bar{y}}(k_p)=[\mathbf{y}^T_1(k_p),\mathbf{y}^T_2(k_p), \cdots, \mathbf{y}^T_{\mathfrak{N}}(k_p)]^T$ and $\mathbf{\bar{x}}(k_p)=[\mathbf{x}^T_1(k_p),$ $\mathbf{x}^T_2(k_p),$ $ \cdots,$ $\mathbf{x}^T_{\mathfrak{N}}(k_p)]^T$ are the transmitted, received and detected data symbol vectors of the UCCA-based systems, respectively, $\mathbf{s}_{\mathfrak{n}}(k_p)$ $=$$[s_{\mathfrak{n}}(\ell_1,k_p),$$s_{\mathfrak{n}}(\ell_2,k_p), $$ \cdots,$$s_{\mathfrak{n}}(\ell_U,k_p)]^T$, $\mathbf{y}_{\mathfrak{m}}(k_p)$$=$$[y_{\mathfrak{m}}(\ell_1,k_p),$$y_{\mathfrak{m}}(\ell_2,k_p),$$\cdots,$$y_{\mathfrak{m}}(\ell_U,k_p)]^T$ and $\mathbf{x}_{\mathfrak{m}}(k_p)$\\$=$$[x_{\mathfrak{m}}(\ell_1,k_p),$$x_{\mathfrak{m}}(\ell_2,k_p),$$\cdots, x_{\mathfrak{m}}(\ell_U,k_p)]^T$ are the corresponding transmitted, received and detected data symbol vectors on the $\mathfrak{n}$th transmit and the $\mathfrak{m}$th receive UCA, $\mathbf{\bar{z}}(k_p)$ $=$ $[\mathbf{z}^T_1(k_p),$ $ \mathbf{z}^T_2(k_p),$ $\cdots,$ $\mathbf{z}^T_{\mathfrak{N}}(k_p)]^T$ is the additive noise vector\\ on the receive UCCA, $\mathbf{z}_{\mathfrak{n}}(k_p)$ $=$ $[z_{\mathfrak{n}}(\ell_1,k_p),$ $z_{\mathfrak{n}}(\ell_2,k_p),$ $\cdots, $ $z_{\mathfrak{n}}(\ell_U,k_p)]^T$, $\mathbf{\hat{\bar{z}}}(k_p)=\mathbf{\bar{D}}(k_p)\mathbf{\bar{z}}(k_p)$, $\mathbf{\bar{D}}(k_p)=\mathbf{\bar{\Gamma}}^{-1}(k_p)\mathbf{I_{\mathfrak{N}}}$ $\otimes\mathbf{F}_{U}\odot\bar{\mathbf{B}}(k_p)$ is the detection matrix for $\mathbf{\bar{y}}(k_p)$, $\mathbf{\bar{\Gamma}}^{-1}(k_p)$ is the matrix designed for eliminating the remaining co-mode interferences between UCAs, and $\mathbf{\bar{\Gamma}}(k_p)=\left[\mathbf{\Gamma}_{\mathfrak{m},\mathfrak{n}}\right]_{\mathfrak{N}\times\mathfrak{N}}$,
%
%
$\mathbf{\Gamma}_{\mathfrak{m},\mathfrak{n}}=\textrm{diag} \{\zeta_{\mathfrak{m},\mathfrak{n}}(\ell_1,k_p),\cdots, \zeta_{\mathfrak{m},\mathfrak{n}}(\ell_U,k_p)\}$ and
\begin{equation*} \label{zeta2}
\zeta_{\mathfrak{m},\mathfrak{n}}(\ell_u,k_p)=\eta(k_p)\frac{N^2}{2^\tau}\frac{i^\tau}{\tau!}\cdot \hat{S}_{\mathfrak{m},\mathfrak{n}}^{\tau}(k_p),
\end{equation*}
$\hat{S}_{\mathfrak{m},\mathfrak{n}}(k_p)=k_pR_{t_\mathfrak{n}}R_{r_\mathfrak{m}}/\hat{r}, \ \mathfrak{m},\mathfrak{n}=1,2,\cdots,\mathfrak{N}$, and $\bm{\mathcal{I}}(k_p)=\left(\mathbf{\bar{\Gamma}}^{-1}(k_p)\mathbf{H'}_\textmd{OAM-MIMO}(k_p)- \mathbf{I}_{\mathfrak{N}U}\right)\mathbf{\bar{s}}(k_p)$ is the remaining interferences induced by the error in detection matrix. The block diagram shown in Fig.\ref{Fig106} can also be applied to the MCMM-OAM-MIMO communication system as long as $\mathbf{s}(k_p)$, $\mathbf{F}_U$, $\mathbf{H}(k_p)$, $\mathbf{z}(k_p)$, $\mathbf{y}(k_p)$, $\mathbf{D}(k_p)$ and $\mathbf{x}(k_p)$ are respectively replaced with $\bar{\mathbf{s}}(k_p)$, $\bar{\mathbf{F}}_U$, $\bar{\mathbf{H}}(k_p)$, $\bar{\mathbf{z}}(k_p)$, $\bar{\mathbf{y}}(k_p)$, $\bar{\mathbf{D}}(k_p)$ and $\bar{\mathbf{x}}(k_p)$, where $\bar{\mathbf{F}}_U=\mathbf{I_{\mathfrak{N}}}\otimes\mathbf{F}_{U}$.

\begin{table}[t]
\small
\setlength{\abovecaptionskip}{0cm}   
\setlength{\belowcaptionskip}{0cm}   
\caption{The complexity comparison between MCMM-OAM and MIMO-OFDM.}
\begin{center}
\setlength{\tabcolsep}{2mm}{
\begin{tabular}{cll}
  \toprule
  \textbf{Scheme}                                                 &\multicolumn{2}{c}{\textbf{Complexity}}\\
  \midrule
  \multirow{3}{*}{MCMM-OAM-MIMO}                                       &AoA estimation        &$\mathcal{O}\left(\widetilde{P}^3\widetilde{U}^3\right)$\\
                                                                  &Beam steering        &$\mathcal{O}\left(PUN^2{\mathfrak{N}}^3\right)$\\
                                                                  &Amplitude detection   &$\mathcal{O}\left(PU^3\mathfrak{N}^3\right)$\\
  \midrule
  \multirow{3}{*}{MIMO-OFDM}                                      &Channel estimation    &$\mathcal{O}\left(PN^3\mathfrak{N}^3\right)$\\
                                                                  &Signal detection      &$\mathcal{O}\left(PN^3\mathfrak{N}^3\right)$ \\
  \bottomrule
  \label{Table2}
\end{tabular}}
\end{center}
\vspace{-1em}
\end{table}

In the presence of interferences and noise, if we define $\bm{R}^\mathcal{I}(k_p)$ and $\bm{R}^{\hat{\bar{z}}}(k_p)$ as the $\mathfrak{N}U\times \mathfrak{N}U$ covariance matrices of $\bm{\mathcal{I}}(k_p)$ and $\mathbf{\hat{\bar{z}}}(k_p)$, i.e.,
\begin{align*} \label{CovarianceMatrix}
\bm{R}^\mathcal{I}(k_p)=&\left(\mathbf{\bar{\Gamma}}^{-1}(k_p)\mathbf{H'}_\textmd{OAM-MIMO}(k_p)- \mathbf{I}\right)\mathbb{E}\left(\mathbf{\bar{s}}(k_p)\mathbf{\bar{s}}^H(k_p)\right)\nonumber\\
&\left(\mathbf{\bar{\Gamma}}^{-1}(k_p)\mathbf{H'}_\textmd{OAM-MIMO}(k_p)- \mathbf{I}\right)^H, \nonumber\\
\bm{R}^{\hat{\bar{z}}}(k_p)=&\mathbf{\bar{D}}(k_p)\mathbb{E}\left(\mathbf{\bar{z}}(k_p) \mathbf{\bar{z}}^H(k_p)\right)\mathbf{\bar{D}}^H(k_p),
\end{align*}
the SINR of the $u$th mode OAM generated by the $\mathfrak{n}$th UCA at the $p$th subcarrier can be formulated as
\begin{equation} \label{SINR2}
\overline{\textrm{SINR}}_{\mathfrak{n}}(\ell_u,k_p)=\frac{\mathbb{E} \left(\left|s_{\mathfrak{n}}(\ell_u,k_p)\right|^2\right)}{\left[\bm{R}^\mathcal{I}(k_p)\right] _{\kappa,\kappa}+ \left[\bm{R}^{\hat{\bar{z}}}(k_p)\right]_{\kappa,\kappa}},
\end{equation}
where $\kappa=(\mathfrak{n}-1)N+u, \mathfrak{n}=1,2,\cdots,\mathfrak{N}$.

Therefore, the SE of the LoS MCMM-OAM-MIMO system can be written as
\begin{align} \label{EAR}
\overline{C} = &\left(1-\frac{T_p\widetilde{P}}{T_cP}\right)\frac{1}{P}\sum_{p=1}^P\sum_{u=1}^U
\log_2\left(1 + \overline{\textrm{SINR}}_{\mathfrak{n}}(\ell_u,k_p)\right)+\nonumber \\
&\frac{1}{P}\sum_{p=1}^P\sum_{u=1}^U\sum_{\mathfrak{n}=2}^\mathfrak{N}\log_2\left(1 + \overline{\textrm{SINR}}_{\mathfrak{n}}(\ell_u,k_p)\right) .
\end{align}

\subsection{Performance and Complexity Discussions}
In this part, we mainly analyze the training overhead and computational complexity of the proposed LoS MCMM-OAM-MIMO system multiplexing $P$ subcarriers and $U$ modalities, which is similar in structure to the traditional MIMO-OFDM system using $P$ subcarriers and $\mathfrak{N} N\times \mathfrak{N}N$  antenna elements. In traditional MIMO-OFDM communication systems especially at sub 6 GHz bands, to correctly recover the transmitted data symbols, channel estimation has to determine the $\mathfrak{N}N\times\mathfrak{N}N$ channel coefficients for each subcarrier, i.e., $PN^2\mathfrak{N}^2$ unknown variables in total. Thus, the required number of training symbols increase proportional to $P$, $N$ and $\mathfrak{N}$ in large-scale antenna systems. However, by taking advantage of the geometrical relationship between the transmit and receive UCCAs, given that the radii and the numbers of UCAs' elements of the transmit UCCA are known to the receiver, the LoS MCMM-OAM-MIMO communication systems only need determining three unknown parameters, i.e, the azimuth angle $\varphi$, the elevation angle $\alpha$ and the distance $r$ to recover the transmitted data symbols. Since $\widetilde{P}$ and $\widetilde{U}$ have only effect on the accuracy of distance and AoA estimation, when $\hat{\varphi}$, $\hat{\alpha}$ and $\hat{r}$ are accurate enough, the number of training symbols required by the LoS MCMM-OAM-MIMO systems do not need to increase with $P$, $N$ and $\mathfrak{N}$ even in large-scale antenna systems. Therefore, although the LoS MCMM-OAM-MIMO system has the same channel capacity as the LoS MIMO-OFDM system, it requires much less training overhead and thus improving the system SE.

\begin{figure}[t]
\setlength{\abovecaptionskip}{-0cm}   
\setlength{\belowcaptionskip}{-0.2cm}   
\begin{center}
\includegraphics[width=9cm,height=7cm]{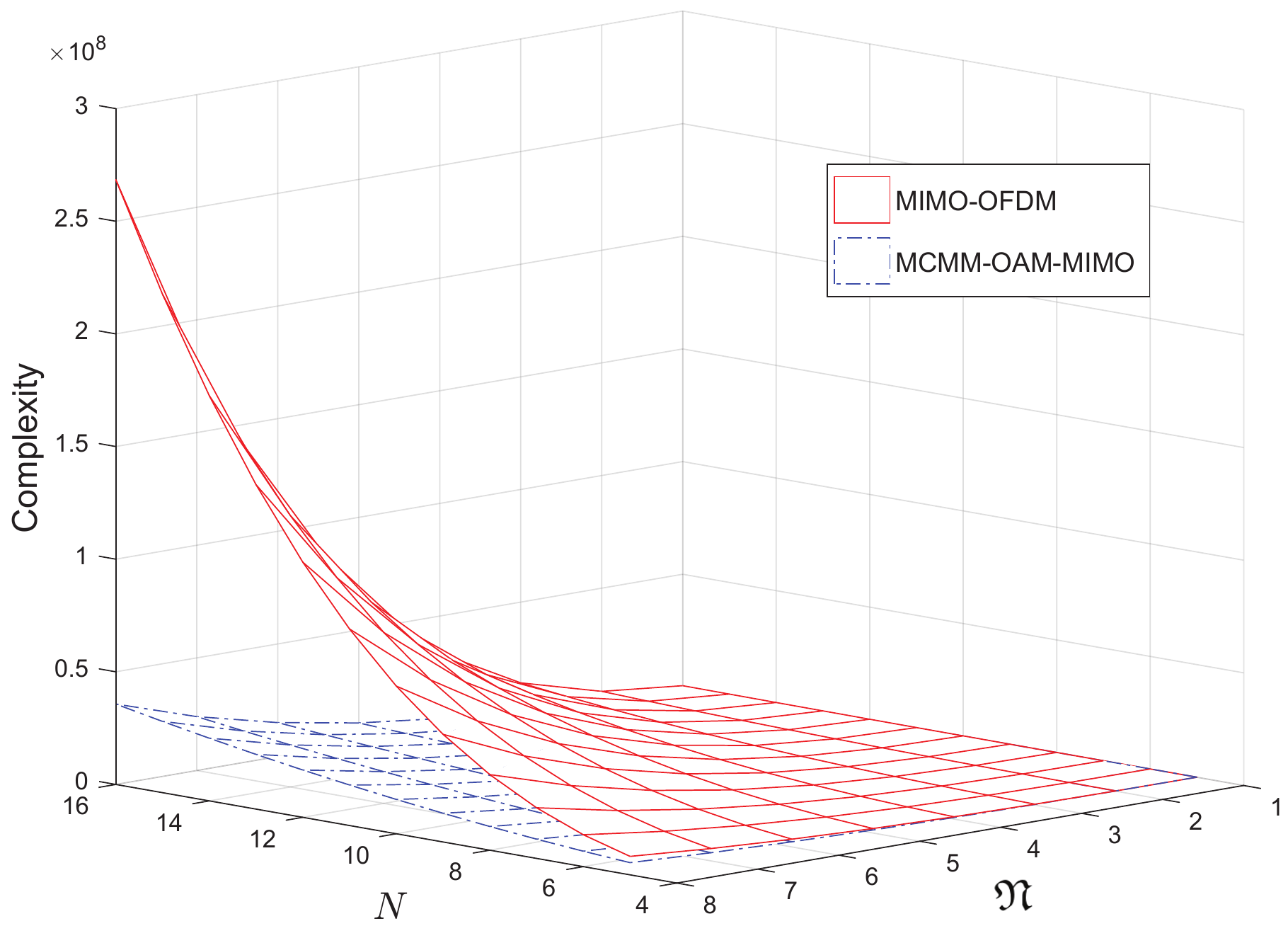}
\end{center}
\caption{The complexities of the MCMM-OAM-MIMO system and MIMO-OFDM system vs. $N$ and $\mathfrak{N}$ at $P=64, \widetilde{P}=8, U=\widetilde{U}=4$.}
\label{Fig108}
\end{figure}

\begin{figure*}[t]
\setlength{\abovecaptionskip}{-0.9cm}   
\setlength{\belowcaptionskip}{-0.0cm}   
\begin{center}
\includegraphics[scale=0.55]{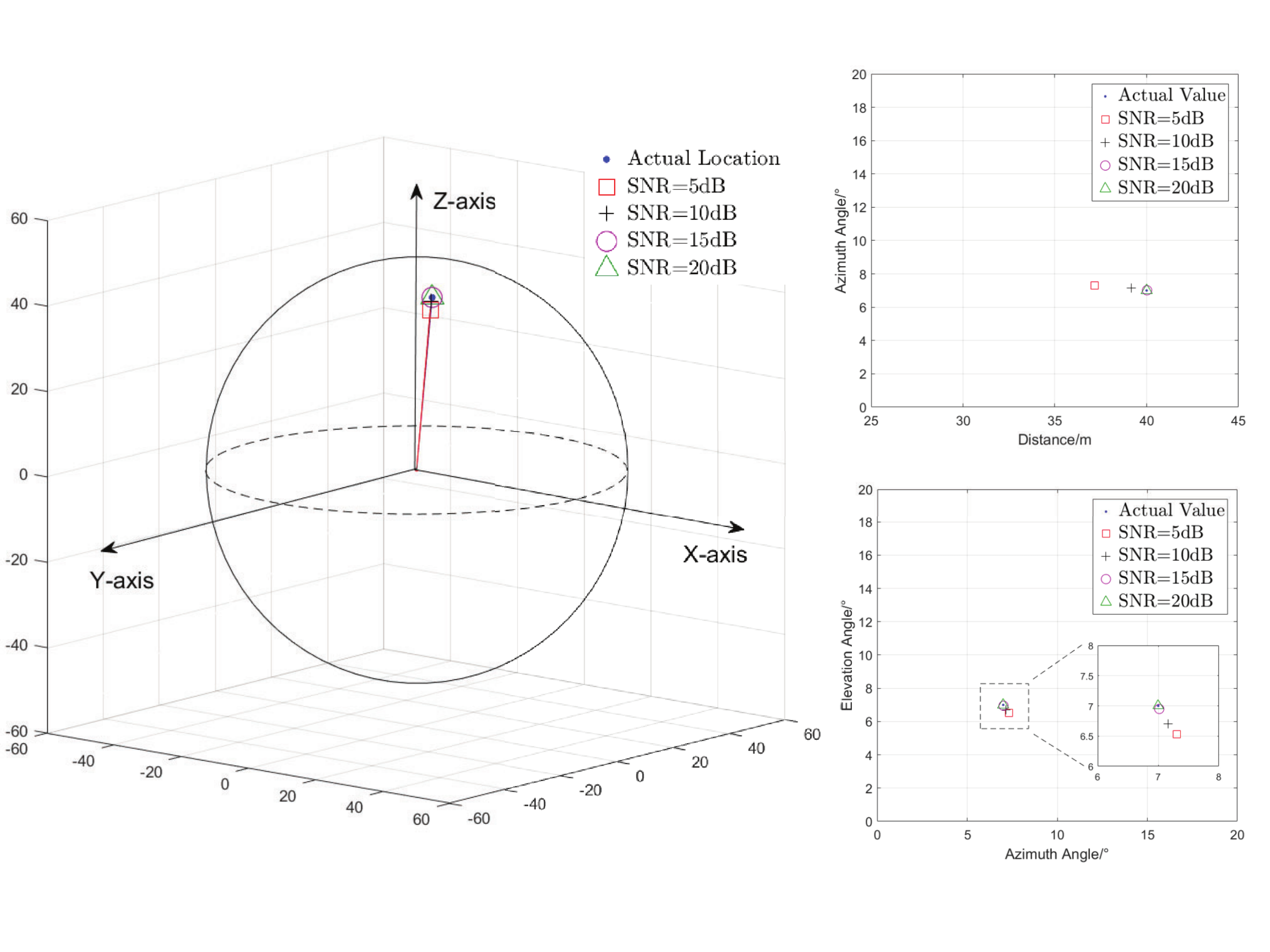}
\end{center}
\caption{The AoA estimation results of the proposed method.}
\label{Fig100}
\end{figure*}

For LoS MCMM-OAM-MIMO communication systems, the complexity of the proposed distance and AoA estimation method is determined by the complexity of the EVD in \eqref{EVD} corresponding to the values of $\widetilde{P}$ and $\widetilde{U}$, the complexity of the proposed beam steering method is determined by the complexity of the calculating $\mathbf{H'}_\textmd{OAM-MIMO}(k_p)$ corresponding to the values of $P$, $U$ $N$ and $\mathfrak{N}$, and the complexity of the proposed amplitude detection method is determined by the complexity of the calculating $\mathbf{\bar{D}}(k_p)$ corresponding to the values of $P$, $U$ and $\mathfrak{N}$. The specific computational complexity of the proposed AoA estimation and signal reception for LoS MCMM-OAM-MIMO communication systems is compared with that of traditional MIMO-OFDM communication systems in the Table \ref{Table2}, and the total computational complexity comparison between the LoS MCMM-OAM-MIMO communication system and MIMO-OFDM communication system is shown in the Fig.\ref{Fig108}. It can be seen from the figure that the complexity of LoS MCMM-OAM-MIMO system is lower than traditional MIMO-OFDM systems when $\mathfrak{N}$ and $N$ are large. For example, when $P=64$, $\widetilde{P}=8$, $U=\widetilde{U}=4$, $N=16$ and $\mathfrak{N}=4$, the LoS MCMM-OAM-MIMO system has lower computational complexity than the traditional MIMO-OFDM system.

\section{Numerical Simulations and Results}
In this section we show the performances of the proposed methods by numerical simulations. We first verify the proposed distance and AoA estimation method at different SNRs, and then compare the BER performances of the proposed OAM signal reception scheme with those of the perfect aligned OAM channel. At last, the SEs of the UCA-based LoS MCMM-OAM system, the UCCA-based LoS MCMM-OAM-MIMO system and the traditional MIMO system are compared under different numbers of OAM training symbols. Unless otherwise stated, the SNRs in all the figures are defined as the ratio of the received signal power versus the noise power.

In Fig. \ref{Fig100}, we choose $\widetilde{P} = 8$ subcarriers from 2.244GHz to 2.578GHz corresponding to the wave numbers $k_1,k_2,\cdots,k_8= 47, 48,\dots, 54$, $N = 9$, ${\widetilde{U}} = 8$ with $\ell_1,\ell_2,\cdots,\ell_8 = -4,-3,\cdots,+3$, $R_t=R_r=15\lambda_1$, $\lambda_1=2\pi/k_1$, $(r, \varphi, \alpha) = (40\textrm{m}, 7^{\circ}, 7^{\circ})$, $[\alpha_a,\alpha_b]=[2^{\circ}, 8^{\circ}]$ and the initial number of intervals $\mathcal{D}=2$. These parameters are selected to ensure the receive UCA being within the main lobe of the transmit UCA at all subcarrier frequencies achieving large enough receive SNRs. Then, by using the proposed distance and AoA estimation method based on 2-D ESPRIT algorithm, the estimated locations of the MCMM-OAM transmitter are shown in Fig. \ref{Fig100}. It is obvious to see from the figure that the estimated distances and AoAs approach to the actual values with the increase of the SNR, e.g., when SNR reaches $20$dB, $(\hat{r}, \hat{\varphi}, \hat{\alpha})=(40.000\textrm{m}, 6.997 ^{\circ}, 7.003^{\circ})$ is very close to the actual location.

\begin{figure}[t]
\setlength{\abovecaptionskip}{-0cm}   
\setlength{\belowcaptionskip}{-0.2cm}   
\begin{center}
\includegraphics[width=8cm,height=7cm]{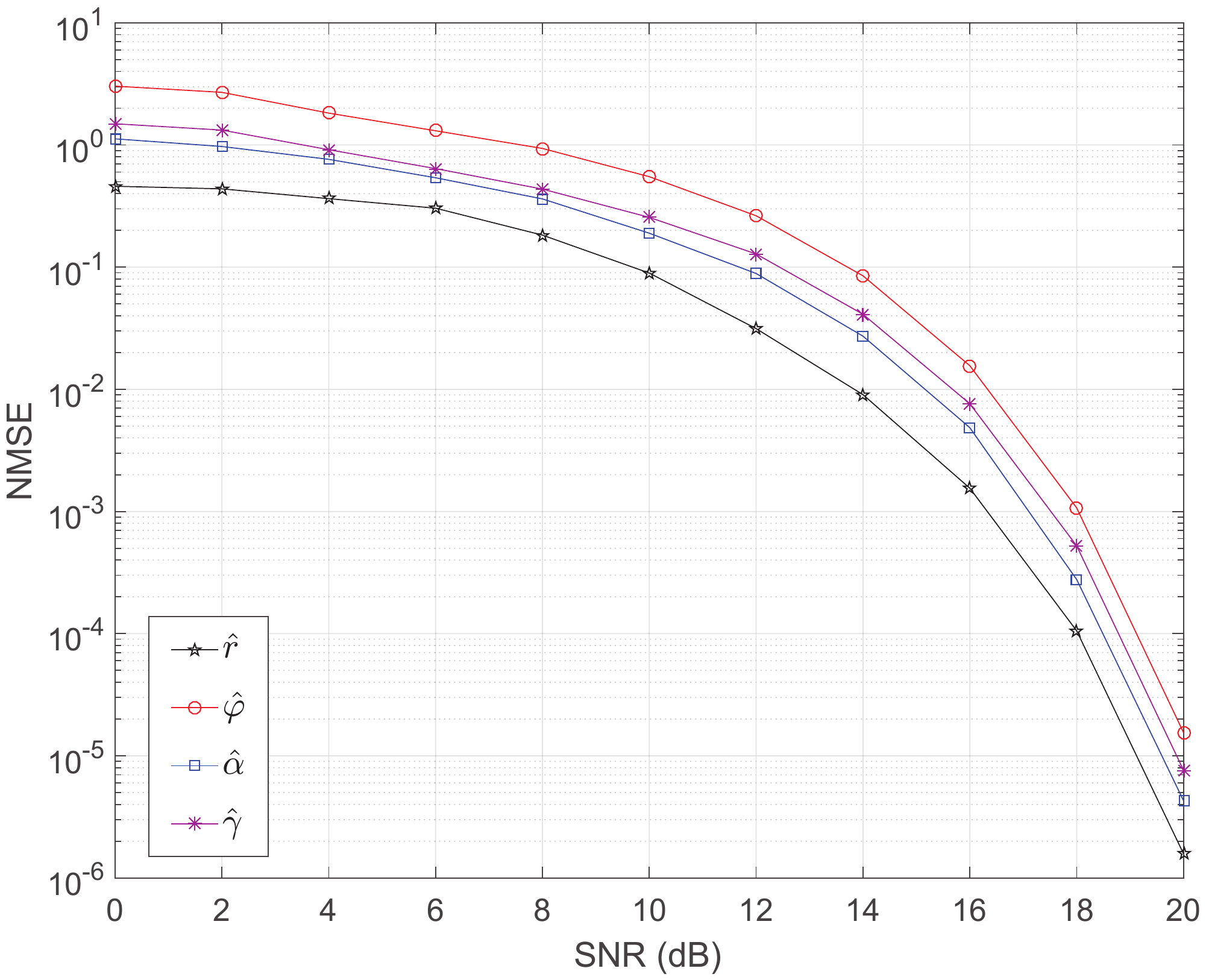}
\end{center}
\caption{The NMSEs of the estimated $\hat{r}$, $\hat{\varphi}$, $\hat{\alpha}$ and $\hat{\gamma}$ vs. SNR at $\widetilde{P}=8$ and $\widetilde{U}=8$.}
\label{Fig101}
\end{figure}

Fig.\ref{Fig101} illustrates the normalized mean-squared errors (NMSEs) of $\hat{r}$, $\hat{\varphi}$, $\hat{\alpha}$ and $\hat{\gamma}$ versus SNR. The NMSE is defined as $\mathbb{E}\{(\hat{x}-x)^2/x^2\}$, where $\hat{x}$ denotes the estimate of $x$. It can be seen from Fig. \ref{Fig101} that as SNR increases all the NMSEs of $\hat{r}$, $\hat{\varphi}$, $\hat{\alpha}$ and $\hat{\gamma}$ decline. Specifically, due to different effects of SNR on the estimations of $r$, $\varphi$, $\alpha$ and $\gamma$, their NMSEs are different, where $\hat{r}$ and $\hat{\varphi}$ have the lowest and the highest NMSE, respectively.

\begin{figure}[t]
\setlength{\abovecaptionskip}{-0cm}   
\setlength{\belowcaptionskip}{-0.2cm}   
\begin{center}
\includegraphics[width=8cm,height=7cm]{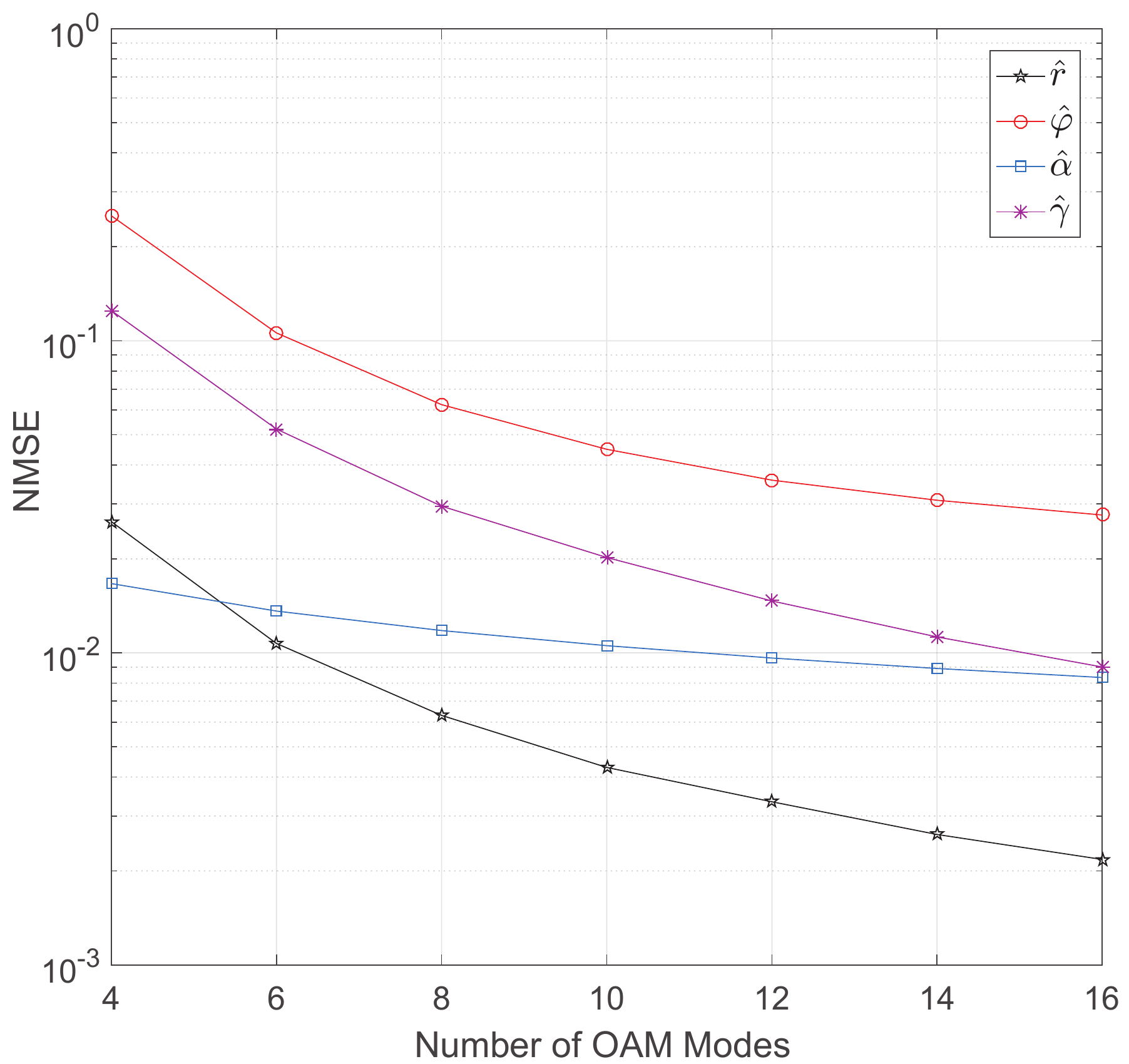}
\end{center}
\caption{The NMSEs of the estimated $\hat{r}$, $\hat{\varphi}$, $\hat{\alpha}$ and $\hat{\gamma}$ vs. the number of OAM modes $\widetilde{U}$ at SNR=15dB.}
\label{Fig102}
\end{figure}
\begin{figure}[t]
\setlength{\abovecaptionskip}{-0cm}   
\setlength{\belowcaptionskip}{-0.2cm}   
\begin{center}
\includegraphics[width=8cm,height=7cm]{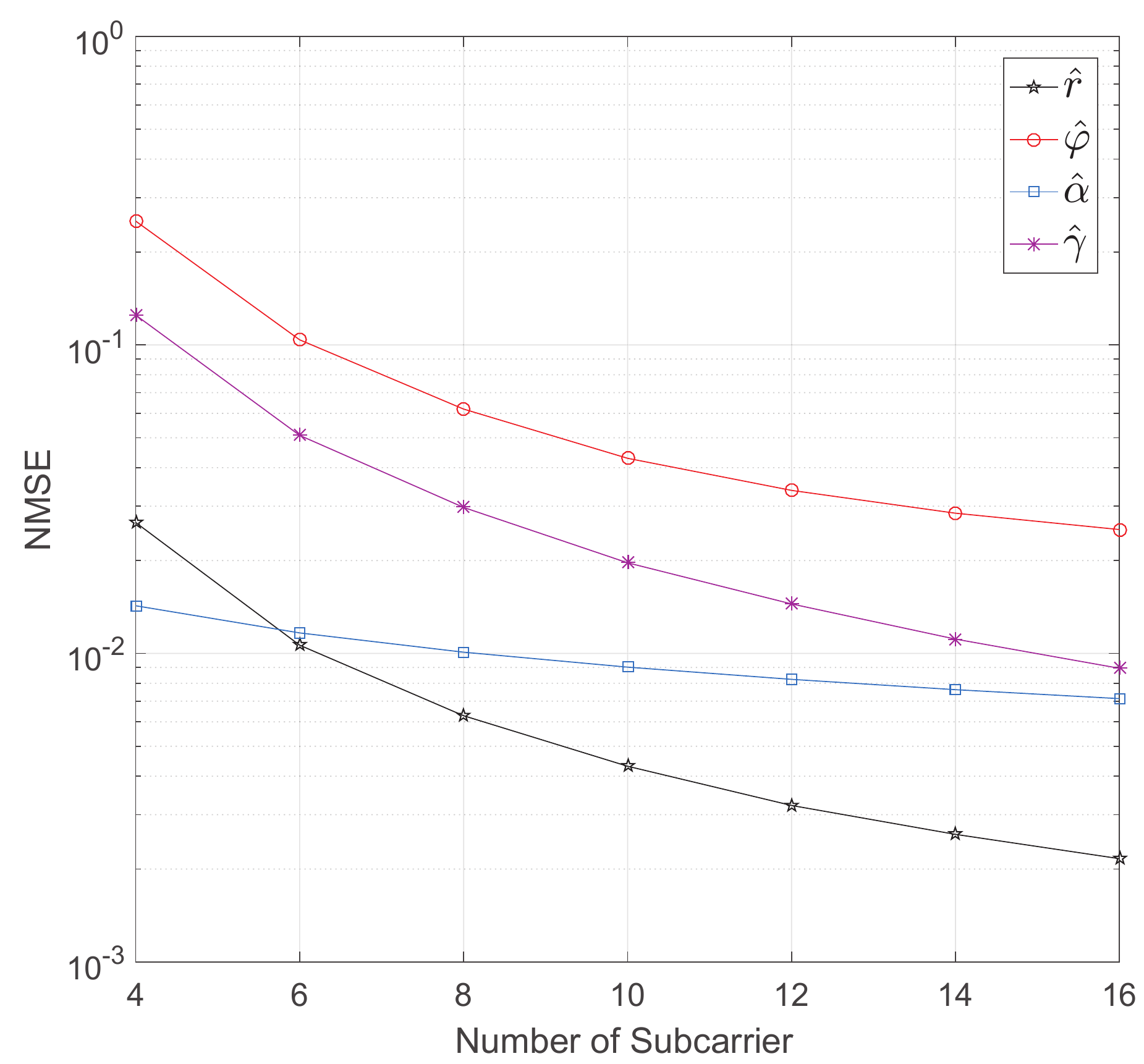}
\end{center}
\caption{The NMSEs of the estimated $\hat{r}$, $\hat{\varphi}$, $\hat{\alpha}$ and $\hat{\gamma}$ vs. the number of subcarriers $\widetilde{P}$ at SNR=15dB.}
\label{Fig105}
\end{figure}
\begin{figure}[t]
\setlength{\abovecaptionskip}{-0cm}   
\setlength{\belowcaptionskip}{-0.2cm}   
\begin{center}
\includegraphics[width=8cm,height=7cm]{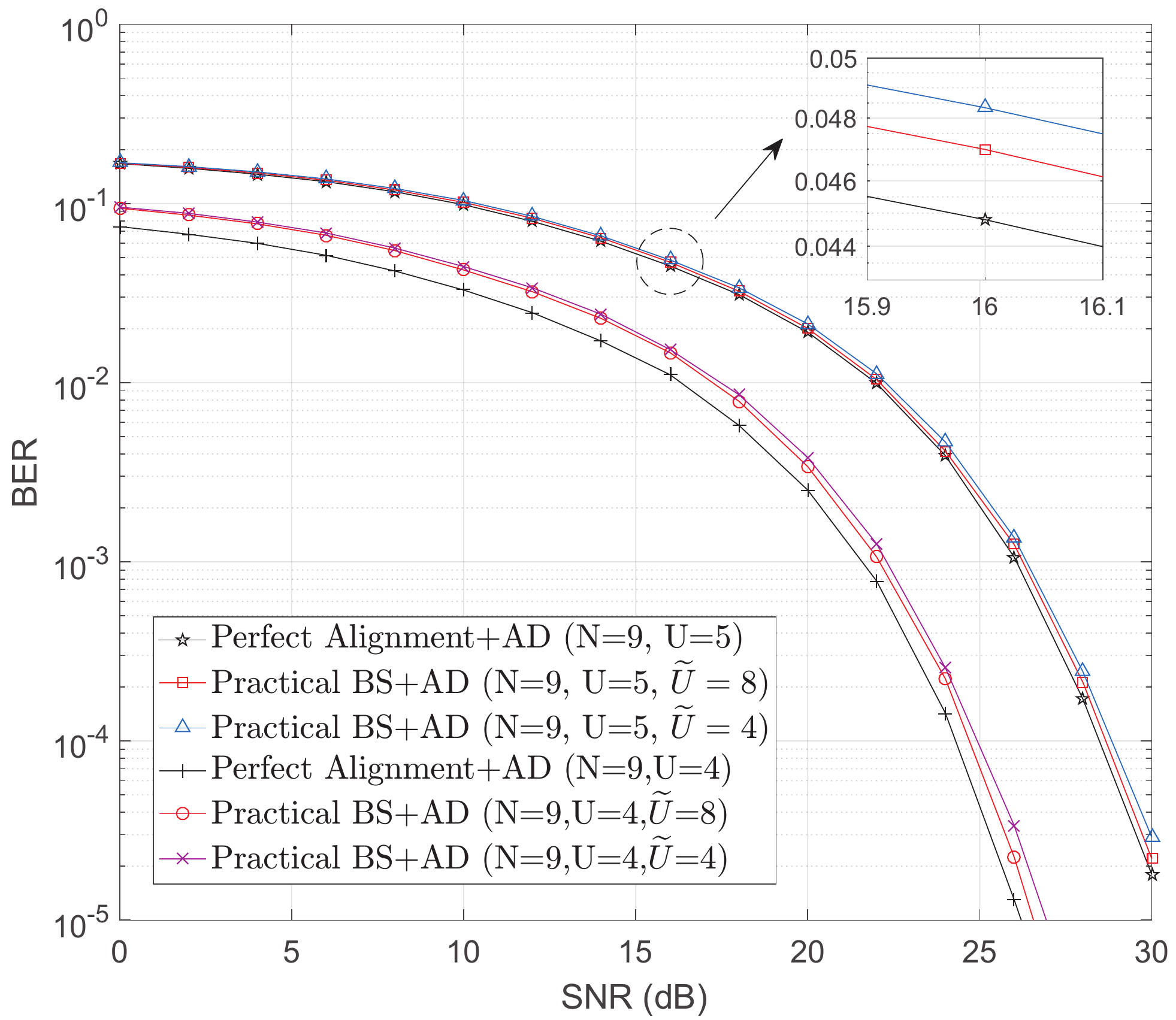}
\end{center}
\caption{The BER comparison of OAM detection. AD: Amplitude Detection}
\label{Fig104}
\end{figure}
\begin{figure}[t]
\setlength{\abovecaptionskip}{-0cm}   
\setlength{\belowcaptionskip}{-0.2cm}   
\begin{center}
\includegraphics[width=8cm,height=7cm]{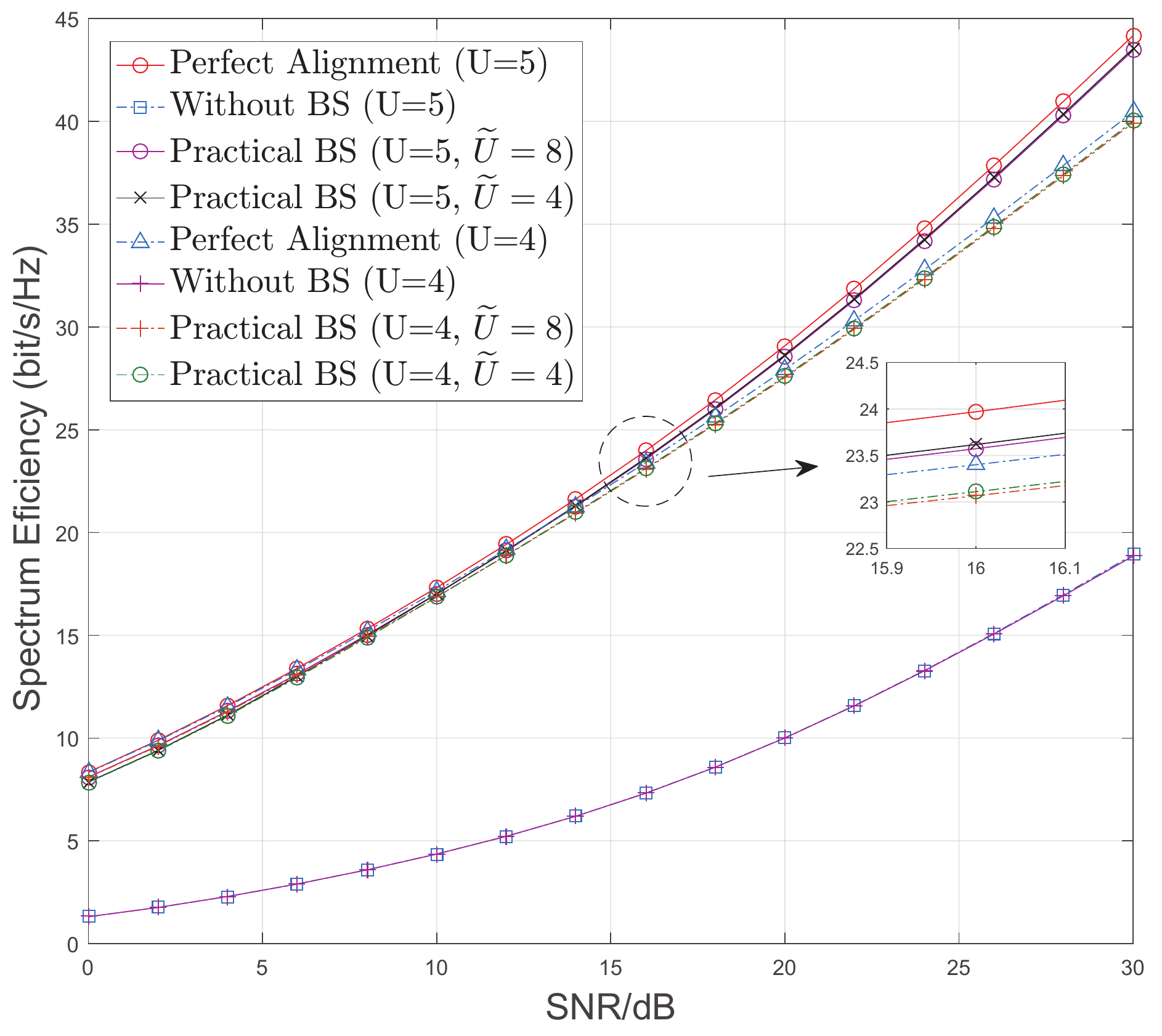}
\end{center}
\caption{The spectrum efficiencies of the UCA-based LoS MCMM-OAM system. Practical BS+AD: Beam steering and amplitude detection with the estimated values $\hat{\gamma}$, $\hat{\alpha}$ and $\hat{r}$.}
\label{Fig107}
\end{figure}

Fig.\ref{Fig102} and Fig.\ref{Fig105} show the NMSEs of $\hat{r}$, $\hat{\varphi}$, $\hat{\alpha}$ and $\hat{\gamma}$ versus the number of OAM modes ${\widetilde{U}}$ and the number of subcarriers $\widetilde{P}$. It can be seen from the almost same two figures that as ${\widetilde{U}}$ and $\widetilde{P}$ increase the NMSEs of $\hat{r}$, $\hat{\varphi}$, $\hat{\alpha}$ and $\hat{\gamma}$ all decreases. The effects of ${\widetilde{U}}$ and $\widetilde{P}$ on the estimation performances of $r$, $\varphi$, $\alpha$ and $\gamma$ are the same due to that ${\widetilde{U}}$ and $\widetilde{P}$ are symmetrical in the 2-D ESPRIT-based estimation method. However, the NMSEs of $\hat{r}$, $\hat{\varphi}$, $\hat{\alpha}$ and $\hat{\gamma}$ are different. In contrast to $\hat{r}$, $\hat{\gamma}$ and $\hat{\varphi}$, $\hat{\alpha}$ changes slowly with the increase of $\widetilde{U}$ and $\widetilde{P}$ since that it mainly depends on the search procedure under \eqref{amplitude}.

After obtaining the estimated distance and AoA, Fig. \ref{Fig104} compares the BERs of the proposed OAM signal detection scheme with those of the amplitude detection in perfect aligned OAM channel under $N=9$ and 16QAM modulation. When $U=5$, the BERs of the OAM signal detection including beam steering (BS) and amplitude detection (AD) are very close to those of the AD in perfect aligned OAM channel. It is worth noting that the BER of the proposed OAM signal detection does not increase significantly when using fewer OAM modes (i.e., $\widetilde{U}=4$ with $\ell_1,\ell_2,\ell_3,\ell_4 = -2,-1,0,+1$) in the distance and AoA estimation. Comparing $U=4$ with $U=5$, both the BERs of the OAM signal detection with $\widetilde{U}=8$ and ${\widetilde{U}}=4$ become much better because of abandoning the high-order OAM modes with smaller gains.

In Fig. \ref{Fig107}, $T_c$ is assumed to be $256$ OFDM symbols, $N=9$. Then, the SEs of the UCA-based LoS MCMM-OAM system under different values of $U$ and $\widetilde{U}$ are shown in Fig. \ref{Fig107}. It can be seen from the figure that in contrast to the reception with only AD and without BS, the SEs of the UCA-based LoS MCMM-OAM system with practical BS and AD are greatly improved in the misaligned OAM channel, which get close to the SEs of the OAM system in perfect aligned channel. Furthermore, the SE increases slightly when 4 OAM modes rather than 8 OAM modes being used in the distance and AoA estimation, because the reduced number of training symbols contributes more to the SE than the detriments induced by the inaccuracy in the distance and AoA estimates. However, too few training symbols may lead to the opposite result. In addition, the SE of the UCA-based LoS MCMM-OAM system with $U=4$ is near to that with $U=5$, which indicates the high-order OAM modes have minor contributions to the SE of the UCA-based OAM systems in long-distance communications.

\begin{figure}[t]
\setlength{\abovecaptionskip}{-0cm}   
\setlength{\belowcaptionskip}{-0.3cm}   
\begin{center}
\includegraphics[width=8cm,height=7cm]{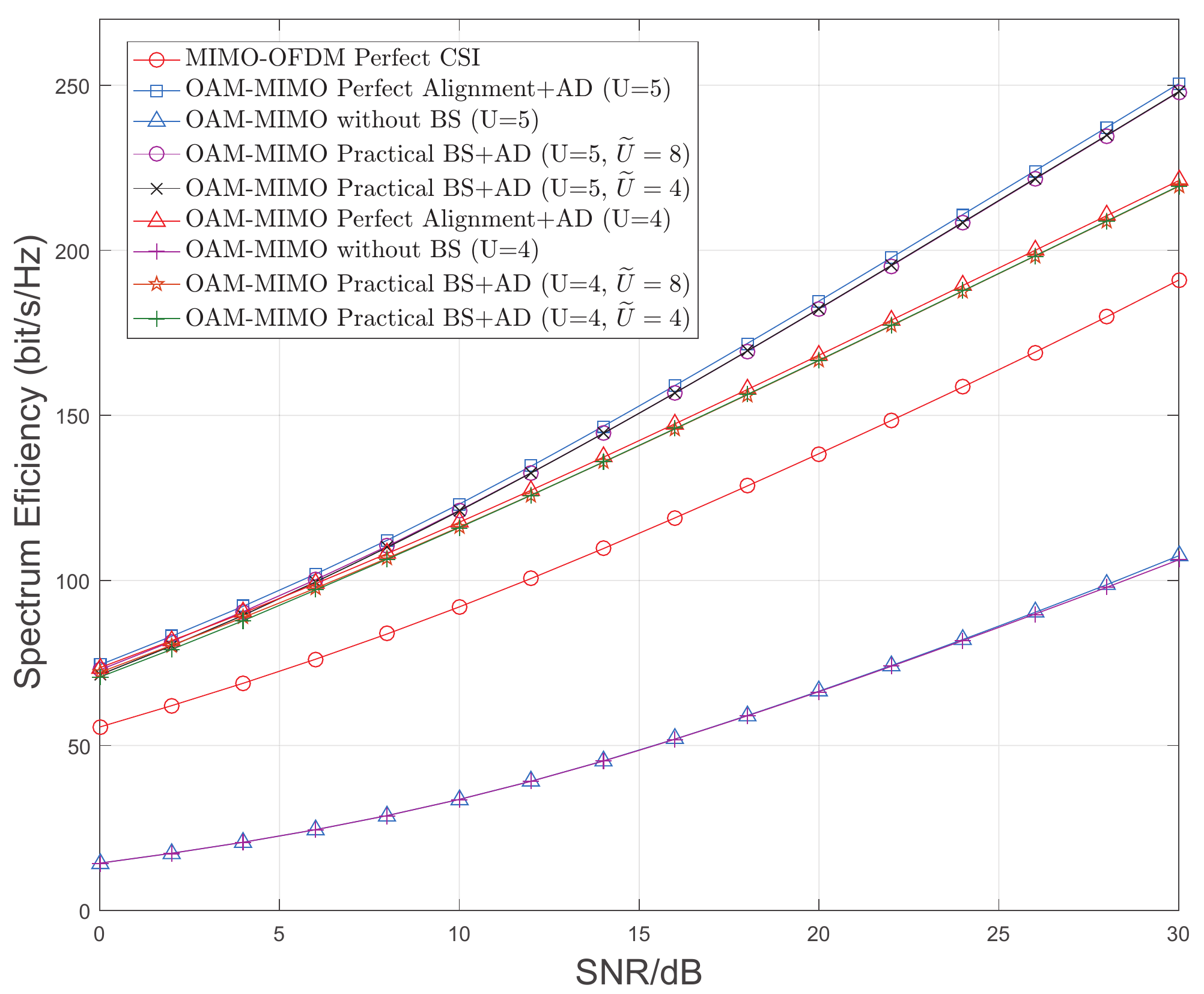}
\end{center}
\caption{The spectrum efficiencies of the UCCA-based LoS MCMM-OAM-MIMO system and the traditional MIMO-OFDM system. Practical BS+AD: Beam steering and amplitude detection with the estimated values $\hat{\gamma}$, $\hat{\alpha}$ and $\hat{r}$.}
\label{Fig103}
\end{figure}

In Fig. \ref{Fig103}, $T_c$ is also assumed to be $256$ OFDM symbols, $N=16$, $\mathfrak{N}=4$, $\widetilde{P}=8$, $P=64$. The SEs of the UCCA-based LoS MCMM-OAM-MIMO system and the traditional MIMO-OFDM system are compared in Fig. \ref{Fig103}. We assume that the traditional channel estimation of the MIMO-OFDM system requires $N\mathfrak{N}=64$ training symbols per subcarrier or subchannel. From the figure one can get the same conclusion as in Fig. \ref{Fig107}, which verifies that the proposed distance and AoA estimation method and the OAM signal reception scheme can be applied to the UCCA-based LoS MCMM-OAM-MIMO system. More importantly, it is obvious that the SEs of the UCCA-based LoS MCMM-OAM-MIMO system are about 20\% higher than the SE of the MIMO-OFDM system with traditional channel estimation due to the greatly reduced number of training symbols.


\section{Conclusions and Future Work}
\subsection{Conclusions}
In this paper, we have proposed a UCA-based LoS MCMM-OAM communication scheme including the generation, the distance and AoA estimation and the reception of the multi-mode OAM beams. In terms of OAM generation, we have verified that the UCA can generate multi-mode OAM radio beam with both the RF analog synthesis method and the baseband digital synthesis method. Then, we proposed a distance and AoA estimation method based on the 2-D ESPRIT algorithm, which is shown being able to accurately estimate the location of the LoS MCMM-OAM transmitter with flexible number of training symbols. Thereafter, we proved that the proposed beam steering method can eliminate inter-mode interferences induced by the misalignment error between the transmit and receive UCAs, and the resulting diagonal elements of the effective OAM channel approaches a function of distance given other parameters are fixed. Therefore, the designed multi-mode OAM reception composed of the beam steering and the diagonal amplitude detection is shown to achieve the performance of an ideally aligned OAM channel and low complexity. At last, the proposed methods are extended to the UCCA-based LoS MCMM-OAM-MIMO system, which exhibits higher SE than the large-scale MIMO-OFDM system with traditional channel estimation.

\subsection{Future Work}
It is worth noting that when multiple OAM modes are transmitted simultaneously, due to the interferometry, the receiving SNR is not uniform on the ring of the main lobe. Therefore, if the receive UCA is rotatable around its axis, the antenna elements of the receive UCA could be set in the maximum SNR positions on the ring of the main lobe to further improve the system performances, which means that another rotation angle within the plane of the receive UCA should be estimated. Although in this manuscript we consider the transmit and receive UCAs being fixed and not able to rotate, we believe the novel UCA-based OAM communication system with rotatable UCAs having better SE performance, which is left for our future work.

\section*{Acknowledgment}
The authors would like to thank the editor and the anonymous reviewers for their careful reading and valuable suggestions that helped to improve the quality of this manuscript.

%
\bibliographystyle{IEEEtran}
\bibliography{IEEEabrv,mutiple_OAM}

\begin{thebibliography}{10}
\providecommand{\url}[1]{#1}
\csname url@samestyle\endcsname
\providecommand{\newblock}{\relax}
\providecommand{\bibinfo}[2]{#2}
\providecommand{\BIBentrySTDinterwordspacing}{\spaceskip=0pt\relax}
\providecommand{\BIBentryALTinterwordstretchfactor}{4}
\providecommand{\BIBentryALTinterwordspacing}{\spaceskip=\fontdimen2\font plus
\BIBentryALTinterwordstretchfactor\fontdimen3\font minus
  \fontdimen4\font\relax}
\providecommand{\BIBforeignlanguage}[2]{{%
\expandafter\ifx\csname l@#1\endcsname\relax
\typeout{** WARNING: IEEEtran.bst: No hyphenation pattern has been}%
\typeout{** loaded for the language `#1'. Using the pattern for}%
\typeout{** the default language instead.}%
\else
\language=\csname l@#1\endcsname
\fi
#2}}
\providecommand{\BIBdecl}{\relax}
\BIBdecl

\bibitem{WRC}
(2018, September) World radiocommunication conference ({WRC}). [Online].
  Available: \url{https://www.itu.int/en/ITU-R/conferences/wrc/
  Pages/default.aspx}.

\bibitem{Allen1992Orbital}
L.~Allen, M.~W. Beijersbergen, R.~J. Spreeuw, and J.~P. Woerdman, ``Orbital
  angular momentum of light and the transformation of {Laguerre-Gaussian} laser
  modes,'' \emph{Phys. Rev. A: At. Mol. Opt. Phys.}, vol.~45, no.~11, pp.
  8185--8189, 1992.

\bibitem{Tamburini2012Encoding}
F.~Tamburini, E.~Mari, A.~Sponselli, B.~Thid\'e, A.~Bianchini, and F.~Romanato,
  ``Encoding many channels in the same frequency through radio vorticity: first
  experimental test,'' \emph{New J. Phys.}, vol.~14, no.~3, p. 033001, 2012.

\bibitem{Yan2014High}
Y.~Yan, G.~Xie, M.~P.~J. Lavery, H.~Huang, N.~Ahmed, C.~Bao, Y.~Ren, Y.~Cao,
  L.~Li, Z.~Zhao, A.~F. Molisch, M.~Tur, M.~J. Padgett, and A.~E. Willner,
  ``High-capacity millimetre-wave communications with orbital angular momentum
  multiplexing,'' \emph{Nature Commun.}, vol.~5, p. 4876, 2014.

\bibitem{Ren2017Line}
Y.~Ren, L.~Li, G.~Xie, Y.~Yan, Y.~Cao, H.~Huang, N.~Ahmed, Z.~Zhao, P.~Liao,
  C.~Zhang, G.~Caire, A.~F. Molisch, M.~Tur, and A.~E. Willner, ``Line-of-sight
  millimeter-wave communications using orbital angular momentum multiplexing
  combined with conventional spatial multiplexing,'' \emph{{IEEE} Trans.
  Wireless Commun.}, vol.~16, no.~5, pp. 3151--3161, 2017.

\bibitem{Zhang2017Mode}
W.~Zhang, S.~Zheng, X.~Hui, R.~Dong, X.~Jin, H.~Chi, and X.~Zhang, ``Mode
  division multiplexing communication using microwave orbital angular momentum:
  An experimental study,'' \emph{{IEEE} Trans. Wireless Commun.}, vol.~16,
  no.~2, pp. 1308--1318, 2017.

\bibitem{Chen2018A}
R.~Chen, W.~Yang, H.~Xu, and J.~Li, ``A 2-{D} {FFT}-based transceiver
  architecture for {OAM-OFDM} systems with {UCA} antennas,'' \emph{{IEEE}
  Trans. Veh. Technol.}, vol.~67, no.~6, pp. 5481--5485, 2018.

\bibitem{Zhang2019Orbital}
C.~{Zhang} and Y.~{Zhao}, ``Orbital angular momentum nondegenerate index
  mapping for long distance transmission,'' \emph{{IEEE} Trans. Wireless
  Commun.}, vol.~18, no.~11, pp. 5027--5036, Nov 2019.

\bibitem{Chen2020}
R.~{Chen}, H.~{Zhou}, M.~{Moretti}, X.~{Wang}, and J.~{Li}, ``Orbital angular
  momentum waves: Generation, detection and emerging applications,'' \emph{IEEE
  Commun. Surv. Tut.}, vol.~99, no.~1, pp. 1--30, Nov. 2019.

\bibitem{Bennis2013Flat}
A.~Bennis, R.~Niemiec, C.~Brousseau, K.~Mahdjoubi, and O.~Emile, ``Flat plate
  for {OAM} generation in the millimeter band,'' in \emph{Proc. 7th Eur. Conf.
  Antennas Propag.}, 2013, pp. 3203--3207.

\bibitem{Cheng2014Generation}
L.~Cheng, W.~Hong, and Z.-C. Hao, ``Generation of electromagnetic waves with
  arbitrary orbital angular momentum modes,'' \emph{Sci. Rep.}, vol.~4, p.
  4814, 2014.

\bibitem{Turnbull1996The}
G.~A. Turnbull, D.~A. Robertson, G.~M. Smith, L.~Allen, and M.~J. Padgett,
  ``The generation of free-space {Laguerre-Gaussian} modes at millimetre-wave
  frequencies by use of a spiral phaseplate,'' \emph{Opt. Commun.}, vol. 127,
  no.~4, pp. 183--188, 1996.

\bibitem{Mari2015Near}
E.~Mari, F.~Spinello, M.~Oldoni, R.~A. Ravanelli, F.~Romanato, and G.~Parisi,
  ``Near-field experimental verification of separation of oam channels,''
  \emph{IEEE Antennas Wireless Propag. Lett.}, vol.~14, pp. 556--558, 2015.

\bibitem{Mohammadi2010system}
S.~M. Mohammadi, L.~K.~S. Daldorff, and et~al., ``Orbital angular momentum in
  radio-a system study,'' \emph{{IEEE} Trans. Antennas Propag.}, vol.~58,
  no.~2, pp. 565--572, Feb 2010.

\bibitem{Liang2016Orbital}
J.~Liang and S.~Zhang, ``Orbital angular momentum {(OAM)} generation by
  cylinder dielectric resonator antenna for future wireless communication,''
  \emph{IEEE Access}, vol.~4, pp. 9570--9574, 2016.

\bibitem{Zhang2016Generation}
Z.~Zhang and S.~Xiao, ``Generation of multiple orbital angular momentum {(OAM)}
  modes with a circularly polarized multimode patch antenna,'' in \emph{Proc.
  IEEE MTT-S Int. Wireless Symp.}, 2016, pp. 1--4.

\bibitem{Zhang2017Four}
W.~Zhang, S.~Zheng, X.~Hui, Y.~Chen, X.~Jin, H.~Chi, and X.~Zhang,
  ``Four-{OAM}-mode antenna with traveling-wave ring-slot structure,''
  \emph{IEEE Antennas Wireless Propag. Lett.}, vol.~16, pp. 194--197, 2017.

\bibitem{Tennant2012Generation}
A.~Tennant and B.~Allen, ``Generation of radio frequency {OAM} radiation modes
  using circular time-switched and phased array antennas,'' in \emph{Proc.
  Loughborough Antennas Propag. Conf.}, 2012.

\bibitem{lee2018experimental}
D.~Lee, H.~Sasaki, H.~Fukumoto, Y.~Yagi, T.~Kaho, H.~Shiba, and T.~Shimizu,
  ``An experimental demonstration of 28 {GHz} band wireless {OAM-MIMO} (orbital
  angular momentum multi-input and multi-output) multiplexing,'' in \emph{Proc.
  IEEE 87th Veh. Technol. Conf.}, 2018.

\bibitem{Yu2016Generating}
S.~Yu, L.~Li, G.~Shi, C.~Zhu, and Y.~Shi, ``Generating multiple orbital angular
  momentum vortex beams using a metasurface in radio frequency domain,''
  \emph{Appl. Phys. Lett.}, vol. 108, no.~24, p. 241901, 2016.

\bibitem{Sanming2018Dual}
Y.~Shen, J.~Yang, H.~Meng, W.~Dou, and S.~Hu, ``Generating millimeter-wave
  {Bessel} beam with orbital angular momentum using reflective-type metasurface
  inherently integrated with source,'' \emph{Appl. phys. Lett.}, vol. 112,
  no.~14, p. 141901, 2018.

\bibitem{XieG2016}
G.~{Xie}, Y.~{Yan}, Z.~{Zhao}, and et~al., ``Tunable generation and angular
  steering of a millimeter-wave orbital-angular-momentum beam using
  differential time delays in a circular antenna array,'' in \emph{Proc. IEEE
  Int. Conf. Commun.}, 2016.

\bibitem{Chen2018Beam}
R.~Chen, H.~Xu, M.~Moretti, and J.~Li, ``Beam steering for the misalignment in
  {UCA}-based {OAM} communication systems,'' \emph{IEEE Wireless Commun.
  Lett.}, vol.~7, no.~4, pp. 582--585, 2018.

\bibitem{Xie2015Performance}
G.~Xie, L.~Li, Y.~Ren, and et~al., ``Performance metrics and design
  considerations for a free-space optical orbital-angular-momentum-multiplexed
  communication link,'' \emph{Optica}, vol.~2, no.~4, pp. 357--365, 2015.

\bibitem{Liu2015Orbital}
K.~Liu, Y.~Cheng, Z.~Yang, H.~Wang, Y.~Qin, and X.~Li,
  ``Orbital-angular-momentum-based electromagnetic vortex imaging,''
  \emph{{IEEE} Antennas Wireless Propag. Lett.}, vol.~14, pp. 711--714, 2015.

\bibitem{Chen2018OAMradar}
R.~Chen, W.-X. Long, Y.~Gao, and J.~Li, ``Orbital angular momentum-based
  two-dimensional super-resolution targets imaging,'' in \emph{Proc. IEEE
  Global Conf. Signal Inf. Process.}, 2018, pp. 1--4.

\bibitem{Edfors2012Is}
O.~Edfors and A.~J. Johansson, ``Is orbital angular momentum {(OAM)} based
  radio communication an unexploited area?'' \emph{{IEEE} Trans. Antennas
  Propag.}, vol.~60, no.~2, pp. 1126--1131, 2012.

\bibitem{Oldoni2015}
M.~Oldoni, F.~Spinello, E.~Mari, G.~Parisi, C.~G. Someda, F.~Tamburini,
  F.~Romanato, R.~A. Ravanelli, P.~Coassini, and B.~Thid¨¦, ``Space-division
  demultiplexing in orbital-angular-momentum-based mimo radio systems,''
  \emph{IEEE Trans. Antennas Propag.}, vol.~63, no.~10, pp. 4582--4587, Oct
  2015.

\bibitem{Thid2007Utilization}
B.~Thid\'e, H.~Then, J.~Sj\"oholm, K.~Palmer, J.~Bergman, T.~D. Carozzi, Y.~N.
  Istomin, N.~H. Ibragimov, and R.~Khamitova, ``Utilization of photon orbital
  angular momentum in the low-frequency radio domain,'' \emph{Phys. Rev.
  Lett.}, vol.~99, no.~8, p. 087701, 2007.

\bibitem{Gaffoglio2016OAM}
R.~Gaffoglio, A.~Cagliero, A.~D. Vita, and B.~Sacco, ``{OAM} multiple
  transmission using uniform circular arrays: Numerical modeling and
  experimental verification with two digital television signals,'' \emph{Radio
  Sci.}, vol.~51, no.~6, pp. 645--658, 2016.

\bibitem{Gong2017Generation}
Y.~{Gong}, R.~{Wang}, Y.~{Deng}, B.~{Zhang}, N.~{Wang}, N.~{Li}, and P.~{Wang},
  ``Generation and transmission of {OAM}-carrying vortex beams using circular
  antenna array,'' \emph{{IEEE} Trans. Antennas Propag.}, vol.~65, no.~6, pp.
  2940--2949, 2017.

\bibitem{Yuan2017Beam}
T.~Yuan, Y.~Cheng, H.~Wang, and Y.~Qin, ``Beam steering for electromagnetic
  vortex imaging using uniform circular arrays,'' \emph{{IEEE} Antennas
  Wireless Propag. Lett.}, vol.~16, pp. 704--707, 2017.

\bibitem{Xu2016Free}
C.~{Xu}, S.~{Zheng}, W.~{Zhang}, Y.~{Chen}, H.~{Chi}, X.~{Jin}, and X.~{Zhang},
  ``Free-space radio communication employing {OAM} multiplexing based on
  {R}otman lens,'' \emph{IEEE Microw. Wireless Compon. Lett.}, vol.~26, no.~9,
  pp. 738--740, Sep. 2016.

\bibitem{Zhang2016Orbital}
W.~{Zhang}, S.~{Zheng}, Y.~{Chen}, X.~{Jin}, H.~{Chi}, and X.~{Zhang},
  ``Orbital angular momentum-based communications with partial arc sampling
  receiving,'' \emph{IEEE Commun. Lett.}, vol.~20, no.~7, pp. 1381--1384, July
  2016.

\bibitem{Bremer2019}
J.~Bremer, ``An algorithm for the rapid numerical evaluation of {Bessel}
  functions of real orders and arguments,'' \emph{Adv. Comput. Math.}, vol.~45,
  no.~1, pp. 173--211, 2019.

\end{thebibliography}
\begin{IEEEbiography}[{\includegraphics[width=1in,height=1.25in,clip,keepaspectratio]{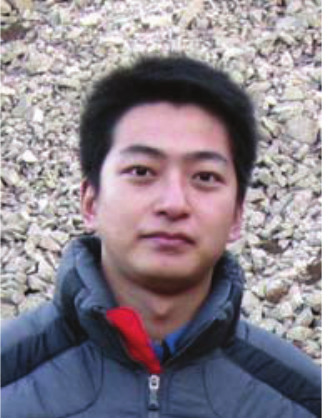}}]{Rui Chen}
(S'08-M'11) received the B.S., M.S. and Ph.D. degrees in Communications and Information Systems from Xidian University, Xi'an, China, in 2005, 2007 and 2011, respectively. From 2014 to 2015, he was a visiting scholar at Columbia University in the City of New York. He is currently an associate professor and Ph.D. supervisor in the school of Telecommunications Engineering at Xidian University. He has published about 50 papers in international journals and conferences and held 10 patents. He is an Associate Editor for International Journal of Electronics, Communications, and Measurement Engineering (IGI Global). His research interests include broadband wireless communication systems, array signal processing and intelligent transportation systems.
\end{IEEEbiography}

\begin{IEEEbiography}[{\includegraphics[width=1in,height=1.25in,clip,keepaspectratio]{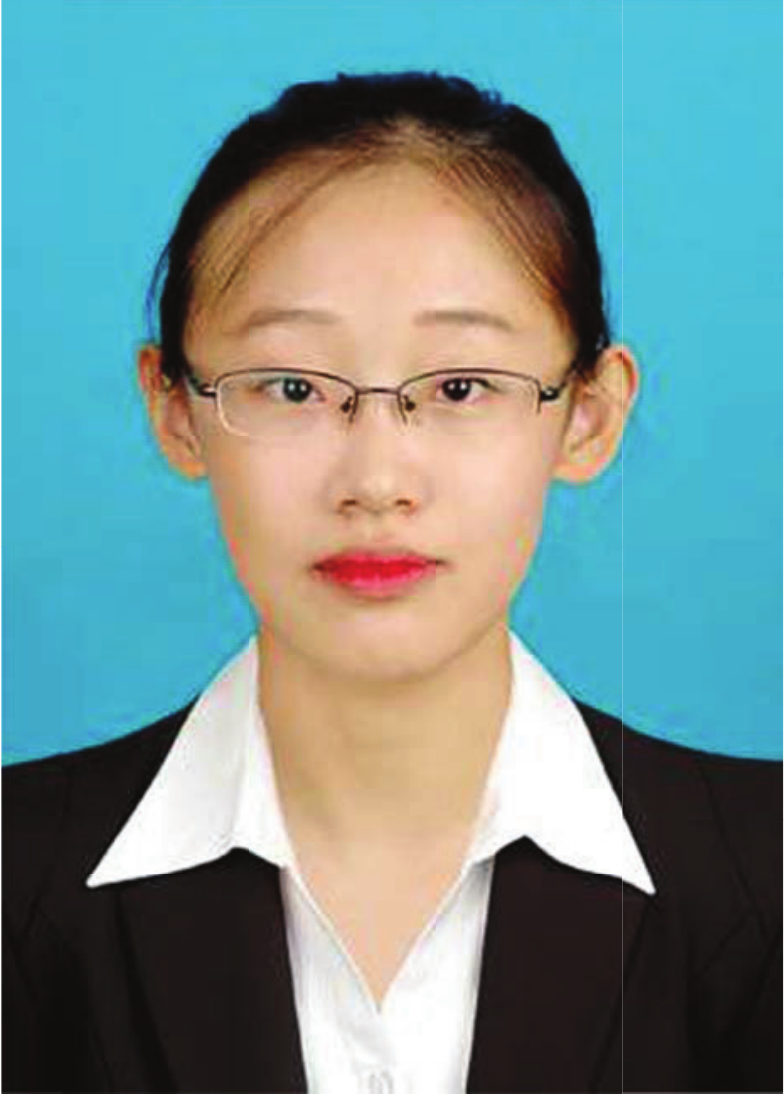}}]{Wen-Xuan Long}
(S'18) received the B.S. degree (with Highest Hons.) in Rail Transit Signal and Control from Dalian Jiaotong University, Dalian, China in 2017. She is currently pursuing a double Ph.D. degree in Communications and Information Systems at Xidian University, China and University of Pisa, Italy. Her research interests include broadband wireless communication systems and array signal processing.
\end{IEEEbiography}

\begin{IEEEbiography}[{\includegraphics[width=1in,height=1.25in,clip,keepaspectratio]{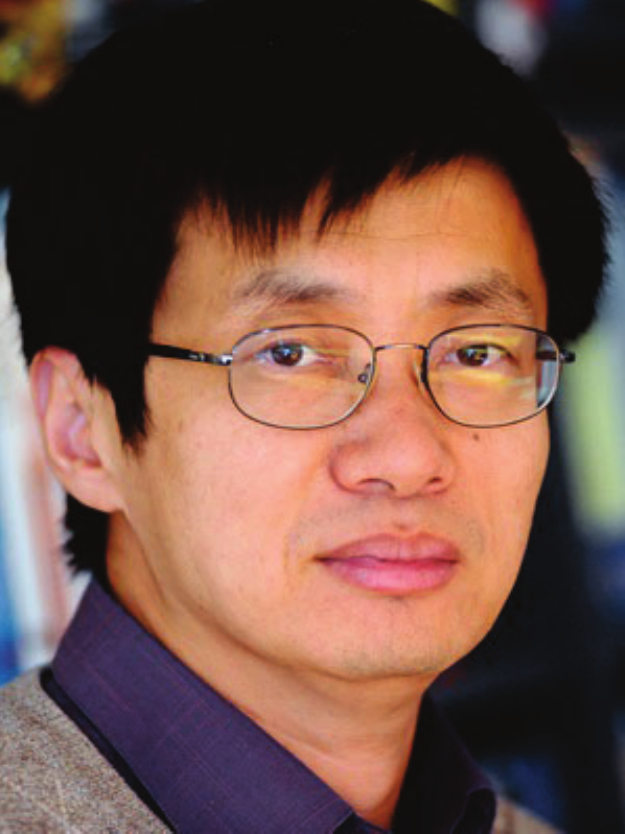}}]{Xiaodong Wang}
(S'98-M'98-SM'04-F'08) received the Ph.D. degree in electrical engineering from Princeton University. He is currently a Professor of electrical engineering with Columbia University, New York, NY, USA. He has authored the book Wireless Communication Systems: Advanced Techniques for Signal Reception (Prentice-Hall, 2003). His research interests include computing, signal processing, and communications, and has published extensively in these areas. His current research interests include wireless communications, statistical signal processing, and genomic signal processing. He was a recipient of the 1999 NSF CAREER Award, the 2001 IEEE Communications Society and Information Theory Society Joint Paper Award, and the 2011 IEEE Communication Society Award for Outstanding Paper on New Communication Topics. He was an Associate Editor for the IEEE TRANSACTIONS ON COMMUNICATIONS, the IEEE TRANSACTIONS ON WIRELESS COMMUNICATIONS, the IEEE TRANSACTIONS ON SIGNAL PROCESSING, and the IEEE TRANSACTIONS ON INFORMATION THEORY. He is listed as an ISI highly cited author.
\end{IEEEbiography}

\begin{IEEEbiography}[{\includegraphics[width=1in,height=1.25in,clip,keepaspectratio]{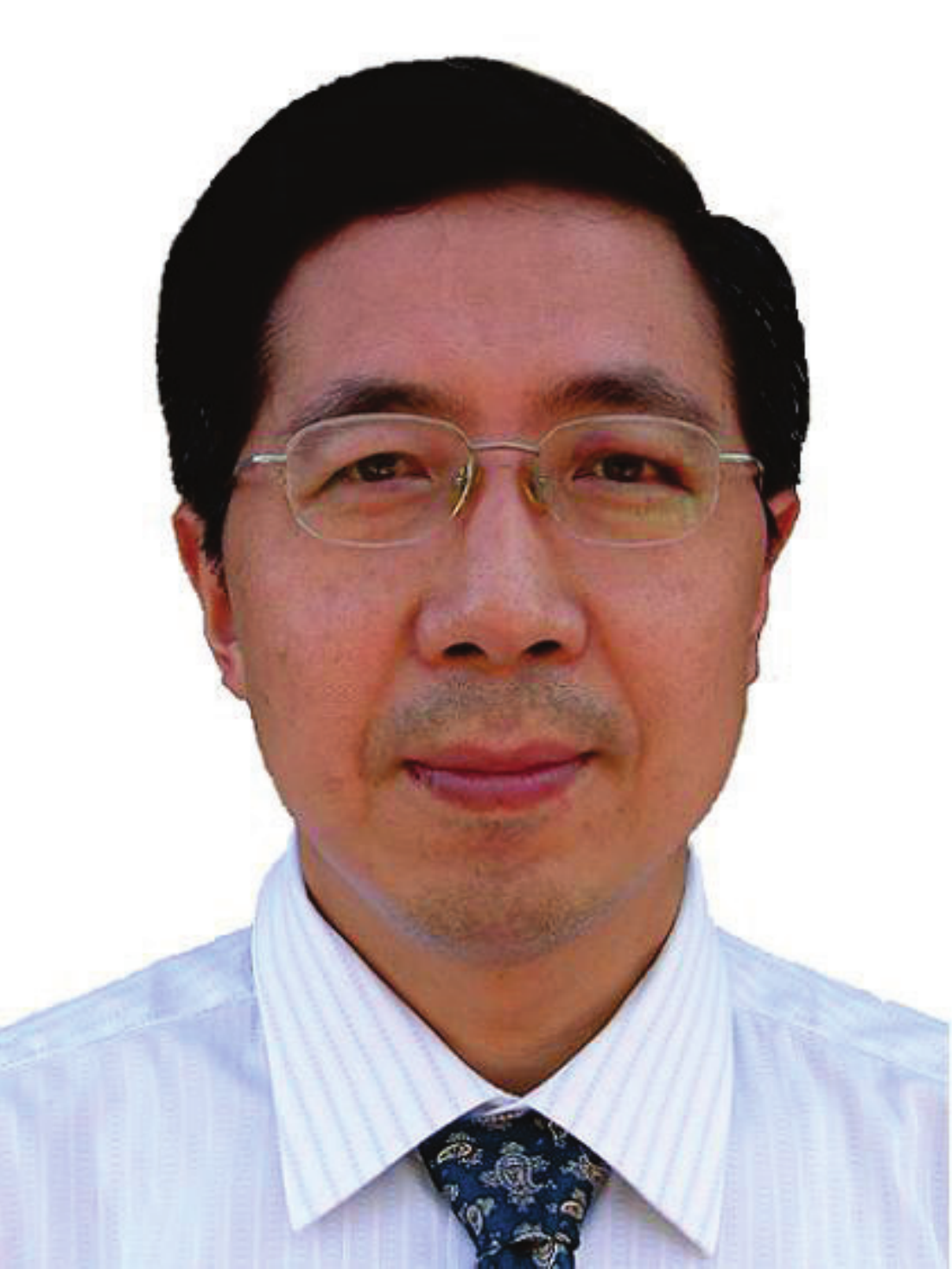}}]{Jiandong Li}
(SM'05) received the B.E., M.S., and Ph.D. degrees in communications engineering from
Xidian University, Xi’an, China, in 1982, 1985, and 1991, respectively. He was a Visiting Professor with the Department of Electrical and Computer Engineering, Cornell University, from 2002 to 2003. He has been a faculty member of the School of Telecommunications Engineering, Xidian University, since 1985, where he is currently a Professor and the Vice Director of the Academic Committee, State Key Laboratory of Integrated Service Networks. His major research interests include wireless communication theory, cognitive radio, and signal processing. He was recognized as a Distinguished Young Researcher by NSFC and a Changjiang Scholar by the Ministry of Education, China, respectively. He served as the General Vice Chair of ChinaCom 2009 and the TPC Chair of the IEEE ICCC 2013.
\end{IEEEbiography}

\end{document}